\documentclass{cernyrep}
\usepackage{texnames}
\usepackage[T1]{fontenc}
\iftrue
\newcommand{\divop}{\nabla \cdot}
\newcommand{\curlop}{\nabla \times}
\newcommand{\gradop}{\nabla}
\fi
\begin{document}
\title{Maxwell's equations for magnets}
 
\author{A. Wolski}

\institute{University of Liverpool, Liverpool, UK and the Cockcroft Institute, 
Daresbury, UK}

\maketitle 

\begin{abstract}
Magnetostatic fields in accelerators are conventionally described in
terms of multipoles.  We show that in two dimensions, multipole fields
do provide solutions of Maxwell's equations, and we consider the distributions of
electric currents and geometries of ferromagnetic materials required (in
idealized situations) to generate specified multipole fields.  Then,
we consider how to determine the multipole components in a given field.
Finally, we show how the two-dimensional multipole description may be
extended to three dimensions; this allows fringe fields, or the main fields
in such devices as undulators and wigglers, to be expressed in terms
of a set of modes, where each mode provides a solution to Maxwell's equations.
\end{abstract}
 
\section{Maxwell's equations}
 
Maxwell's equations may be written in differential form as follows:
\begin{eqnarray}
\divop \vec{D} & = & \rho, \label{eq:maxwell1} \\
\divop \vec{B} & = & 0, \label{eq:maxwell2} \\
\curlop \vec{H} & = & \vec{J} + \frac{\partial \vec{D}}{\partial t}, \label{eq:maxwell4} \\
\curlop \vec{E} & = & -\frac{\partial \vec{B}}{\partial t}. \label{eq:maxwell3}
\end{eqnarray}
The fields $\vec{B}$ (magnetic flux density) and $\vec{E}$ (electric
field strength) determine the force on a particle of charge $q$ travelling with
velocity $\vec{v}$ (the Lorentz force equation):
\begin{equation}
\vec{F} = q\left( \vec{E} + \vec{v} \times \vec{B} \right). \nonumber 
\end{equation}
The electric displacement $\vec{D}$ and magnetic intensity $\vec{H}$ are
related to the electric field and magnetic flux density by
\begin{eqnarray}
\vec{D} & = & \varepsilon \vec{E}, \nonumber \\
\vec{B} & = & \mu \vec{H}. \nonumber
\end{eqnarray}
The electric permittivity $\varepsilon$ and magnetic permeability $\mu$
depend on the medium within which the fields exist.  The values of these
quantities in vacuum are fundamental physical constants.  In SI units:
\begin{eqnarray}
\mu_0 & = & 4\pi \times 10^{-7}\,\textrm{Hm}^{-1}, \nonumber \\
\varepsilon_0 & = & \frac{1}{\mu_0 c^2}, \nonumber
\end{eqnarray}
where $c$ is the speed of light in vacuum.  The permittivity and permeability
of a material characterize the response of that material to electric and
magnetic fields.  In simplified models, they are often regarded as constants
for a given material; however, in reality the permittivity and permeability
can have a complicated dependence on the fields that are present.  Note that
the \emph{relative permittivity} $\varepsilon_r$ and the
\emph{relative permeability} $\mu_r$ are frequently used.  These are
dimensionless quantities, defined by
\begin{equation}
\varepsilon_r = \frac{\varepsilon}{\varepsilon_0}, \quad
\mu_r = \frac{\mu}{\mu_0}.
\end{equation}
That is, the relative permittivity is the permittivity of a material relative
to the permittivity of free space, and similarly for the relative permeability.

The quantities $\rho$ and $\vec{J}$ are respectively the electric charge
density (charge per unit volume) and electric current density
($\vec{J} \cdot \vec{n}$ is the charge crossing unit area perpendicular
to unit vector $\vec{n}$ per unit time).  Equations (\ref{eq:maxwell2})
and (\ref{eq:maxwell3}) are independent of $\rho$ and $\vec{J}$, and are
generally referred to as the `homogeneous' equations; the other two
equations, (\ref{eq:maxwell1}) and (\ref{eq:maxwell4}) are dependent
on $\rho$ and $\vec{J}$, and are generally referred to as the
`inhomogeneous' equations.  The charge density and current density
may be regarded as \emph{sources} of electromagnetic fields.  When the
charge density and current density are specified (as functions of space,
and, generally, time), one can integrate Maxwell's equations
(\ref{eq:maxwell1})--(\ref{eq:maxwell4}) to find possible electric and
magnetic fields in the system.  Usually, however, the solution one finds
by integration is not unique: for example, the field within an accelerator
dipole magnet may be modified by propagating an electromagnetic wave
through the magnet.  However, by imposing certain constraints (for
example, that the fields within a magnet are independent of time) it
is possible to obtain a unique solution for the fields in a given system
of electric charges and currents.

Most realistic situations are sufficiently complicated that solutions
to Maxwell's equations cannot be obtained analytically.  A variety of
computer codes exist to provide solutions numerically, once the charges,
currents, and properties of the materials present are all specified,
see, for example, Refs. \cite{bib:vectorfields, bib:cst, bib:radia}.
Solving for the fields in realistic (three-dimensional) systems often
requires a reasonable amount of computing power; some sophisticated
techniques have been developed for solving Maxwell's equations
numerically with good efficiency \cite{bib:russenschuck}.  We do not
consider such techniques here, but focus instead on the analytical
solutions that may be obtained in idealized situations.  Although the
solutions in such cases may not be sufficiently accurate to complete
the design of a real accelerator magnet, the analytical solutions do
provide a useful basis for describing the fields in real magnets, and
provide also some important connections with the beam dynamics in an
accelerator.

An important feature of Maxwell's equations is that, for systems containing
materials with constant permittivity and permeability (i.e., permittivity and
permeability that are independent of the fields present), the equations are
\emph{linear} in the fields and sources.  That is, each term in the equations 
involves a field or a source to (at most) the first power, and products of
fields or sources do not appear.  As a consequence, the \emph{principle of
superposition} applies: if $\vec{B}_1$ and $\vec{B}_2$ are solutions
of Maxwell's equations with the current densities $\vec{J}_1$ and
$\vec{J}_2$, then the field $\vec{B}_T = \vec{B}_1 + \vec{B}_2$
will be a solution of Maxwell's equations, with the source given by
the total current density $\vec{J}_T = \vec{J}_1 + \vec{J}_2$.
This means that it is possible to represent complicated fields as
superpositions of simpler fields.  An important and widely used
analysis technique for accelerator magnets is to decompose
the field (determined from either a magnetic model, or from measurements
of the field in an actual magnet) into a set of multipoles.  While it
is often the ideal to produce a field consisting of a single multipole
component, this is never perfectly achieved in practice: the multipole
decomposition indicates the extent to which components other than the
`desired' multipole are present.  Multipole decompositions also
produce useful information for modelling the beam dynamics.  Although
the principle of superposition strictly only applies in systems where
the permittivity and permeability are independent of the fields, it is
always possible to perform a multipole decomposition of the fields in
free space (e.g., in the interior of a vacuum chamber), since in that
region the permittivity and permeability are constants.  However, it should
be remembered that for nonlinear materials (where the permeability, for
example, depends on the magnetic field strength), the field inside the
material comprising the magnet will not necessarily be that expected if
one were simply to add together the fields corresponding to the multipole
components.

Solutions to Maxwell's equations lead to a rich diversity of
phenomena, including the fields around charges and currents in certain
simple configurations, and the generation, transmission and absorption of
electromagnetic radiation.  Many existing texts cover these phenomena
in detail; see, for example, the authoritative text by Jackson
\cite{bib:jackson}.  Therefore, we consider only briefly the electric field around
a point charge and the magnetic field around a long straight wire carrying
a uniform current: our main purpose here is to remind the reader of two
important integral theorems (Gauss's theorem, and Stokes's theorem), of which
we shall make use later. In the following sections, we discuss
analytical solutions to Maxwell's equations for situations relevant to
some of the types of magnets commonly used in accelerators.  These include
multipoles (dipoles, quadrupoles, sextupoles, and so on), solenoids, and
insertion devices (undulators and wigglers).  We consider only static fields.
We begin with two-dimensional fields, that is fields that are
independent of one coordinate (generally, the coordinate representing
the direction of motion of the beam).  We will show that multipole
fields are indeed solutions of Maxwell's equations, and we will
derive the current distributions needed to generate `pure' multipole
fields.  We then discuss multipole decompositions, and compare
techniques for determining the multipole components present in a given
field from numerical field data (from a model, or from measurements).
Finally, we consider how the two-dimensional multipole decomposition
may be extended to three-dimensional fields, to include (for example)
insertion devices, and fringe fields in multipole magnets.

\section{Integral theorems and the physical interpretation of Maxwell's equations
\label{sec:integraltheorems}}

\subsection{Gauss's theorem and Coulomb's law}

Gauss's theorem states that for any smooth vector field $\vec{a}$:
\begin{equation}
\int_V \divop \vec{a} \,dV = \int_{\partial V} \vec{a} \cdot d\vec{S}, \nonumber
\end{equation}
where $V$ is a volume bounded by the closed surface $\partial V$.  Note that
the area element $d\vec{S}$ is oriented to point \emph{out} of $V$.

Gauss's theorem is helpful for obtaining physical interpretations of two
of Maxwell's equations, (\ref{eq:maxwell1}) and (\ref{eq:maxwell2}).
First, applying Gauss's theorem to (\ref{eq:maxwell1}) gives:
\begin{equation}
\int_V \divop \vec{D} \,dV = \int_{\partial V} \vec{D} \cdot d\vec{S} = q,
\label{eq:coulomb1}
\end{equation}
where $q = \int_V \rho \, dV$ is the total charge enclosed by $\partial V$.

\begin{figure}[t]
\centering
\includegraphics[width=0.7\linewidth]{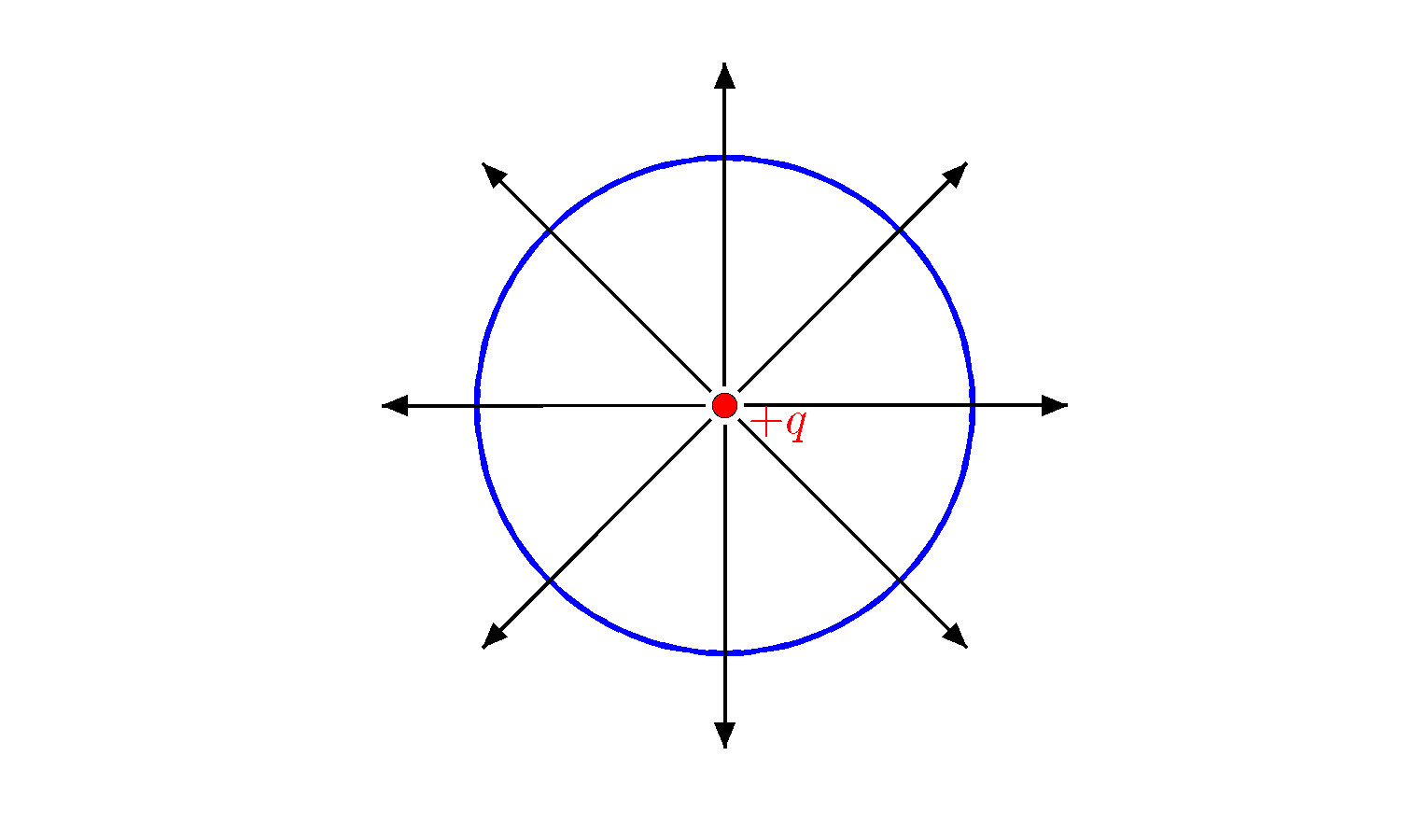}
\caption{Electric field lines from a point charge $q$.  The field lines are
everywhere perpendicular to a spherical surface centred on the charge.
\label{fig:coulombslaw}}
\end{figure}

Suppose that we have a single isolated point charge in an homogeneous, isotropic
medium with constant permittivity $\varepsilon$.  In this case, it
is interesting to take $\partial V$ to be a sphere of radius $r$.  By symmetry,
the magnitude of the electric field must be the same at all points on $\partial V$,
and must be normal to the surface at each point.  Then, we can perform the
surface integral in (\ref{eq:coulomb1}):
\begin{equation}
\int_{\partial V} \vec{D} \cdot d\vec{S} = 4\pi r^2 D. \nonumber
\end{equation}
This is illustrated in Fig.~\ref{fig:coulombslaw}: the outer circle represents a
cross-section of a sphere ($\partial V$) enclosing volume $V$, with the charge $q$
at its centre.  The black arrows in Fig.~\ref{fig:coulombslaw} represent the
electric field lines, which are everywhere perpendicular to the surface $\partial V$.
Since $\vec{D} = \varepsilon \vec{E}$, we
find Coulomb's law for the magnitude of the electric field around a point
charge:
\begin{equation}
E = \frac{q}{4\pi \varepsilon r^2}. \nonumber
\end{equation}

Applied to Maxwell's equation (\ref{eq:maxwell2}), Gauss' theorem leads to:
\begin{equation}
\int_V \divop \vec{B}\, dV = \int_{\partial V} \vec{B}\cdot d\vec{S} = 0. \nonumber
\end{equation}
In other words, the magnetic flux integrated over any closed surface must
equal zero---at least, until we discover magnetic monopoles.  Lines of
magnetic flux occur in closed loops; whereas lines of electric field
can start (and end) on electric charges.

\subsection{Stokes's theorem and Amp\`ere's law}

\begin{figure}[t]
\centering
\includegraphics[width=0.7\linewidth]{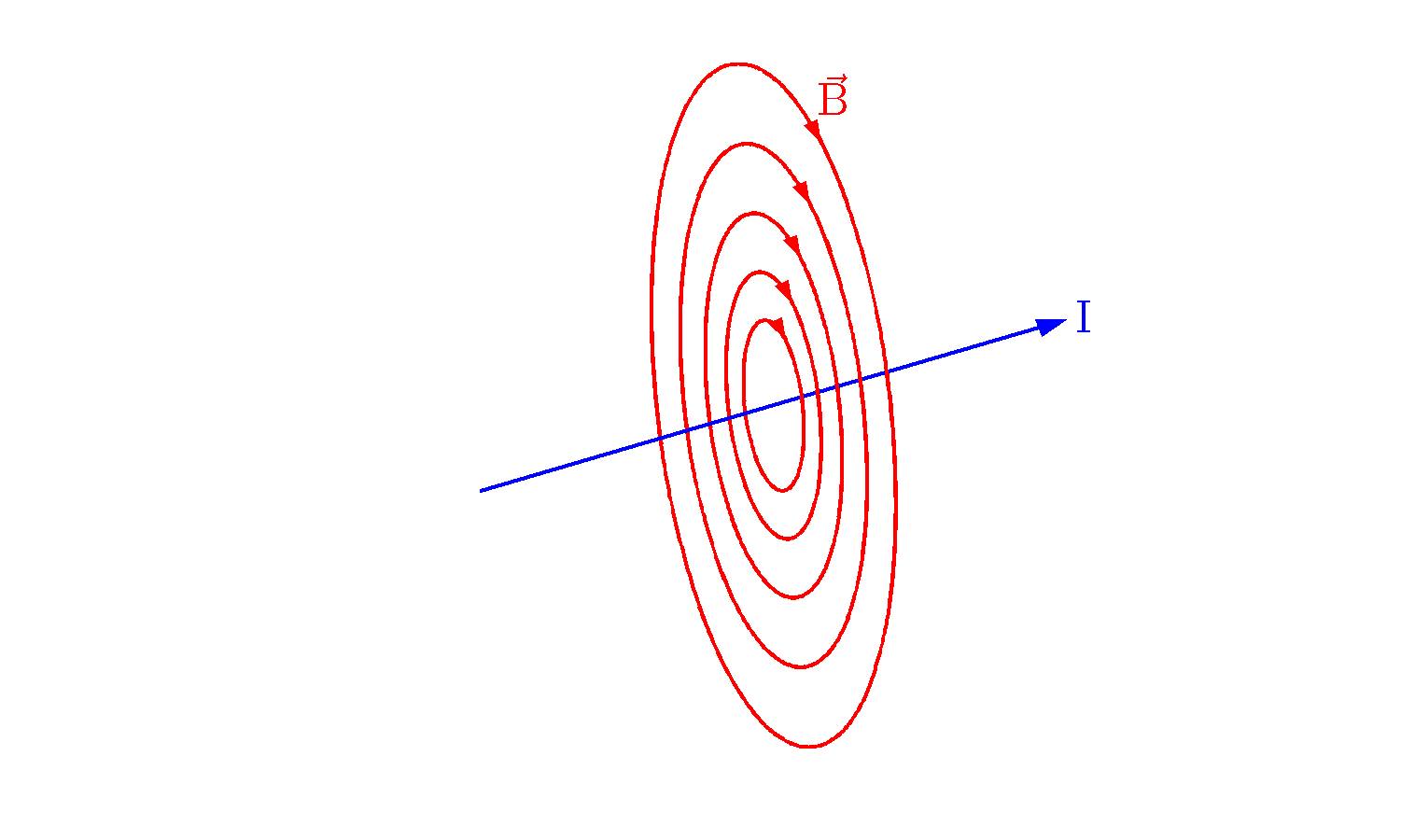}
\caption{Magnetic field lines around a long straight wire carrying a current $I$
\label{fig:ampereslaw}}
\end{figure}

Stokes's theorem states that for any smooth vector field $\vec{a}$:
\begin{equation}
\int_S \curlop \vec{a} \cdot d\vec{S} = \int_{\partial S} \vec{a} \cdot d\vec{l},
\label{eq:stokestheorem}
\end{equation}
where the loop $\partial S$ bounds the surface $S$.  Applied to Maxwell's
equation (\ref{eq:maxwell4}), Stokes's theorem leads to
\begin{equation}
\int_{\partial S} \vec{H} \cdot d\vec{l} = \int_S \vec{J} \cdot d\vec{S},
\label{eq:ampere1}
\end{equation}
which is Amp\`ere's law.  From Amp\`ere's law, we can derive an expression for
the strength of the magnetic field around a long, straight wire carrying
current $I$.  The magnetic field must have rotational symmetry around the wire.
There are two possibilities: a radial field, or a field consisting of closed
concentric loops centred on the wire (or some superposition of these fields).
A radial field would violate Maxwell's equation (\ref{eq:maxwell2}).  Therefore,
the field must consist of closed concentric loops; and by considering a
circular loop of radius $r$, we can perform the integral in Eq.~(\ref{eq:ampere1}):
\begin{equation}
2\pi r H = I, \nonumber
\end{equation}
where $I$ is the total current carried by the wire.  In this case, the line integral
is performed around a loop $\partial S$ centred on the wire, and in a plane
perpendicular to the wire: essentially, this corresponds to one of the magnetic
field lines, see Fig.~\ref{fig:ampereslaw}.  The total current passing through
the surface $S$ bounded by the loop $\partial S$ is simply the total current $I$.

In an homogeneous, isotropic
medium with constant permeability $\mu$, $\vec{B} = \mu_0 \vec{H}$, we obtain
the expression for the magnetic flux density at distance $r$ from the wire:
\begin{equation}
B = \frac{\mu I}{2\pi r}. \label{eq:ampere2}
\end{equation}
This result will be useful when we come to consider how to generate specified
multipole fields from current distributions.

Finally, applying Stokes's theorem to the homogeneous Maxwell's equation
(\ref{eq:maxwell3}), we find
\begin{equation}
\int_{\partial S} \vec{E} \cdot d\vec{l} = -\frac{\partial}{\partial t}
\int_S \vec{B} \cdot d\vec{S}. \label{eq:faraday1}
\end{equation}
Defining the electromotive force $\mathscr{E}$ as the integral of the
electric field around a closed loop, and the magnetic flux $\Phi$ as the
integral of the magnetic flux density over the surface bounded by the loop,
Eq.~(\ref{eq:faraday1})
gives
\begin{equation}
\mathscr{E} = -\frac{\partial \Phi}{\partial t},
\end{equation}
which is Faraday's law of electromagnetic induction.  Faraday's law is
significant for magnets with time-dependent fields, such as pulsed
magnets (used for injection and extraction), and magnets that are
`ramped' (for example, when changing the beam energy in a storage
ring).  The change in magnetic field will induce a voltage across the
coil of the magnet that must be taken into account when designing the
power supply.  Also, the induced voltages can induce eddy currents in
the core of the magnet, or in the coils themselves, leading to heating.
This is an issue for superconducting magnets, which must be ramped
slowly to avoid quenching \cite{bib:bottura}.

\subsection{Boundary conditions}

Gauss's theorem and Stokes's theorem can be applied to Maxwell's equations
to derive constraints on the behaviour of electromagnetic fields at
boundaries between different materials.  Here, we shall focus on the
boundary conditions on the magnetic field: these conditions will be
useful when we consider multipole fields in iron-dominated magnets.

\begin{figure}[t]
\centering
\includegraphics[width=0.4\linewidth]{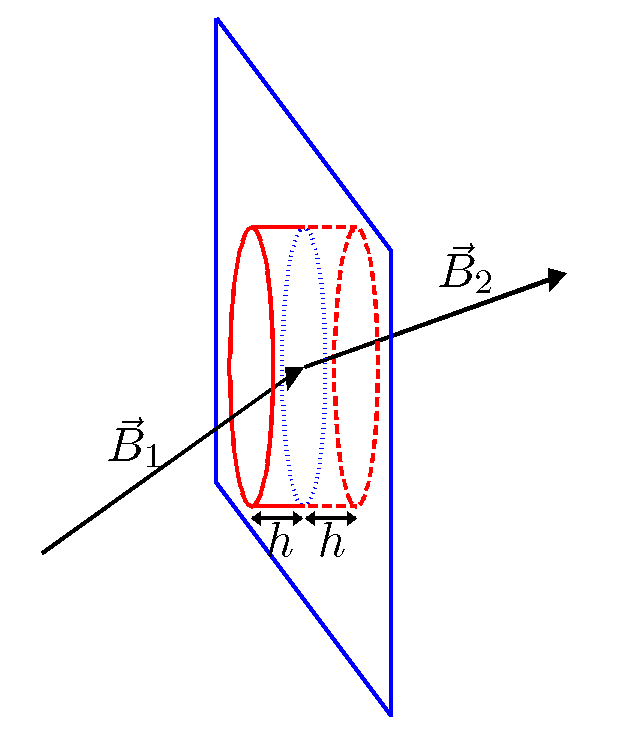}
\hspace{0.06\linewidth}
\includegraphics[width=0.4\linewidth]{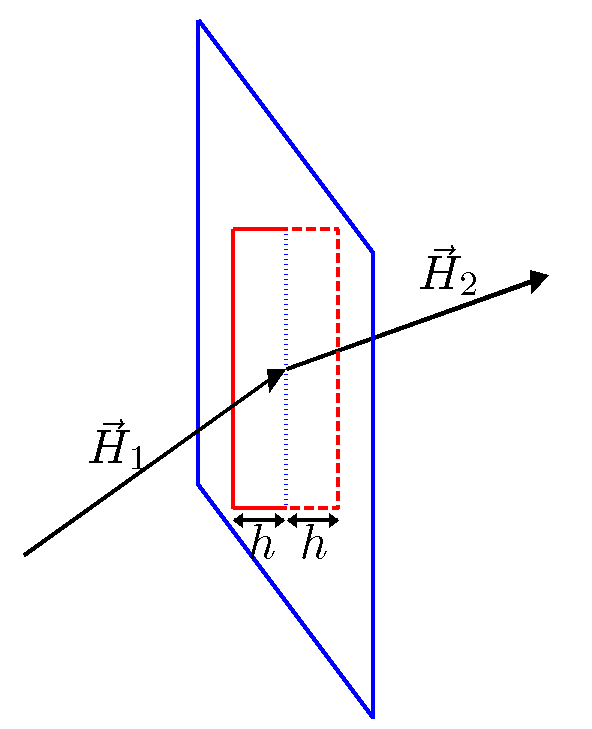}
\caption{(a) Left: `pill box' surface for derivation of the boundary
conditions on the normal component of the magnetic flux density at the
interface between two media.  (b) Right: geometry for derivation of the
boundary conditions on the tangential component of the magnetic
intensity at the interface between two media.
\label{fig:boundaryconditionb}}
\end{figure}

Consider first a short cylinder or `pill box' that crosses the boundary
between two media, with the flat ends of the cylinder parallel to the
boundary, see Fig.~\ref{fig:boundaryconditionb}\,(a).  Applying Gauss's
theorem to Maxwell's equation (\ref{eq:maxwell2}) gives
\begin{equation}
\int_V \divop \vec{B}\, dV = \int_{\partial V} \vec{B} \cdot d\vec{S} = 0,
\nonumber
\end{equation}
where the boundary $\partial V$ encloses the volume $V$ within the cylinder.
If we take the limit where the length of the cylinder ($2h$) approaches zero,
then the only contributions to the surface integral come from the flat ends;
if these have infinitesimal area $dS$, then since the orientations of these
surfaces are in opposite directions on opposite sides of the boundary, and
parallel to the normal component of the magnetic field, we find
\begin{equation}
- B_{1\perp}\, dS + B_{2\perp}\, dS = 0,
\nonumber
\end{equation}
where $B_{1\perp}$ and $B_{2\perp}$ are the normal components of the
magnetic flux density on either side of the boundary.  Hence
\begin{equation}
B_{1\perp} = B_{2\perp}.
\label{eq:boundarynormalb}
\end{equation}
In other words, the normal component of the magnetic flux density is
continuous across a boundary.

A second boundary condition, this time on the component of the magnetic
field parallel to a boundary, can be obtained by applying Stokes's theorem
to Maxwell's equation (\ref{eq:maxwell4}).
In particular, we consider a surface $S$ bounded by a loop $\partial S$ that
crosses the boundary of the material, see
Fig.~\ref{fig:boundaryconditionb}\,(b).  If we integrate both sides of 
Eq.~(\ref{eq:maxwell4}) over that surface, and apply Stokes's theorem 
(\ref{eq:stokestheorem}), we find
\begin{equation}
\int_S \curlop \vec{H} \cdot d\vec{S}
 = \int_{\partial S} \vec{H} \cdot d\vec{l}
 = I + \frac{\partial}{\partial t} \int_S \vec{D} \cdot d\vec{S},
\nonumber
\end{equation}
where $I$ is the total current flowing through the surface $S$.
Now, let the surface $S$ take the form of a thin strip, with the short
ends perpendicular to the boundary, and the long ends parallel to the
boundary.  In the limit that the length of the short ends goes to
zero, the area of $S$ goes to zero: both the current flowing
through the surface $S$, and the electric displacement integrated
over $S$ become zero.  However, there are still contributions to
the integral of $\vec{H}$ around $\partial S$ from the long sides
of the strip.  Thus we find that
\begin{equation}
H_{1\parallel} = H_{2\parallel}, \label{eq:boundaryparallelh}
\end{equation}
where $H_{1\parallel}$ is the component of the magnetic intensity
parallel to the boundary at a point on one side of the boundary,
and $H_{2\parallel}$ is the component of the magnetic intensity
parallel to the boundary at a nearby point on the other side of the
boundary.  In other words, the \emph{tangential} component of the
magnetic intensity $\vec{H}$ is continuous across a boundary.

We can derive a stronger constraint on the magnetic field at a
boundary in the case where the material on one side of the boundary
has infinite permeability (which can provide a reasonable model for
some ferromagnetic materials).  Since $\vec{B} = \mu \vec{H}$, it
follows from (\ref{eq:boundaryparallelh}) that
\begin{equation}
\frac{B_{1\parallel}}{\mu_1} = \frac{B_{2\parallel}}{\mu_2}, \nonumber
\end{equation}
and in the limit $\mu_2 \to \infty$, while $\mu_1$ remains finite,
we must have
\begin{equation}
B_{1\parallel} = 0. \label{eq:boundaryparallelb}
\end{equation}
In other words, the magnetic flux density at the surface of a material
of infinite permeability must be perpendicular to that surface.  Of
course, the permeability of a material characterizes its response to
an applied external magnetic field: in the case where the permeability
is infinite, a material placed in an external magnetic field
acquires a magnetization that exactly cancels any component of the
external field at the surface of the material.

\section{Two-dimensional multipole fields}

Consider a region of space free of charges and currents; for example, the
interior of an accelerator vacuum chamber (at least, in an ideal case, and
when the beam is not present).  If we further exclude propagating electromagnetic
waves, then any magnetic field generated by steady currents outside the vacuum
chamber must satisfy
\begin{eqnarray}
\divop \vec{B} & = & 0, \label{eq:magnetostatic1} \\
\curlop \vec{B} & = & 0. \label{eq:magnetostatic2}
\end{eqnarray}
Equation~(\ref{eq:magnetostatic1}) is just Maxwell's equation (\ref{eq:maxwell2}),
and Eq.~(\ref{eq:magnetostatic2}) follows from Maxwell's equation
(\ref{eq:maxwell4}) given that $\vec{J} = 0$, $\vec{B} = \mu_0 \vec{H}$, and
derivatives with respect to time vanish.

We shall show that a magnetic field $\vec{B} = (B_x, B_y, B_z)$ with $B_z$
constant, and $B_x$, $B_y$ given by
\begin{equation}
B_y + i B_x = C_n (x + iy)^{n-1} \label{eq:multipolefield}
\end{equation}
where $i = \sqrt{-1}$ and $C_n$ is a (complex) constant, satisfies
Eqs.~(\ref{eq:magnetostatic1}) and (\ref{eq:magnetostatic2}).  Note that the
field components $B_x$ and $B_y$ are real, and are obtained from the imaginary
and real parts of the right-hand side of Eq.~(\ref{eq:multipolefield}).
To show that the above field satisfies Eqs.~(\ref{eq:magnetostatic1}) and (\ref{eq:magnetostatic2}), we apply the differential operator
\begin{equation}
\frac{\partial}{\partial x} + i \frac{\partial}{\partial y} \label{eq:diffop}
\end{equation}
to each side of Eq.~(\ref{eq:multipolefield}).  Applied to the left-hand side,
we find
\begin{equation}
\left( \frac{\partial}{\partial x} + i \frac{\partial}{\partial y} \right)
\left( B_y + i B_x \right) =
\left( \frac{\partial B_y}{\partial x} - \frac{\partial B_x}{\partial y} \right)
+ i \left( \frac{\partial B_x}{\partial x} + \frac{\partial B_y}{\partial y} \right).
\label{eq:diffmultipolelhs}
\end{equation}
Applied to the right-hand side of Eq.~(\ref{eq:multipolefield}), the differential
operator (\ref{eq:diffop}) gives
\begin{equation}
\left( \frac{\partial}{\partial x} + i \frac{\partial}{\partial y} \right)
C_n (x + iy)^{n-1} =
C_n (n-1)(x + iy)^{n-2} + i^2 C_n (n-1)(x + iy)^{n-2} = 0.
\label{eq:diffmultipolerhs}
\end{equation}
Combining Eqs.~(\ref{eq:multipolefield}), (\ref{eq:diffmultipolelhs}) and
(\ref{eq:diffmultipolerhs}), we find
\begin{eqnarray}
\frac{\partial B_x}{\partial x} + \frac{\partial B_y}{\partial y} & = & 0, \nonumber \\
\frac{\partial B_y}{\partial x} - \frac{\partial B_x}{\partial y} & = & 0. \nonumber
\end{eqnarray}
Finally, we note that $B_z$ is constant, so any derivatives of $B_z$
vanish; furthermore, $B_x$ and $B_y$ are independent of $z$, so any derivatives
of these coordinates with respect to $z$ vanish.  Thus we conclude that for
the field (\ref{eq:multipolefield})
\begin{eqnarray}
\divop \vec{B} & = & 0, \label{eq:divbzero} \\
\curlop \vec{B} & = & 0, \label{eq:curlbzero}
\end{eqnarray}
and that this field is therefore a solution to Maxwell's equations within
the vacuum chamber.  Of course, this analysis tells us only that the field
is a \emph{possible} physical field: it does not tell us how to generate
such a field.  The problem of generating a field of the form
Eq.~(\ref{eq:multipolefield}) we shall consider in
Section~\ref{sec:generatingmultipoles}.

Fields of the form (\ref{eq:multipolefield}) are known as \emph{multipole
fields}.  The index $n$ (an integer) indicates the \emph{order} of the
multipole: $n=1$ is a dipole field, $n=2$ is a quadrupole field, $n=3$ is
a sextupole field, and so on.  A solenoid field has $C_n = 0$ for all $n$,
and $B_z$ non-zero; usually, a solenoid field is not considered a multipole
field, and we assume (unless stated otherwise) that $B_z = 0$ in a multipole
magnet.  Note that we can apply the principle of superposition to deduce
that a more general magnetic field can be constructed by adding together
a set of multipole fields:
\begin{equation}
B_y + i B_x = \sum_{n=1}^\infty C_n (x + iy)^{n-1}. \label{eq:multipolesum}
\end{equation}
A `pure' multipole field of order $n$ has $C_n \neq 0$ for only that one
value of $n$.

\begin{figure}
\centering
\begin{tabular}{cc}
\includegraphics[width=0.4\linewidth]{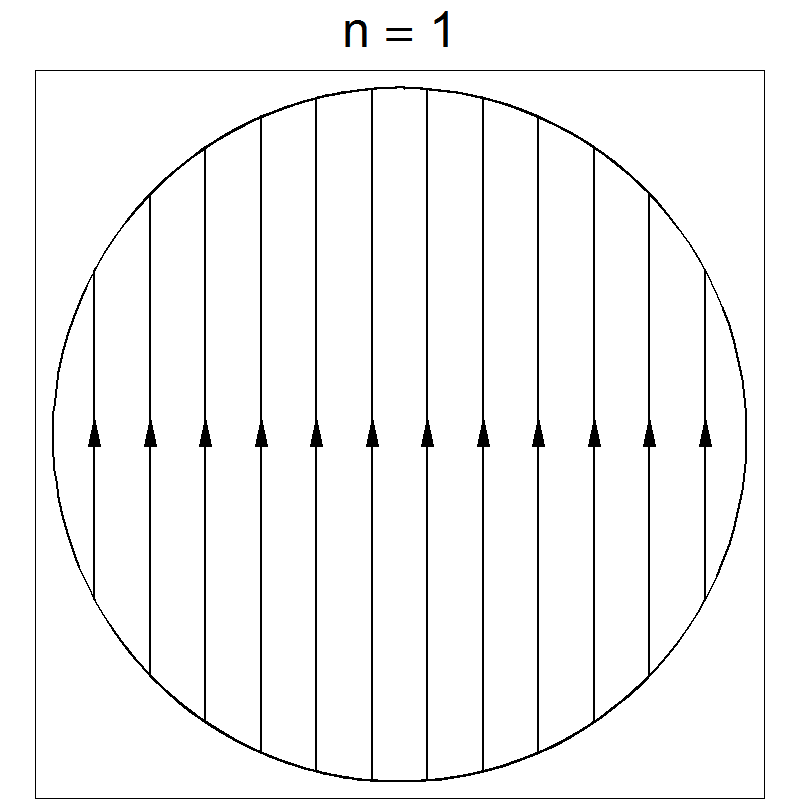} & 
\includegraphics[width=0.4\linewidth]{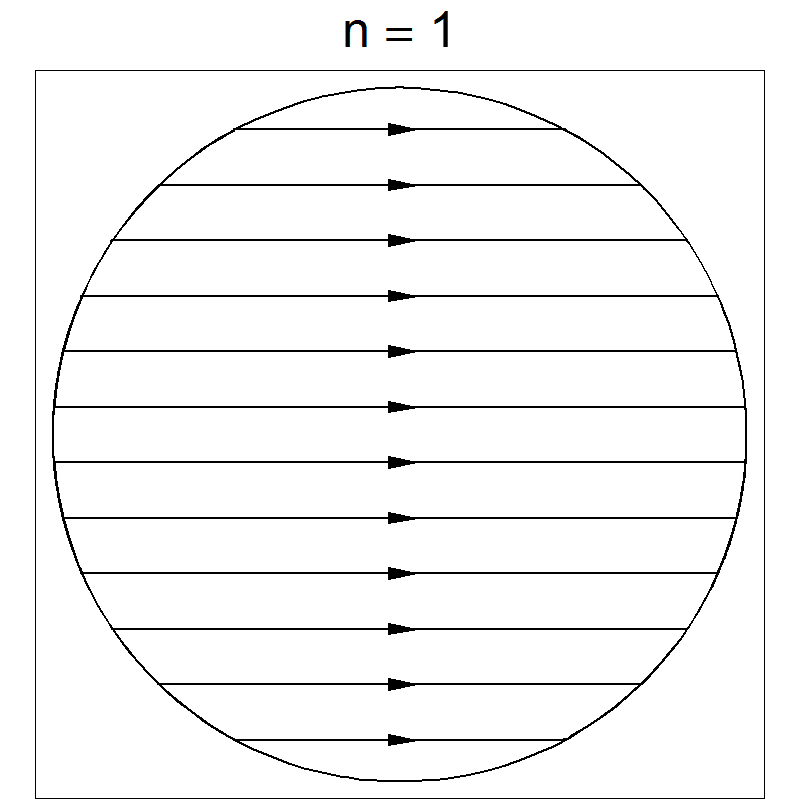} \\
\includegraphics[width=0.4\linewidth]{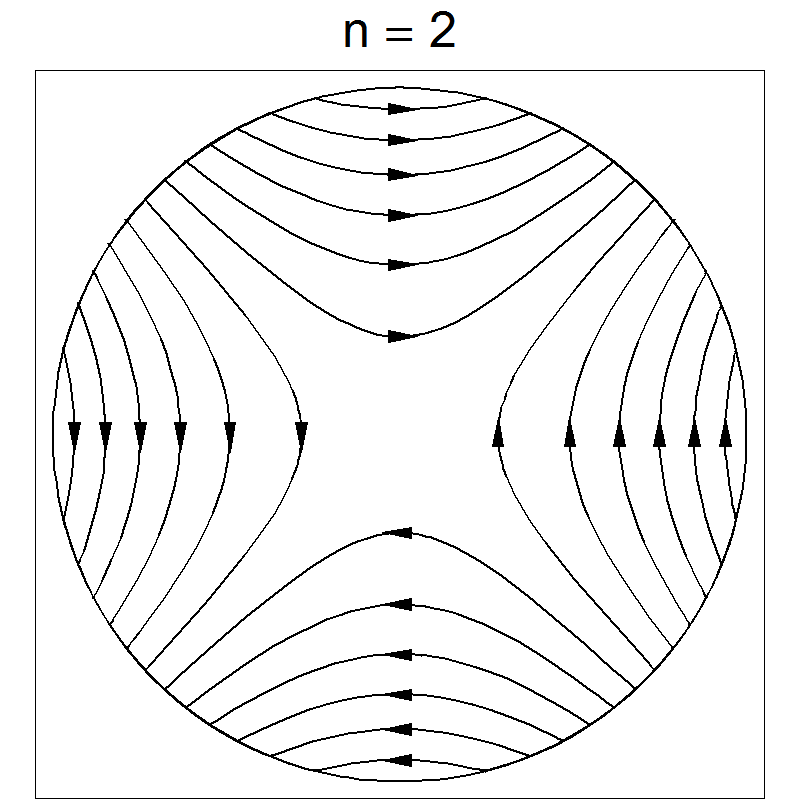} & 
\includegraphics[width=0.4\linewidth]{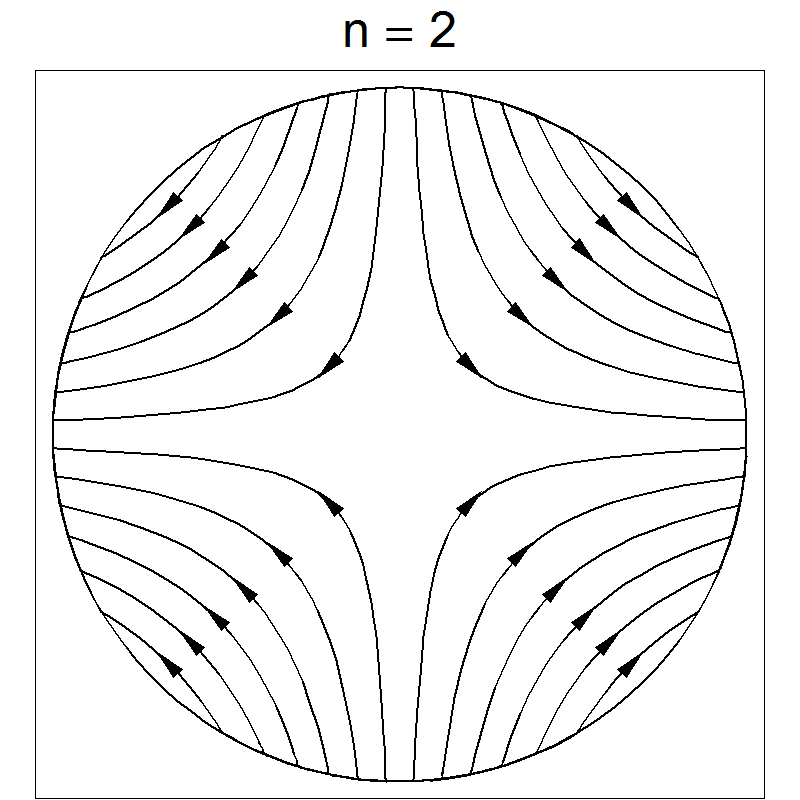} \\
\includegraphics[width=0.4\linewidth]{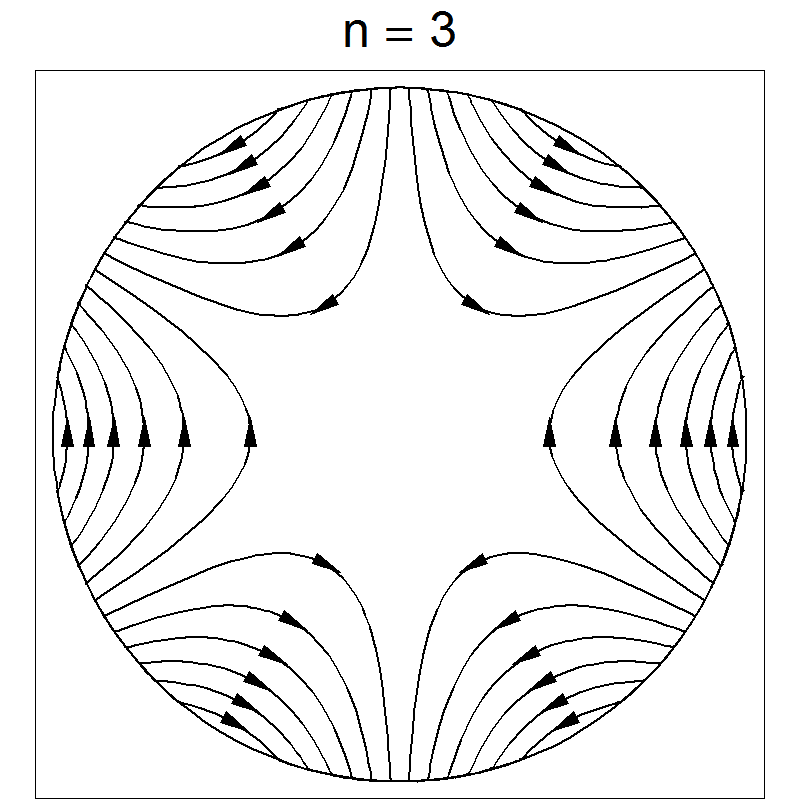} & 
\includegraphics[width=0.4\linewidth]{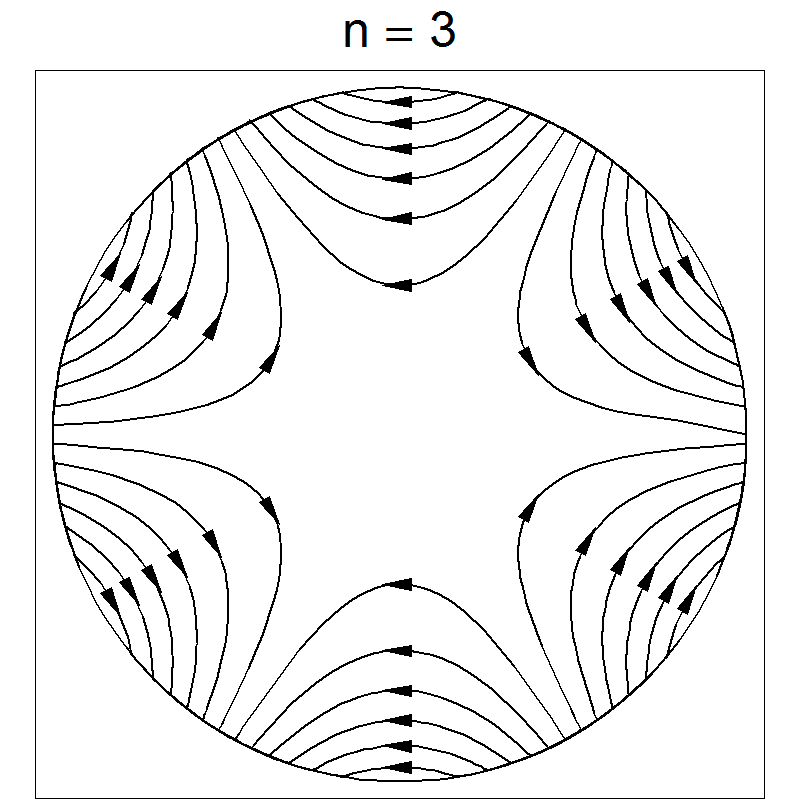}
\end{tabular}
\caption{`Pure' multipole fields.  Top: dipole.  Middle: quadrupole.
Bottom: sextupole.  Fields on the left are normal ($a_n$ positive);
those on the right are skew ($b_n$ positive).  The positive $y$ axis is
vertically up; the positive $x$ axis is horizontal and to the right.}
\label{fig:puremultipole}
\end{figure}

The coefficients $C_n$ in Eq.~(\ref{eq:multipolesum}) characterize the strength
and orientation of each multipole component in a two-dimensional magnetic
field.  It is sometimes more convenient to express the field using polar
coordinates, rather than Cartesian coordinates.  Writing $x = r \cos \theta$
and $y = r\sin \theta$, we see that Eq.~(\ref{eq:multipolesum}) becomes
\begin{equation}
B_y + i B_x = \sum_{n=1}^\infty C_n r^{n-1} e^{i(n-1)\theta}. \nonumber
\end{equation}
By writing the multipole expansion in this form, we see immediately that
the strength of the field in a pure multipole of order $n$ varies as
$r^{n-1}$ with distance from the magnetic axis.  We can go a stage further,
and express the field in terms of polar components:
\begin{equation}
B_y + i B_x = B_r \sin \theta + B_\theta \cos \theta
              + i B_r \cos \theta - i B_\theta \sin \theta
            = \left( B_\theta + i B_r \right) e^{-i\theta}, \nonumber
\end{equation}
thus:
\begin{equation}
B_\theta + i B_r = \sum_{n=1}^\infty C_n r^{n-1} e^{in\theta}.
\label{eq:multipolepolarcoords}
\end{equation}
By writing the field in this form, we see that for a pure multipole
of order $n$, rotation of the magnet through $\pi/n$ around the $z$ axis
simply changes the sign of the field.  We also see that if we write
\begin{equation}
C_n = \left| C_n \right|\, e^{in\phi_n} \nonumber
\end{equation}
then the value of $\phi_n$ (the phase of $C_n$) determines the orientation
of the field.  Conventionally, a pure multipole with $\phi_n = 0$ is known
as a `normal' multipole, while a pure multipole with $\phi_n = \pi/2n$
is known as a `skew' multipole (Fig.~\ref{fig:puremultipole}).

The units of $C_n$ depend on the order of the multipole.  In SI units,
for a dipole, the units of $C_1$ are tesla (T); for a quadrupole, the units
of $C_2$ are Tm$^{-1}$; for a sextupole, the units of $C_3$ are Tm$^{-2}$,
and so on.  It is sometimes preferred to specify multipole components in
dimensionless units.  In that case, we introduce a reference field,
$B_\textrm{ref}$, and a reference radius, $R_\textrm{ref}$.  The multipole
expansion is then written
\begin{equation}
B_y + i B_x = B_\textrm{ref} \sum_{n=1}^\infty (a_n + i b_n)
\left( \frac{x+iy}{R_\textrm{ref}} \right)^{n-1}.
\label{eq:multipoleexpansioncartesian}
\end{equation}
This is a standard notation for multipole fields, see, for
example, Ref.~\cite{bib:chaotigner}.  In polar coordinates
\begin{equation}
B_y + i B_x = B_\textrm{ref} \sum_{n=1}^\infty (a_n + i b_n)
\left( \frac{r}{R_\textrm{ref}} \right)^{n-1} e^{i(n-1)\theta}.
\label{eq:multipoleexpansionpolar}
\end{equation}
The reference field and reference radius can be chosen arbitrarily, but
must be specified if the coefficients $a_n$ and $b_n$ are to be interpreted
fully.

Note that for a pure multipole field of order $n$, the coefficients
$a_n$ and $b_n$ are related to the derivates of the field components
with respect to the $x$ and $y$ coordinates.  Thus, for a normal multipole:
\begin{equation}
\frac{\partial^{n-1}B_y}{\partial x^{n-1}} =
  (n-1)! \frac{B_\textrm{ref}}{R_\textrm{ref}^{n-1}} a_n, \nonumber
\end{equation}
and for a skew multipole:
\begin{equation}
\frac{\partial^{n-1}B_x}{\partial x^{n-1}} =
  (n-1)! \frac{B_\textrm{ref}}{R_\textrm{ref}^{n-1}} b_n. \nonumber
\end{equation}
A normal dipole has a uniform vertical field; a normal quadrupole has
a vertical field for $y=0$, that increases linearly with $x$; a
normal sextupole has a vertical field for $y=0$ that increases as the
square of $x$; and so on.

\section{Generating multipole fields \label{sec:generatingmultipoles}}

Given a system of electric charges and currents, we can integrate
Maxwell's equations to find the electric and magnetic fields generated
by those charges and currents.  In general, the integration must be
done numerically; but for simple systems it is possible to find
analytical solutions.  We considered two such cases in
Section~\ref{sec:integraltheorems}: the electric field around an
isolated point charge, and the magnetic field around a long straight
wire carrying a constant current.

It turns out that we can combine the magnetic fields from long,
straight, parallel wires to generate pure multipole fields.  It is
also possible to generate pure multipole fields using high-permeability
materials with the appropriate geometry.  We consider both methods in
this section.  For the moment, we deal with `idealized' geometries
without practical constraints.  We discuss the impact of some of the
practical limitations in later sections.

\subsection{Current distribution for a multipole field}

Our goal is to determine a current distribution that will generate
a pure multipole field of specified order.  As a first step, we
derive the multipole components in the field around a long straight
wire carrying a uniform current.  We already know from Amp\`ere's
law~(\ref{eq:ampere2}) that the field at distance $r$ from a long
straight wire carrying current $I$ in free space has magnitude given by
\begin{equation}
B = \frac{\mu_0 I}{2\pi r}, \nonumber
\end{equation}
and that the direction of the field describes a circle centred on the
wire.  To derive the multipole components in the field, we first
derive an expression for the field components at an arbitrary point
$(x,y)$ from a wire carrying current $I$, passing through a point
$(x_0,y_0)$ and parallel to the $z$ axis.

\begin{figure}[t!]
\centering
\includegraphics[width=0.6\textwidth]{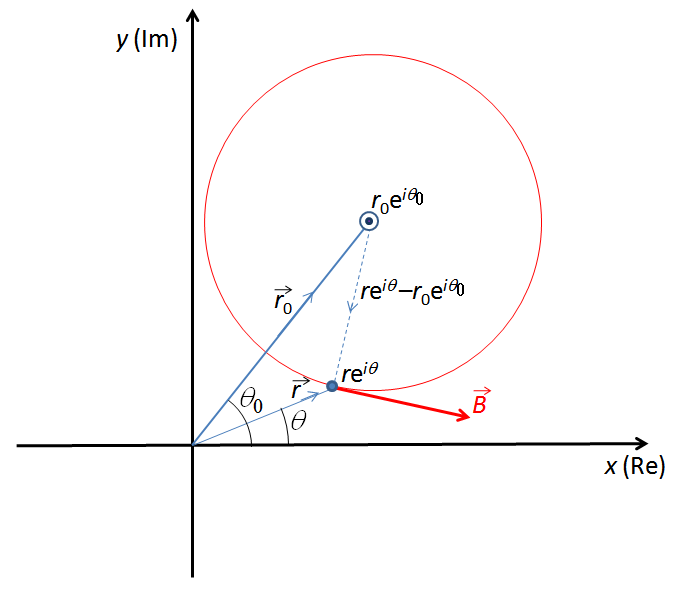}
\caption{Geometry for calculation of multipole components in the field
around a long, straight wire carrying a uniform current.  The wire
passes through $r_0 e^{i\theta_0}$, and is parallel to the $z$ axis
(the direction of the current is pointing out of the page).
\label{fig:singlewiremultipole}}
\end{figure}

Since we are working in two dimensions, we can represent the
components of a vector by the real and imaginary parts of a
complex number.  Thus, the vector from $(x_0,y_0)$ to a point
$(x,y)$ is given by $r e^{i\theta} - r_0 e^{i\theta_0}$, and the
magnitude of the field at $(x,y)$ is
\begin{equation}
B = \frac{\mu_0 I}{2\pi} \, \frac{1}{\left| r e^{i\theta} - r_0 e^{i\theta_0} \right|}.
\nonumber
\end{equation}
The geometry is shown in Fig.~\ref{fig:singlewiremultipole}.
The direction of the field is perpendicular to the line from
$(x_0,y_0)$ to $(x,y)$.  Since a rotation through 90$^\circ$ can be
represented by a multiplication by $i$, we can write
\begin{equation}
B_x + i B_y = \frac{\mu_0 I}{2\pi} \,
  \frac{i \left( r e^{i\theta} - r_0 e^{i\theta_0} \right)}
       {  \left| r e^{i\theta} - r_0 e^{i\theta_0} \right|^2},
\nonumber
\end{equation}
and hence
\begin{equation}
B_y + i B_x = \frac{\mu_0 I}{2\pi} \,
  \frac{\left( r e^{-i\theta} - r_0 e^{-i\theta_0} \right)}
       {\left| r e^{i\theta} - r_0 e^{i\theta_0} \right|^2}
       = \frac{\mu_0 I}{2\pi} \,
  \frac{1}
       {r e^{i\theta} - r_0 e^{i\theta_0}}.
\nonumber
\end{equation}

Now, we write the magnetic field as
\begin{equation}
B_y + i B_x = - \frac{\mu_0 I}{2\pi r_0} \,
  \frac{e^{-i\theta_0}}
       {1 - \frac{r}{r_0} e^{i(\theta - \theta_0)}},
\nonumber
\end{equation}
and use the Taylor series expansion for $(1 - \zeta )^{-1}$, where $\zeta$
is a complex number with $|\zeta|<1$:
\begin{equation}
\frac{1}{1 - \zeta} = \sum_{n=0}^{\infty} \zeta^n, \nonumber
\end{equation}
to write
\begin{equation}
B_y + i B_x = - \frac{\mu_0 I}{2\pi r_0} \, e^{-i\theta_0} \,
  \sum_{n=1}^\infty \left( \frac{r}{r_0} \right)^{n-1}
  e^{i(n-1)(\theta - \theta_0)}.
\label{eq:singlewiremultipolefield}
\end{equation}
Equation~(\ref{eq:singlewiremultipolefield}) is valid for $r<r_0$.
Comparing with the standard multipole expansion,
Eq.~(\ref{eq:multipoleexpansionpolar}), we see that if we choose
for the reference field $B_\textrm{ref}$ and the reference radius
$R_\textrm{ref}$
\begin{eqnarray}
B_\textrm{ref} & = & \frac{\mu_0 I}{2\pi r_0}, \nonumber \\
R_\textrm{ref} & = & r_0, \nonumber
\end{eqnarray}
then the coefficients for the multipole components in the field
are given by
\begin{equation}
b_n + i a_n = -e^{-in\theta_0}. \nonumber
\end{equation}
The field around a long straight wire can be represented as an
infinite sum over all multipoles.

\begin{figure}
\centering
\begin{tabular}{cc}
\includegraphics[width=0.4\linewidth]{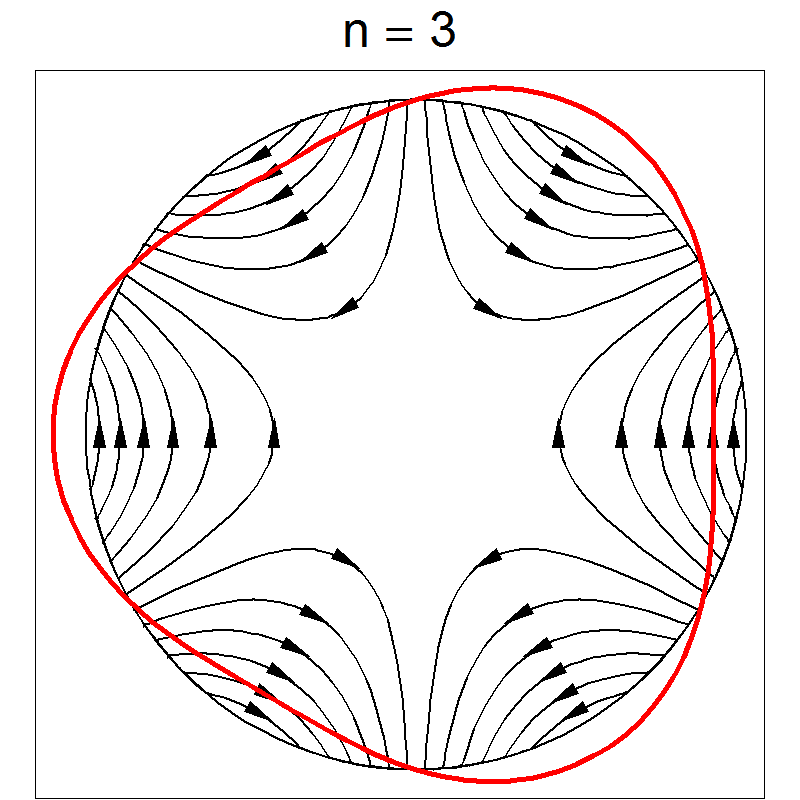} & 
\includegraphics[width=0.4\linewidth]{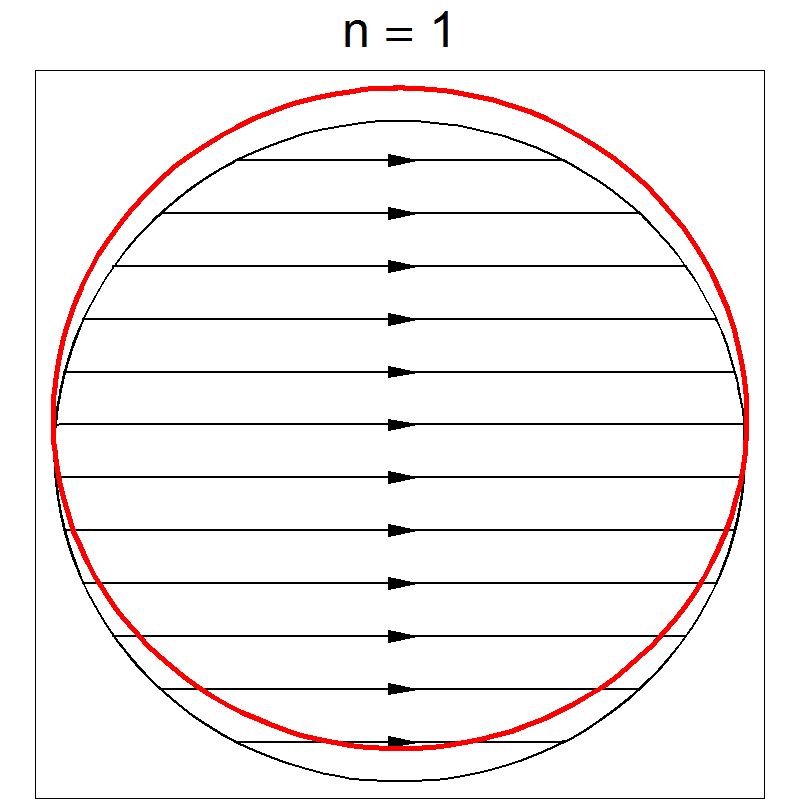} \\
\includegraphics[width=0.4\linewidth]{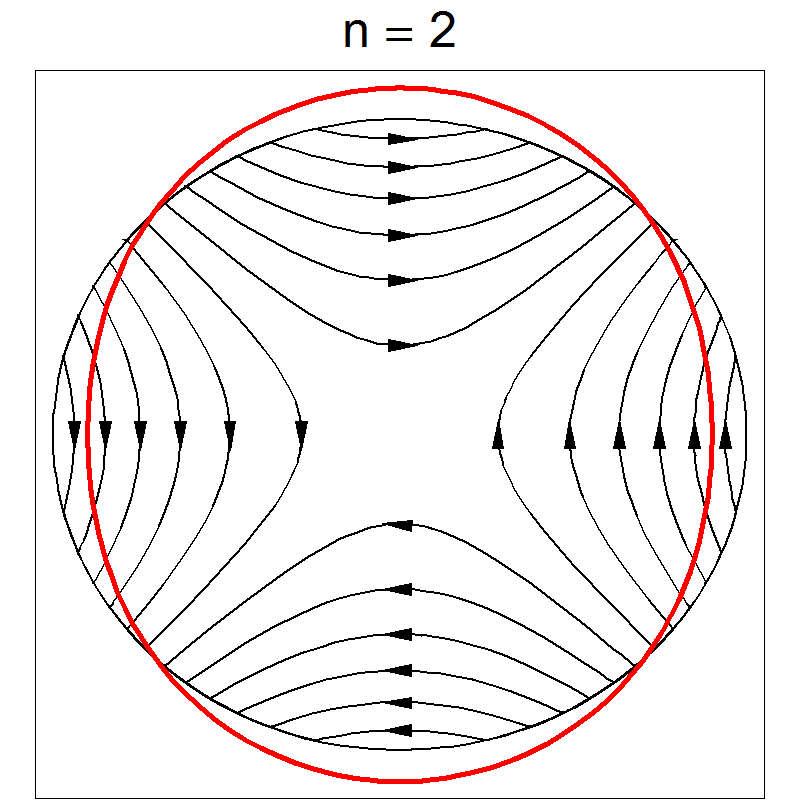} & 
\includegraphics[width=0.4\linewidth]{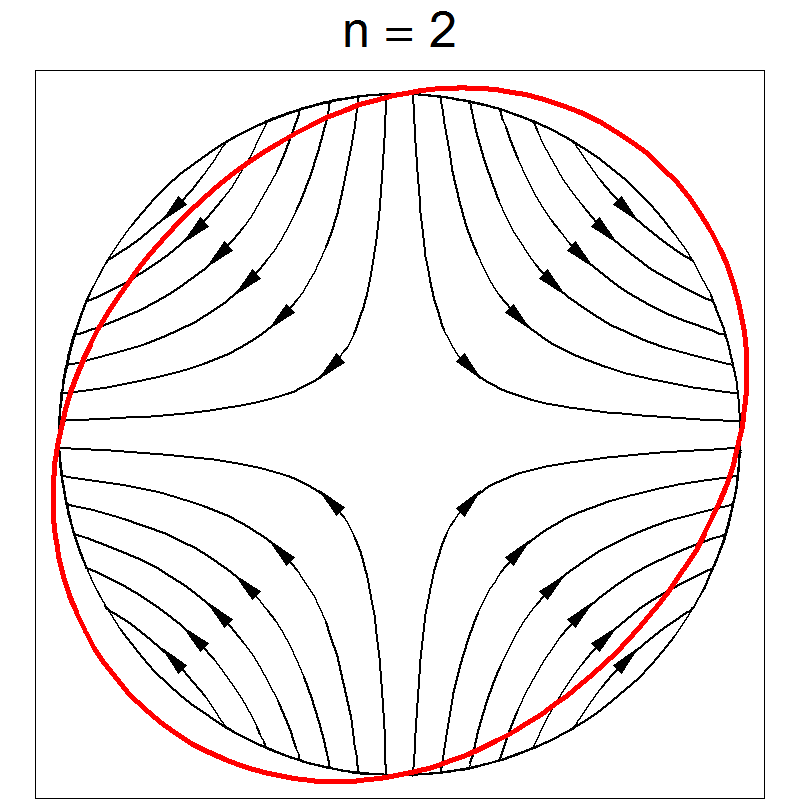} \\
\includegraphics[width=0.4\linewidth]{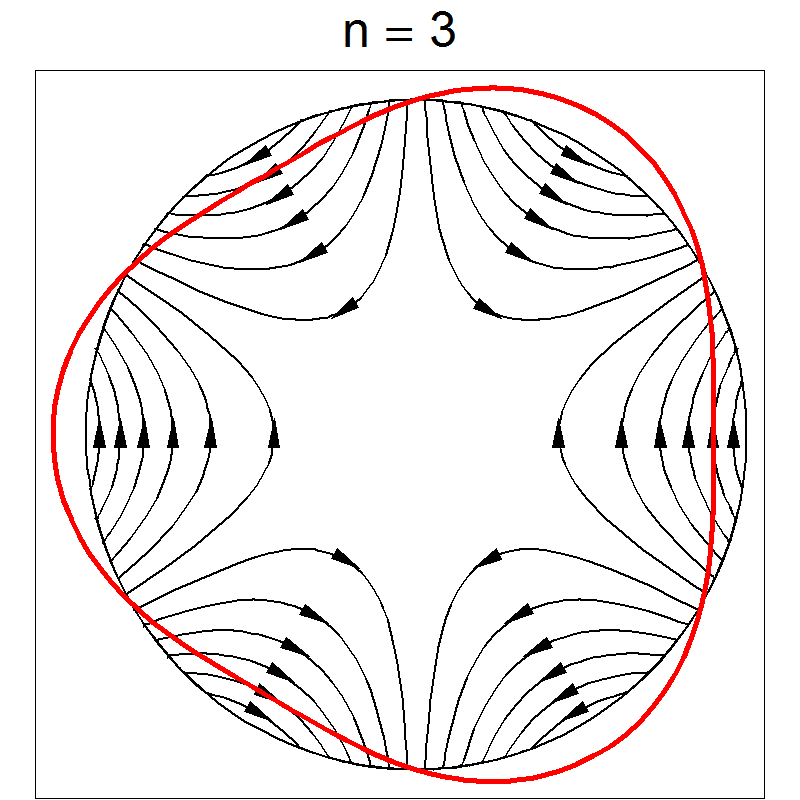} & 
\includegraphics[width=0.4\linewidth]{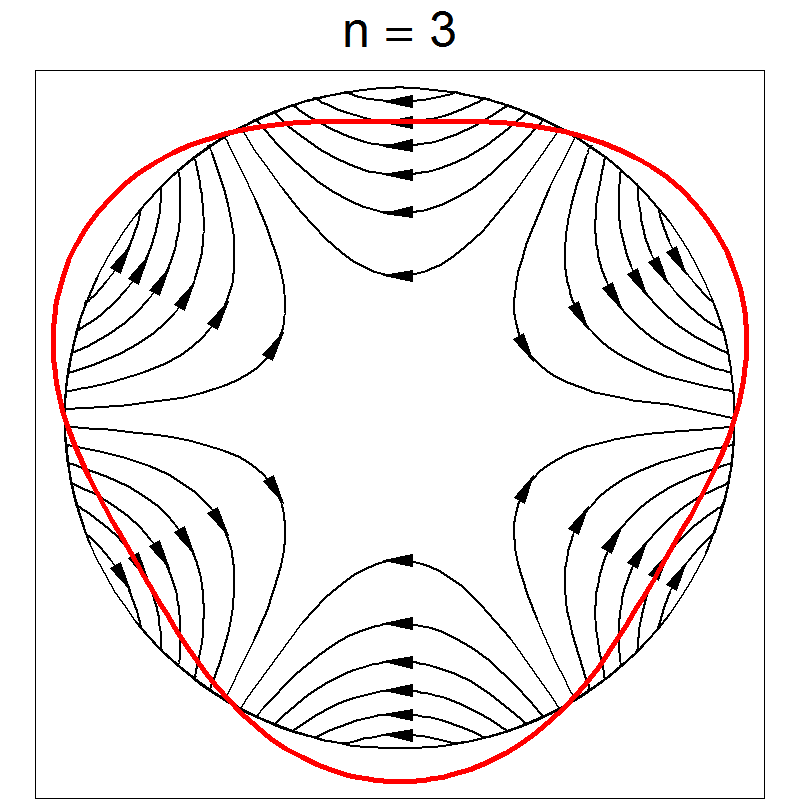}
\end{tabular}
\caption{Current distributions for generating pure multipole fields.
Top: dipole.  Middle: quadrupole.
Bottom: sextupole.  Fields on the left are normal ($a_n$ positive);
those on the right are skew ($b_n$ positive).  The positive $y$ axis is
vertically up; the positive $x$ axis is horizontal and to the right.
The deviation of the red line from the circular boundary shows the
local current density.  Current is flowing in the positive $z$
direction (out of the page) for increased radius, and in the negative
$z$ direction for reduced radius.}
\label{fig:current distributions}
\end{figure}

Now we consider a current flowing on the surface of a cylinder
of radius $r_0$.  Suppose that the current flowing in a section of
the cylinder at angle $\theta_0$ and subtending angle $d\theta_0$ at
the origin is $I\!(\theta_0)\, d\theta_0$.  By the principle of
superposition, we can obtain the total field by summing the
contributions from the currents at all values of $\theta_0$:
\begin{equation}
B_y + i B_x = - \frac{\mu_0}{2\pi r_0} \,
  \sum_{n=1}^\infty
  \left( \frac{r}{r_0} \right)^{n-1}
  e^{i(n-1)\theta}
  \int_0^{2\pi} e^{-in\theta_0} \, I\!(\theta_0) 
   \, d\theta_0.
\label{eq:currentdistributionmultipoles}
\end{equation}
We see that the multipole components are related to the Fourier
components in the current distribution over the cylinder of radius
$r_0$.  In particular, if we consider a current distribution with
just a single Fourier component
\begin{equation}
I\!(\theta_0) = I_0 \cos \left( n_0 \theta_0 - \phi \right),
\label{eq:puremultipolecurrentdistribution}
\end{equation}
the integral in the right-hand side of
Eq.~(\ref{eq:currentdistributionmultipoles}) vanishes except for
$n = n_0$, and we find
\begin{equation}
B_y + i B_x = - \frac{\mu_0 I_0}{2 \pi r_0} \,
  \left( \frac{r}{r_0} \right)^{n_0-1}
  e^{i(n_0-1)\theta}
  \pi e^{-i\phi}.
\nonumber
\end{equation}
The current distribution (\ref{eq:puremultipolecurrentdistribution})
generates a pure multipole field of order $n_0$.  If we choose,
as before
\begin{eqnarray}
B_\textrm{ref} & = & \frac{\mu_0 I_0}{2\pi r_0}, \nonumber \\
R_\textrm{ref} & = & r_0, \nonumber
\end{eqnarray}
then the multipole coefficients are
\begin{equation}
b_n + i a_n = - \pi e^{-i\phi}. \nonumber
\end{equation}
The parameter $\phi$ gives the `angle' of the current distribution.
For $\phi = 0$ or $\phi = \pi$, the current generates a normal
multipole; for $\phi = \pm \pi/2$, the current generates a skew
multipole (Fig~\ref{fig:current distributions}).

\begin{figure}[t!]
\centering
\includegraphics[width=0.6\textwidth]{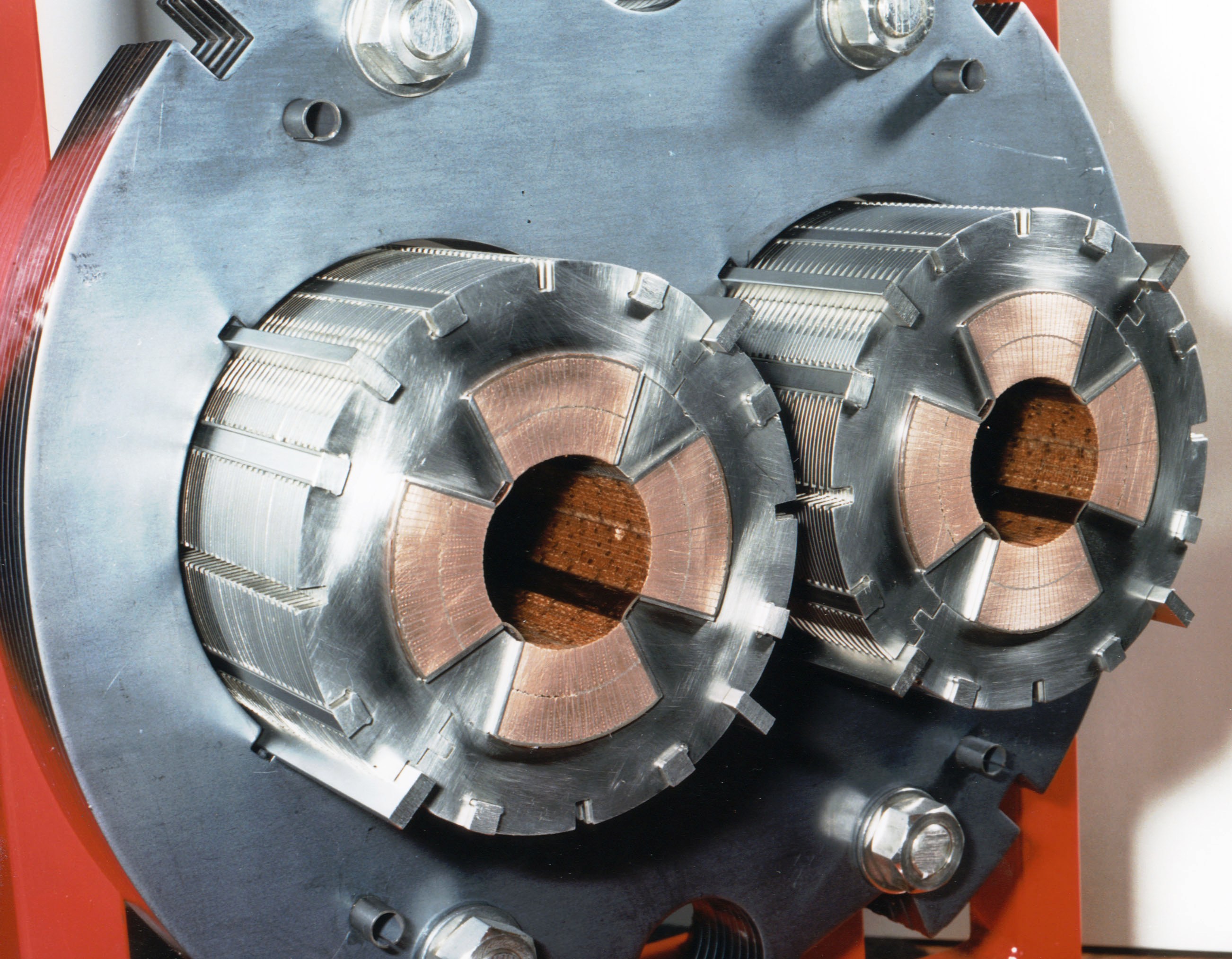}
\caption{Superconducting quadrupoles in the LHC
\label{fig:superconductingquad}}
\end{figure}

The fact that a sinusoidal current distribution on a cylinder can generate
a pure multipole field is not simply of academic interest.  By winding wires
in an appropriate pattern on a cylinder, it is possible to approximate
a sinusoidal current distribution closely enough to produce a multipole
field of acceptable quality for many applications.  Usually, several layers
of windings are used with a different pattern of wires in each layer, to
improve the approximation to a sinusoidal current distribution.  Superconducting
wires can be used to achieve strong fields: an example of superconducting
quadrupoles in the LHC is shown in Fig.~\ref{fig:superconductingquad}.

\subsection{Geometry of an iron-dominated multipole magnet
\label{sec:irondominatedmultipoles}}

Normal-conducting magnets usually use iron cores to increase the flux
density achieved by a given current.  In such a magnet, the shape of
the magnetic field depends mainly on the geometry of the iron.  In this
section, we shall derive the geometry required to generate a pure multipole
of given order.  To simplify the problem, we will make some approximations:
in particular, we shall assume that the iron core has uniform cross-section
and infinite extent along $z$; that there are no limits to the iron in
$x$ or $y$; and that the iron has infinite permeability.  The field in a 
more realistic magnet will generally need to be calculated numerically;
however, the characteristics derived from our idealized model are often a
good starting point for the design of an iron-dominated multipole magnet.

We base our analysis on the magnetic scalar potential, $\varphi$, which is
related to the magnetic field $\vec{B}$ by
\begin{equation}
\vec{B} = - \gradop \varphi. \label{eq:scalarpotential}
\end{equation}
Note that the curl of the field in this case is zero, for any function
$\varphi$: this is a consequence of the mathematical properties of the
grad and curl operators.  Therefore, it follows from Maxwell's
equation~(\ref{eq:maxwell4}) that a magnetic field can only be derived
from a scalar potential if: (i) there is no current density at the location
where the field is to be calculated; (ii) there is no time-dependent
electric displacement at the location where the field is to be calculated.
Where there exists an electric current or a time-dependent electric field,
it is more appropriate to use a vector potential (in which
case, the magnetic flux density is found from the curl of the vector
potential).  However, for multipole fields, we have already shown that
both the divergence and the curl of the field vanish, Eqs.~(\ref{eq:divbzero})
and (\ref{eq:curlbzero}).  Since the curl of the grad of any function is
identically zero, Eq.~(\ref{eq:curlbzero}) is automatically satisfied
for any field $\vec{B}$ derived using (\ref{eq:scalarpotential}).
From Eq.~(\ref{eq:divbzero}), we find
\begin{equation}
\nabla^2 \varphi = 0, \label{eq:poisson}
\end{equation}
where $\nabla^2$ is the Laplacian operator.  Equation~(\ref{eq:poisson})
is Poisson's equation: the scalar potential in a particular case is found
by solving this equation with given boundary conditions.

To determine the geometry of iron required to generate a pure multipole
field, we shall start by writing down the scalar potential for a pure
multipole field.  Since the magnetic flux density $\vec{B}$ is obtained
from the gradient of the scalar potential, the flux density at any point
must be perpendicular to a surface of constant scalar potential.  However,
we already know from Eq.~(\ref{eq:boundaryparallelb}) that the magnetic
flux density at the surface of a material with infinite permeability must
be perpendicular to that surface.  Hence to generate a pure multipole
field in a magnet containing material of infinite permeability, we just need
to shape the material so that its surface follows a surface of constant
magnetic scalar potential for the required field.

We therefore look for a potential $\varphi$ that satisfies
\begin{equation}
-\left( \frac{\partial}{\partial y} + i \frac{\partial}{\partial x} \right) \varphi =
 B_y + i B_x = C_n (x + i y)^{n-1}.
\nonumber
\end{equation}
As we shall now show, an appropriate solution is
\begin{equation}
\varphi = - \left| C_n \right| \frac{r^n}{n} \sin (n\theta - \phi_n)
\label{eq:multipolepotential}
\end{equation}
where
\begin{equation}
x + iy = re^{i\theta},
\nonumber
\end{equation}
and so
\begin{eqnarray}
x & = & r \cos \theta, \nonumber \\
y & = & r \sin \theta. \nonumber
\end{eqnarray}
That Eq.~(\ref{eq:multipolepotential}) is indeed the potential for
a pure multipole of order $n$ can be shown as follows.  In polar
coordinates, the gradient can be written
\begin{equation}
\gradop \varphi = \hat{r} \frac{\partial \varphi}{\partial r}
         + \frac{\hat{\theta}}{r} \frac{\partial \varphi}{\partial \theta},
\label{eq:gradpolar}
\end{equation}
where $\hat{r}$ and $\hat{\theta}$ are unit vectors in the directions of
increasing $r$ and $\theta$, respectively.  Using
\begin{eqnarray}
\hat{r}      & = &  \hat{x} \cos \theta + \hat{y} \sin \theta, \nonumber \\
\hat{\theta} & = & -\hat{x} \sin \theta + \hat{y} \cos \theta, \nonumber 
\end{eqnarray}
it follows from Eq.~(\ref{eq:gradpolar}) that
\begin{eqnarray}
-\gradop \varphi & = & (\hat{x} \cos \theta + \hat{y} \sin \theta)
                          \left| C_n \right| r^{n-1} \sin (n\theta - \phi_n)
                        - (\hat{x} \sin \theta - \hat{y} \cos \theta)
                          \left| C_n \right| r^{n-1} \cos (n\theta - \phi_n),
\nonumber \\
 & = &   \hat{x} \sin \! \left( (n-1)\theta - \phi_n \right) \left| C_n \right| r^{n-1}
       + \hat{y} \cos \! \left( (n-1)\theta - \phi_n \right) \left| C_n \right| r^{n-1}.
\nonumber
\end{eqnarray}
Thus the field derived from the potential (\ref{eq:multipolepotential}) can
be written
\begin{equation}
B_y + i B_x = \left| C_n \right| e^{-i\phi_n} r^{n-1}e^{i(n-1)\theta}.
\nonumber
\end{equation}
Therefore, if
\begin{equation}
C_n = \left| C_n \right| e^{-i\phi_n},
\nonumber
\end{equation}
then
\begin{equation}
B_y + i B_x = C_n r^{n-1} e^{i(n-1)\theta} = C_n (x + i y)^{n-1},
\nonumber
\end{equation}
and we see that the potential (\ref{eq:multipolepotential}) does indeed
generate a pure multipole field of order $n$.

\begin{figure}
\centering
\begin{tabular}{cc}
\includegraphics[width=0.4\linewidth]{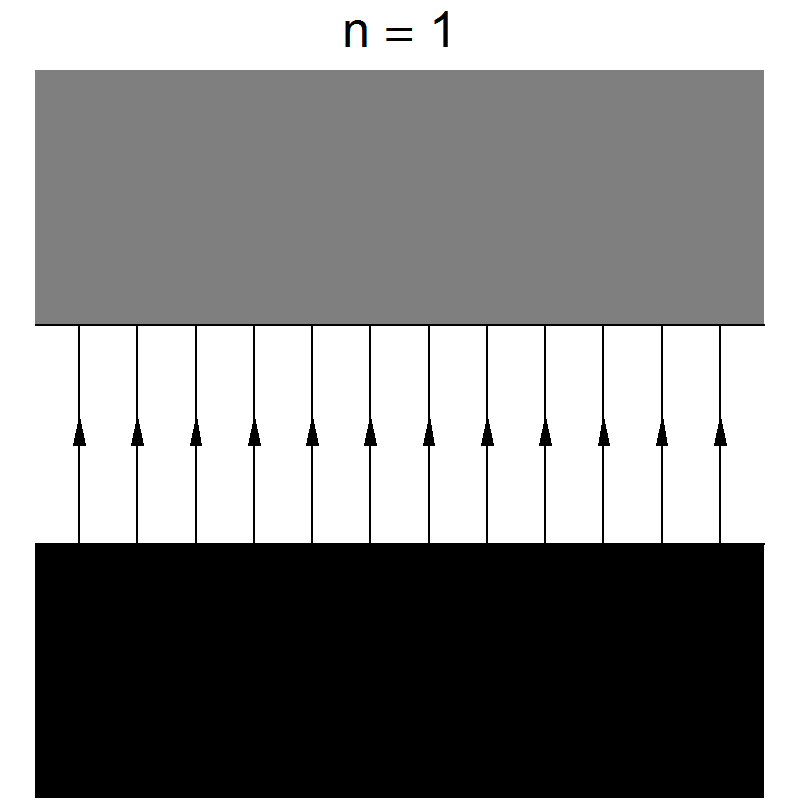} & 
\includegraphics[width=0.4\linewidth]{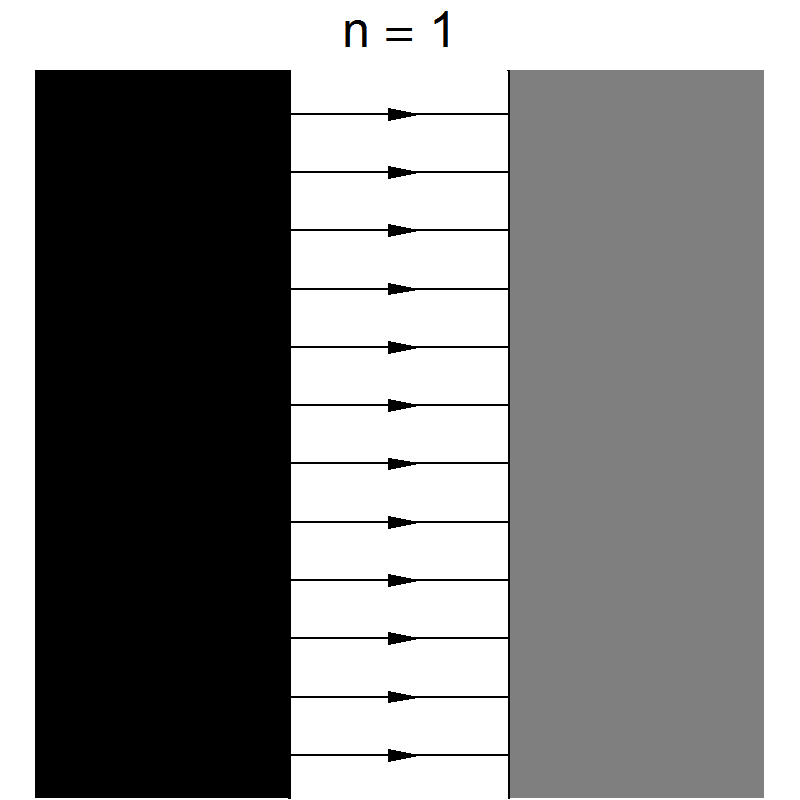} \\
\includegraphics[width=0.4\linewidth]{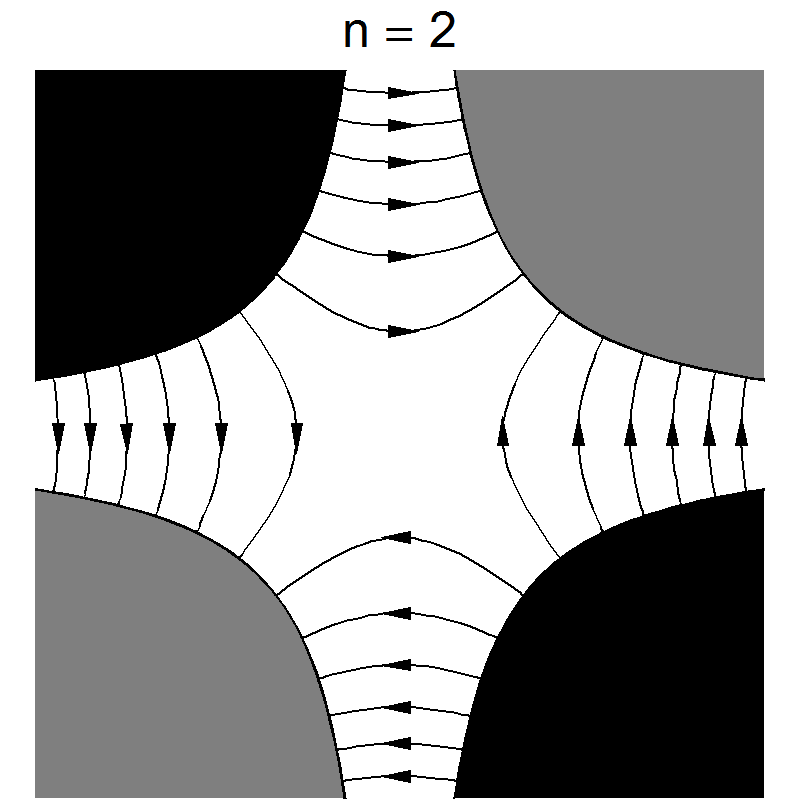} & 
\includegraphics[width=0.4\linewidth]{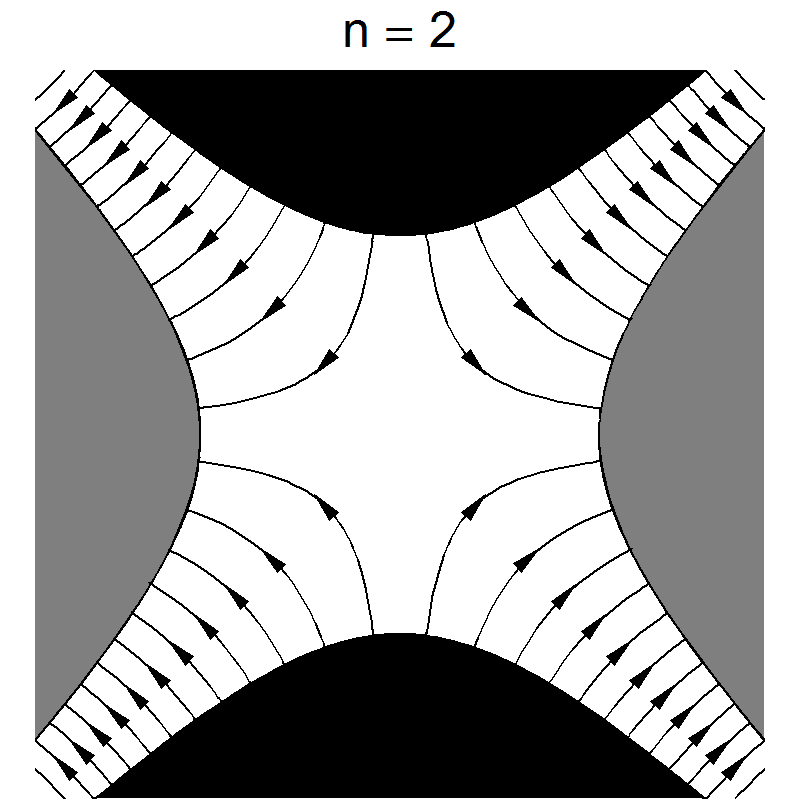} \\
\includegraphics[width=0.4\linewidth]{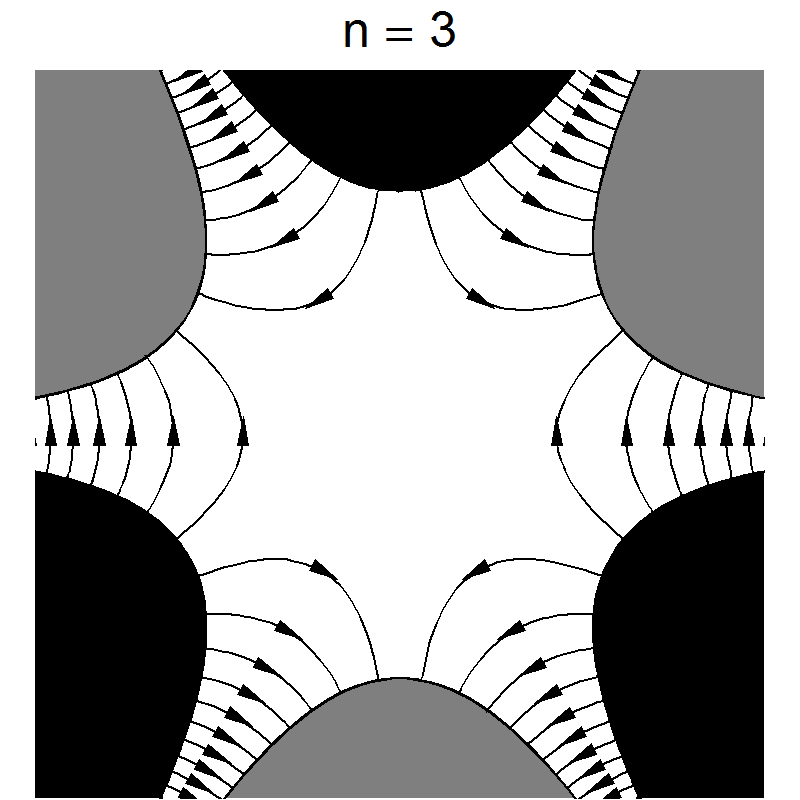} & 
\includegraphics[width=0.4\linewidth]{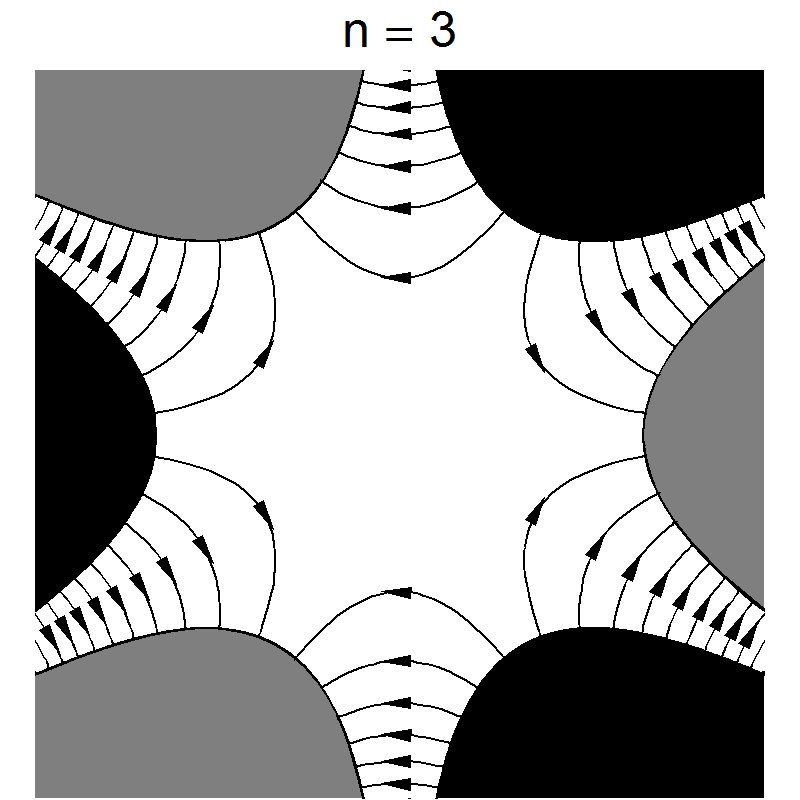}
\end{tabular}
\caption{Pole shapes for generating pure multipole fields.
Top: dipole.  Middle: quadrupole.
Bottom: sextupole.  Fields on the left are normal ($a_n$ positive);
those on the right are skew ($b_n$ positive).  The positive $y$ axis is
vertically up; the positive $x$ axis is horizontal and to the right.
The poles, shown as black (north) or grey (south), are constructed from
material with infinite permeability.
\label{fig:multipolefieldiron}}
\end{figure}

From the above argument, we can immediately conclude that to generate a
pure multipole field, we can shape a high permeability material such that
the surface of the material follows the curve given (in parametric form,
with parameter $\theta$) by
\begin{equation}
r^n \sin (n\theta - \phi_n) = r_0^n, \label{eq:ironpoleshape}
\end{equation}
where $r_0$ is a constant giving the minimum distance between the surface
of the material and the origin.  The cross-sections of iron-dominated
multipole magnets of orders 1, 2 and 3 are shown in
Fig.~\ref{fig:multipolefieldiron}.  Note that $r \to \infty$ for
$n\theta - \phi_n \to \textrm{integer} \times \pi$.  Treating each
region between infinite values of $r$ as a separate pole, we see that
a pure multipole of order $n$ has 2$n$ poles.  We also see that the potential
changes sign when moving from one pole to either adjacent pole: that is,
poles alternate between `north' and `south'.  The field must be generated
by currents flowing along wires between the poles, parallel to the $z$ axis:
to avoid direct contribution from the field around the wires, these wires
should be located a large (in fact, infinite) distance from the origin.

\begin{figure}
\centering
\includegraphics[width=0.4\linewidth]{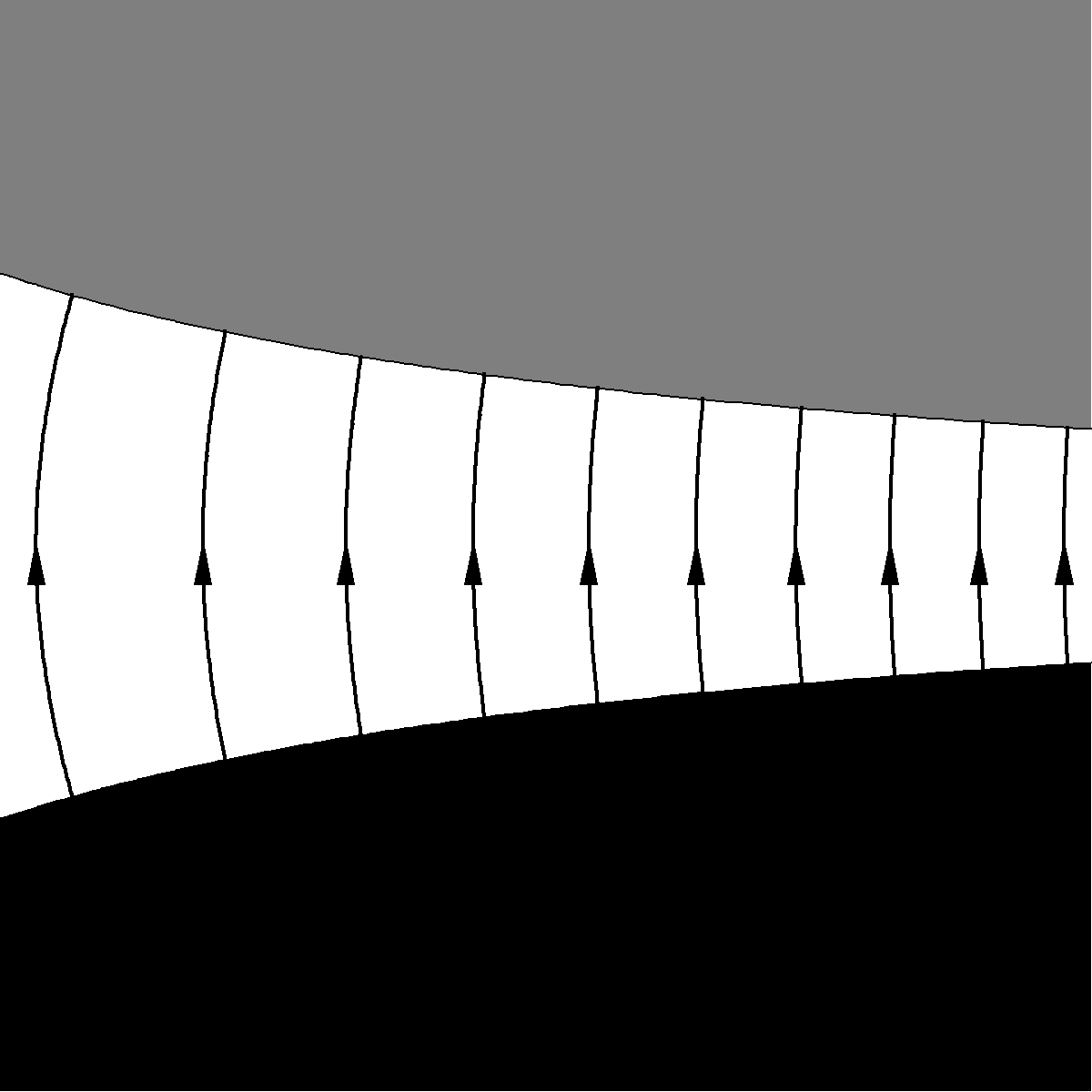}
\caption{Pole shapes for dipole magnet with additional quadrupole component
\label{fig:combinedfunctionbend}}
\end{figure}

Note that it is possible to determine the shape of the pole
face for a magnet containing any specified set of multipoles by
summing the potentials for the different multipole components, and
then solving for $r$ as a function of $\theta$, for a fixed value
of the scalar potential.  Magnets designed to
have more than one multipole component are often known as `combined
function' magnets.  Perhaps the most common type of combined function
magnet is a dipole with a quadrupole component: such magnets can be
used to steer and focus a beam simultaneously.  The shape of the
pole faces and the field lines in a dipole with (strong) quadrupole
component is shown in Fig.~\ref{fig:combinedfunctionbend}.

\begin{figure}
\centering
\includegraphics[width=0.4\linewidth]{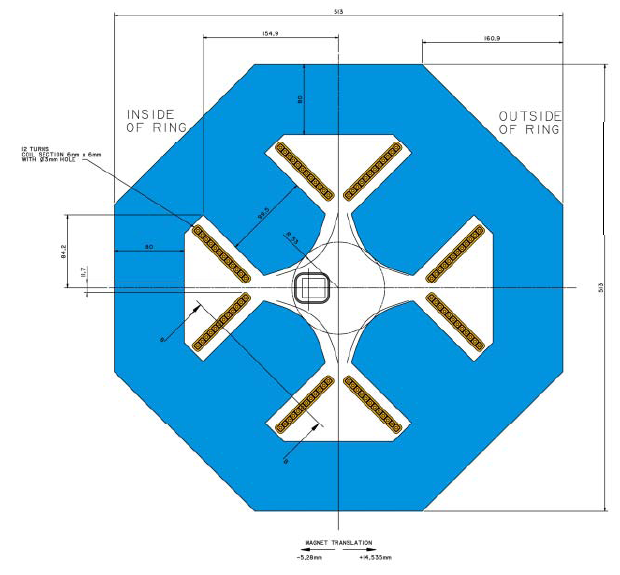}
\hspace{0.1\linewidth}
\includegraphics[width=0.4\linewidth]{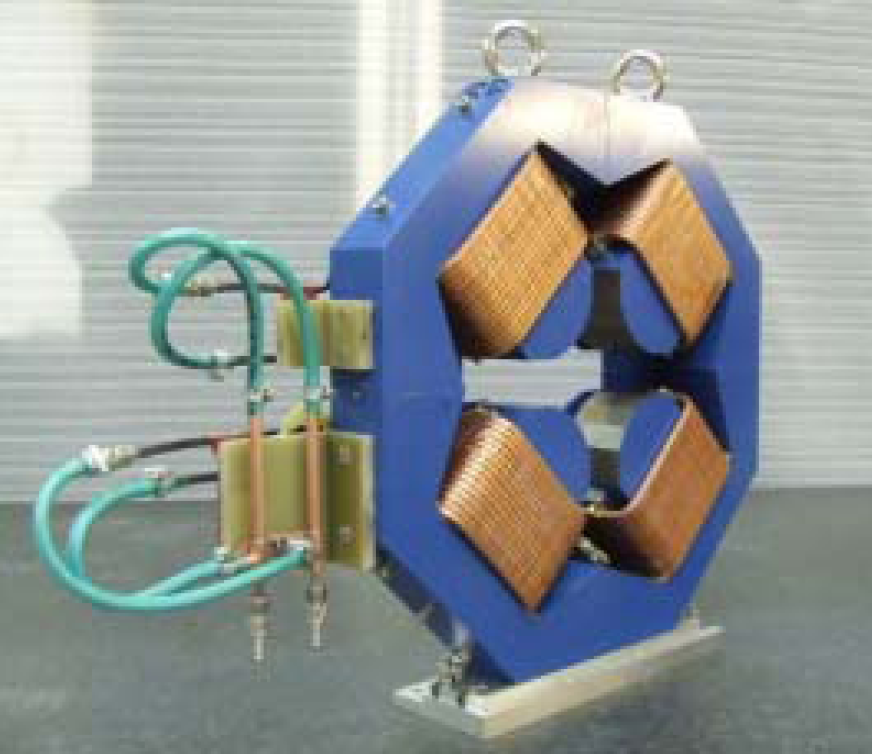}
\caption{Iron-dominated quadrupole magnet for the EMMA Fixed-Field
Alternating Gradient accelerator at Daresbury Laboratory.
Left: magnet cross-section \cite{bib:emmaquads1}.
Right: magnet prototype \cite{bib:emmaquads2}.
\label{fig:quadrupolemagnet}}
\end{figure}

In practice, some variation from the `ideal' geometry is needed
to account for the fact that the material used in the magnet has
finite permeability, and finite extent transversely and longitudinally.
The wires carrying the current that generates the magnetic flux are
arranged in coils around each pole; as we shall see, the strength
of the field is determined by the number of ampere-turns in each coil.
An iron-dominated electromagnetic quadrupole is shown in
Fig.~{\ref{fig:quadrupolemagnet}}.

\begin{figure}
\centering
\includegraphics[width=0.6\linewidth]{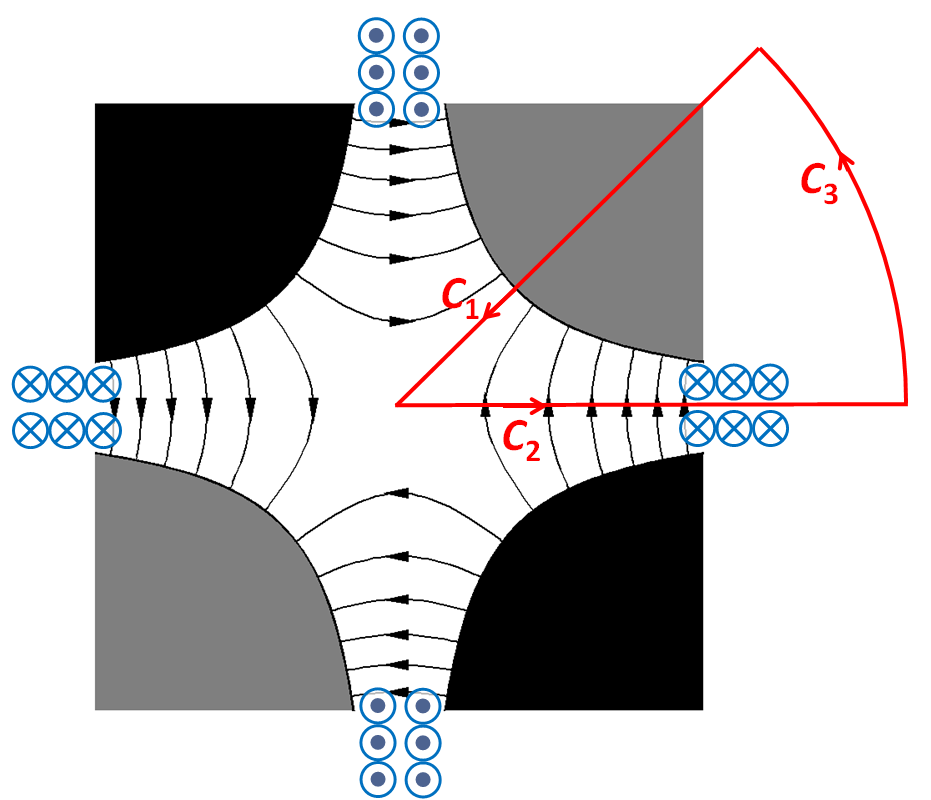}
\caption{Contour for line integral used to calculate the field strength
in an iron-dominated quadrupole\label{fig:quadlineintegral}}
\end{figure}

To complete our discussion of methods to generate multipole fields, we
derive an expression for the field strength in an iron-dominated magnet
with a given number of ampere-turns in the coil around each pole.  To do
this, we consider a line integral as shown in Fig.~\ref{fig:quadlineintegral}.
In the figure, we show a quadrupole; however the generalization to other
orders of multipole is straightforward.  Note that, in principle, the
coils carrying the electric current, and the line segment $C_3$, are an
infinite distance from the origin (the centre of the magnet).

Using Maxwell's equation (\ref{eq:maxwell4}), with constant (zero)
electric displacement, and integrating over the surface $S$ bounded
by the curve $C_1 + C_2 + C_3$ gives
\begin{equation}
\int_S \curlop \vec{H} \cdot d\vec{S} = \int_S \vec{J} \cdot d\vec{S}
  = -NI.
\nonumber
\end{equation}
Note that the surface is oriented so that the normal is parallel to
the positive $z$ axis; and the coil around each pole consists of $N$
turns of wire carrying current $I$.  Applying Stokes's
theorem~(\ref{eq:stokestheorem}) gives
\begin{equation}
\int_{C_1} \vec{H} \cdot d\vec{l}
+ \int_{C_2} \vec{H} \cdot d\vec{l}
+ \int_{C_3} \vec{H} \cdot d\vec{l}
  = -NI.
\nonumber
\end{equation}
We know, from Eq.~(\ref{eq:boundarynormalb}), that the normal component
of the magnetic flux density $\vec{B}$
is continuous across a boundary.  Then, since $\vec{B} = \mu \vec{H}$, it
follows that for a finite field between the poles, and for
$\mu \to \infty$, the magnetic intensity $\vec{H}$ vanishes within
the poles.  Also, the field is perpendicular to the line segment
$C_2$.  Thus, the only part of the integral that makes a non-zero
contribution, is the integral along $C_1$ from the face of the pole
to the origin.  Hence
\begin{equation}
\int_0^{r_0} \frac{B_r(r)}{\mu_0} dr = NI.
\label{eq:multipolelineintegral}
\end{equation}
The contour $C_1$ is chosen so that along this contour the
field has only a radial component, parallel to the contour.
From Eq.~(\ref{eq:multipolepolarcoords}), we see that for a multipole
of order $n$, along this contour we have
\begin{equation}
B_r = \left| C_n \right| r^{n-1}.
\nonumber
\end{equation}
Thus we find by performing the integral in
Eq.~(\ref{eq:multipolelineintegral})
\begin{equation}
\left| C_n \right| = \mu_0 NI \frac{n}{r_0^n}.
\nonumber
\end{equation}
For a normal multipole, the field is given by
\begin{equation}
B_y + i B_x = \frac{\mu_0 n NI}{r_0} \left( \frac{x + iy}{r_0} \right)^{n-1}.
\nonumber
\end{equation}
For example, in a normal quadrupole ($n=2$), the field gradient is given by
\begin{equation}
\frac{\partial B_y}{\partial x} = \frac{2\mu_0 NI}{r_0^2}.
\end{equation}

\section{Multipole decomposition}

In the previous section, we derived the current density distributions
and material geometries needed to generate a pure multipole field of a
given order.  However, the distributions and geometries required are not
perfectly achievable in practice: the currents and materials have infinite
longitudinal extent; and we require either a current that exists purely
on the surface of a cylinder, or infinite permeability materials with
infinite transverse extent.

Real multipole magnets, therefore, will not consist of a single multipole
component, but a superposition of (in general) an infinite number of
multipole fields.  The exact shape of the field can have a significant
impact on the beam dynamics in an accelerator.  In many simulation codes
for accelerator beam dynamics, the magnets are specified by the multipole
coefficients: this is because simple techniques exist for approximating
the effect, for example, of sextupole, octopule, and other higher-order
components in the field of a quadrupole magnet.  The question then arises
how to determine the multipole components in a given magnetic field.

At this point, we can make a distinction between the \emph{design} field
of a magnet, and the field that exists within a fabricated magnet.  The
design field is one that is still in some sense `ideal'; though the
design field for a quadrupole magnet (for example) will contain other
multipole components, because the design has to respect practical
constraints, i.e., the magnet will have finite longitudinal and transverse
extent, any currents will flow in wires of non-zero dimension, and any
materials present will have finite (and often non-linear) permeability.
Usually, one attempts to optimize the design to minimize the strengths of
the multipole components apart from the one required: the residual
strengths are generally known as \emph{systematic} multipole errors.
These errors will be present in any fabricated magnet, although, because
of construction tolerances, the errors will vary between any two magnets
of the same nominal design.  The differences between the multipole
components in the design field and the components in a particular magnet
are known as \emph{random} multipole errors.

The effects of systematic and random multipole errors on an accelerator,
and hence the specification of upper limits on these quantities, can
usually only be properly understood by running beam dynamics simulations.
Therefore, accelerator magnet (and lattice) design often proceeds
iteratively.  Some initial estimate of the limits on the errors is often
needed to guide the magnet design; but then any design that is developed
must be studied by further beam dynamics simulations to determine whether
improvements are needed.

It is therefore important to be able to determine the multipole components
in a magnetic field from numerical field data: these data may come from
either a magnetic model (i.e., from the design of a magnet), or from
measurements on a real device.  There are different procedures that can
be used to achieve the `decomposition' of a field into its multipole
components.  In this section, we shall consider methods based on
Cartesian and polar representations of two-dimensional fields (i.e., fields
that are independent of the longitudinal coordinate).  In
Section~\ref{sec:threedfields} we shall consider decompositions of
three-dimensional fields (i.e., fields that have explicit dependence on
longitudinal as well as transverse coordinates).  However, we first
consider an important concept in the discussion of multipole
field errors, namely how the symmetry of a multipole magnet leads to
`allowed' and `forbidden' higher-order multipoles.

\subsection{Multipole symmetry, `allowed' and `forbidden'
higher-order multipoles
\label{sec:symmetryallowedforbidden}}

A pure multipole field of order $n$ can be written
\begin{equation}
B_y + i B_x = \left| C_n \right| e^{-i\phi_n} r^{n-1} e^{i(n-1)\theta}.
\label{eq:puremultipoleordern}
\end{equation}
The parameter $\phi_n$ characterizes the angular orientation of the
magnet around the $z$ axis.  In particular, from Eq.~(\ref{eq:ironpoleshape}),
we see that a change in $\phi_n$ by $n\alpha$ is equivalent to a rotation of
the coordinates (a change in $\theta$) by $-\alpha$.  Thus, a rotation of a
magnet around the $z$ axis by angle $\alpha$ may be represented by a change
in $\phi_n$ by $n\alpha$.  In particular, if the magnet is rotated by $\pi/n$,
then from Eq.~(\ref{eq:puremultipoleordern}), we see that the field at any
point simply changes sign:
\begin{equation}
\textrm{if } \phi_n \mapsto \phi_n + \pi, \textrm{ then }
\vec{B} \mapsto -\vec{B}.
\label{eq:symmetryconstraint}
\end{equation}
This property of the magnetic field is imposed by the symmetry of the magnet.
In a real magnet, it will not be satisfied exactly, because random variations
in the geometry will break the symmetry.  However, it is possible to maintain
the symmetry exactly in the design of the magnet; this means that although
higher order multipoles will in general be present, only those multipoles
satisfying the symmetry constraint (\ref{eq:symmetryconstraint}) can be
present.  These are the `allowed' multipoles.  Other multipoles, which must
be completely absent, are the `forbidden' multipoles.

\begin{figure}
\centering
\includegraphics[width=0.4\linewidth]{MultipoleFieldIron2Normal.png}
\includegraphics[width=0.4\linewidth]{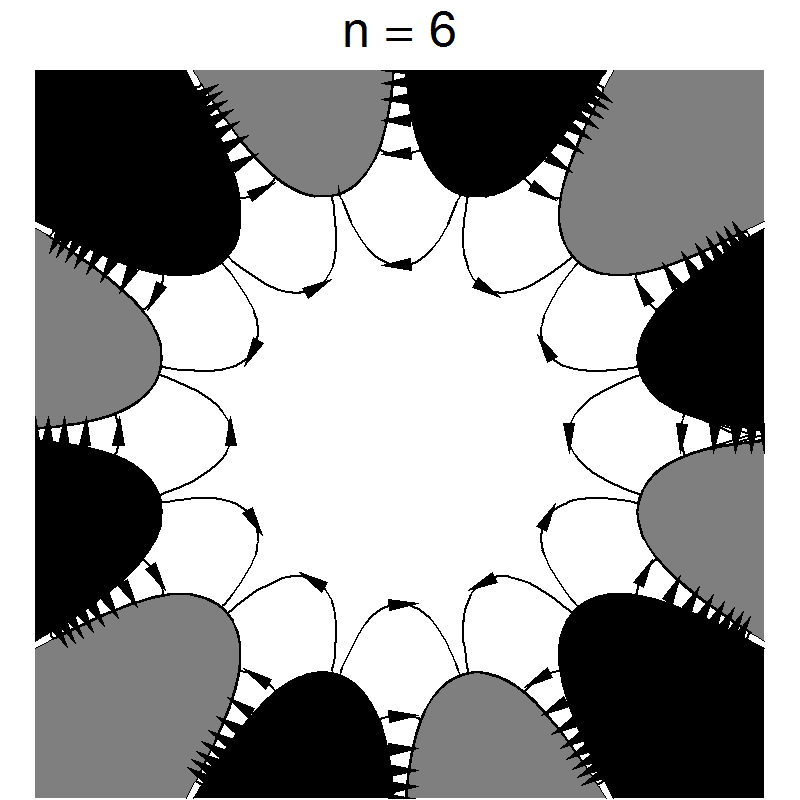}
\caption{Normal quadrupole field (left) and dodecapole field (right).  The
dodecapole is the first higher-order multipole with the same rotational
symmetry as the quadrupole (under a rotation by $\pi /2$, north and
south poles interchange).
\label{fig:allowedhighermultipole}}
\end{figure}

We can derive a simple expression for the allowed multipoles in a magnet
designed with symmetry for a multipole of order $n$.  Consider an additional
multipole (a `systematic error') in this field, of order $m$.  By the
principle of superposition, the total field can be written as
\begin{equation}
B_y + i B_x = \left| C_n \right| e^{-i\phi_n} r^{n-1} e^{i(n-1)\theta}
            + \left| C_m \right| e^{-i\phi_m} r^{m-1} e^{i(m-1)\theta}.
\nonumber
\end{equation}
The geometry is such that under a rotation about the $z$ axis through $\pi /n$,
the magnet looks the same, except that all currents have reversed direction:
therefore the field simply changes sign.  Under this rotation $\phi_n \mapsto 
\phi_n + \pi$; however, $\phi_m \mapsto \phi_m + m\pi/n$.  This means that
we must have:
\begin{equation}
e^{-i\frac{m}{n}\pi} = -1.
\nonumber
\end{equation}
Therefore $m/n$ must be an odd integer.  Assuming that $m \neq n$ (i.e., the
multipole error is of a different order than the `main' multipole field), then
\begin{equation}
\frac{m}{n} = 3,5,7,\dots
\end{equation}
Thus, for a dipole, the allowed higher order multipoles are sextupole,
decapole, etc.; for a quadrupole, the allowed higher order multipoles are
dodecapole, 20-pole, etc.  The fact that the allowed higher order multipoles
have an order given by an odd integer multiplied by the order of the main
multipole is a consequence of the fact that magnetic poles always occur
in north--south pairs.  This is illustrated for a quadrupole in Fig.~\ref{fig:allowedhighermultipole}; here we see that to maintain the
correct rotational symmetry (with the field changing sign under a rotation
through $\pi /2$) the first higher-order multipole must be constructed
by `splitting' each main pole into three, then into five, and so on.

The field in a real magnet will contain all higher order multipoles, not
just the ones allowed by symmetry.  However, it is often the case that the
allowed multipoles dominate over the forbidden multipoles.

\subsection{Fitting multipoles: Cartesian basis\label{sec:multipolefitcartnbasis}}

Suppose we have obtained a set of numerical field data, either from a magnetic
model, or from measurements on a real magnet.  To determine the effect of the
field on the beam dynamics in an accelerator, it is helpful to know the
multipole components in the field.  One way to compute the multipole components
is to fit a polynomial to the field data.  For example, if we consider a 
normal multipole (coefficients $C_n$ are all real), the vertical field along
the $x$ axis (i.e., for $y=0$) is given by
\begin{equation}
B_y = \sum_{n=1}^{\infty} C_n x^{n-1}.
\end{equation}
The number of data points determines the highest order multipole that can be
fitted.  Fitting may be achieved using, for example, a routine that minimizes
the squares of the residuals between the data and the fitted function.
However, although this procedure can, in principle, produce good results, it
is not very robust.  In particular, the presence of multipoles of higher order
than those included in the fit can affect the values determined for those
multipoles that are included in the fit.  We can illustrate this as follows.

\begin{figure}
\centering
\includegraphics[width=0.45\linewidth]{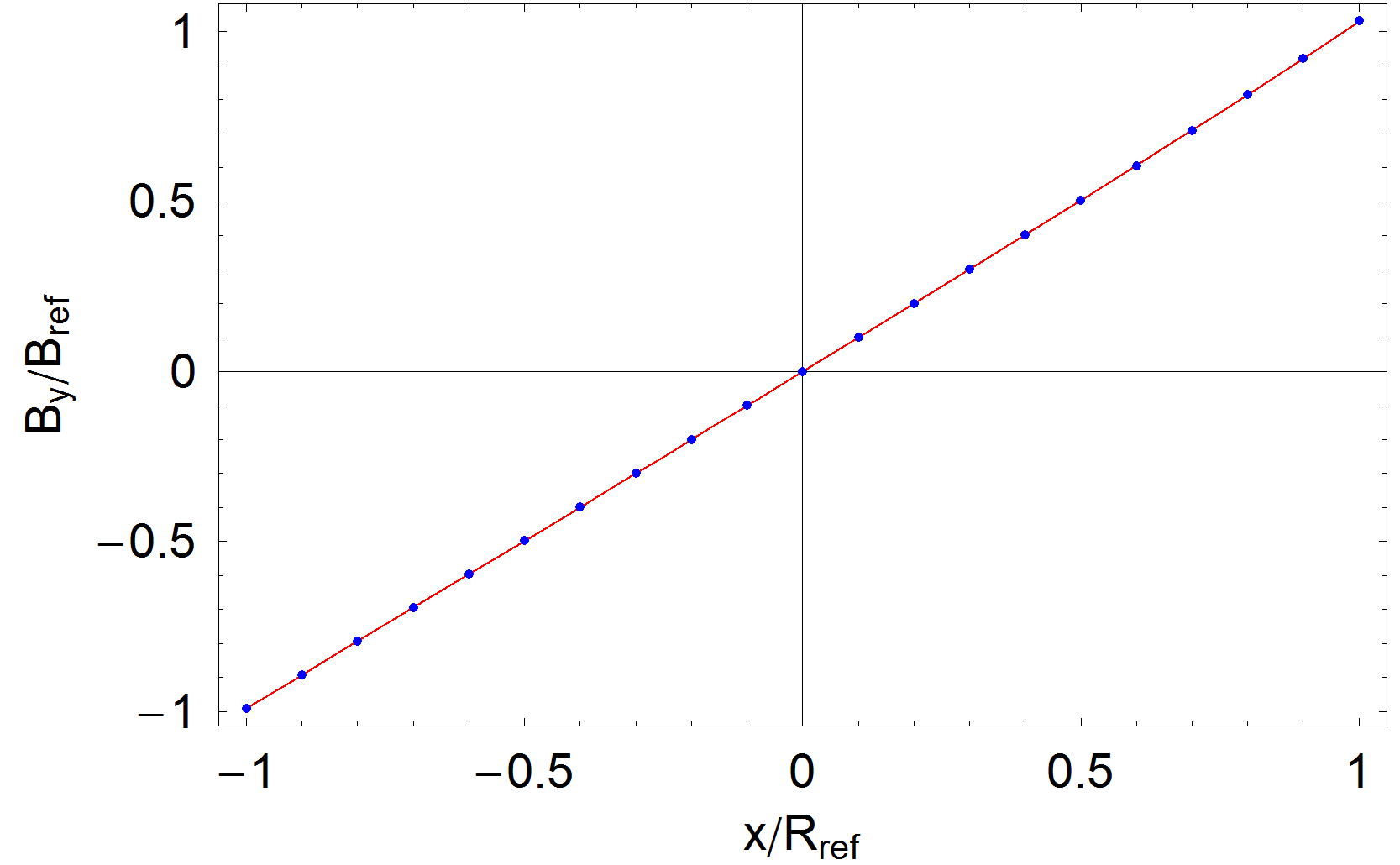}
\hspace{0.05\linewidth}
\includegraphics[width=0.45\linewidth]{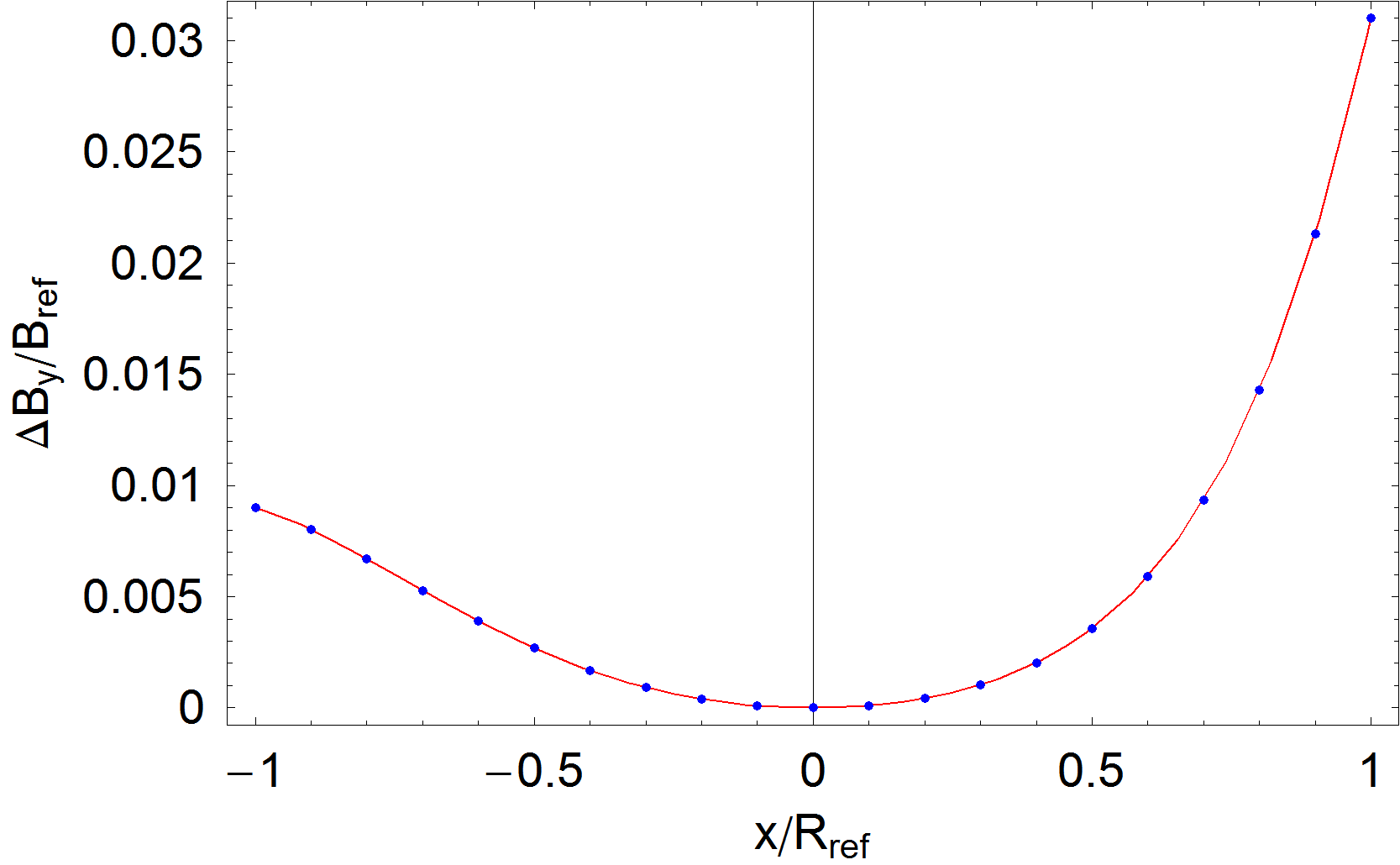}
\caption{Measured (points) and fitted (line) field in a quadrupole
with higher-order multipole errors of order 3, 4, 5 and 6.  Multipoles
up to order 6 are fitted.  Left: total field.  Right: deviation from
quadrupole field.
\label{fig:constructedfield6}}
\end{figure}

Let us construct a quadrupole field ($n = $2), and add to it higher order
multipoles of order 3, 4, 5 and 6.  The values of the coefficients $a_n$
(actual values, and fitted values in two different cases) are given in
Table~\ref{tab:constructedfield}.  The field $B_y / B_\textrm{ref}$ is plotted
as a function of $x / R_\textrm{ref}$ in Fig.~\ref{fig:constructedfield6}:
the field data are shown as points, while the fit, including multipoles up to
order 6, is shown as a line.  Also shown is the deviation
$\Delta B_y / B_\textrm{ref}$ from an ideal quadrupole field, i.e.
$\Delta B_y$ is the contribution of the higher order multipoles.  We see
that if we base the fit on all the multipoles that are present (i.e. up
to order 6), then we obtain accurate values for all multipole coefficients.

\begin{table}[b!]
\centering
\caption{Actual and fitted multipole values for a quadrupole field with
artificially constructed multipole errors
\label{tab:constructedfield}}
\begin{tabular}{cccc}
\hline
$n$ & actual coefficient $a_n$ & 
\multicolumn{2}{c}{fitted coefficient $a_n$}  \\
    &                          & $(n \leq 6)$ &
                                 $(n \leq 5)$ \\
\hline
2 & 1.000 & 1.000 & 0.9972 \\
3 & 0.010 & 0.010 & 0.0100 \\
4 & 0.001 & 0.001 & 0.0131 \\
5 & 0.010 & 0.010 & 0.0100 \\
6 & 0.010 & 0.010 & --- \\
\hline
\end{tabular}
\end{table}

\begin{figure}
\centering
\includegraphics[width=0.45\linewidth]{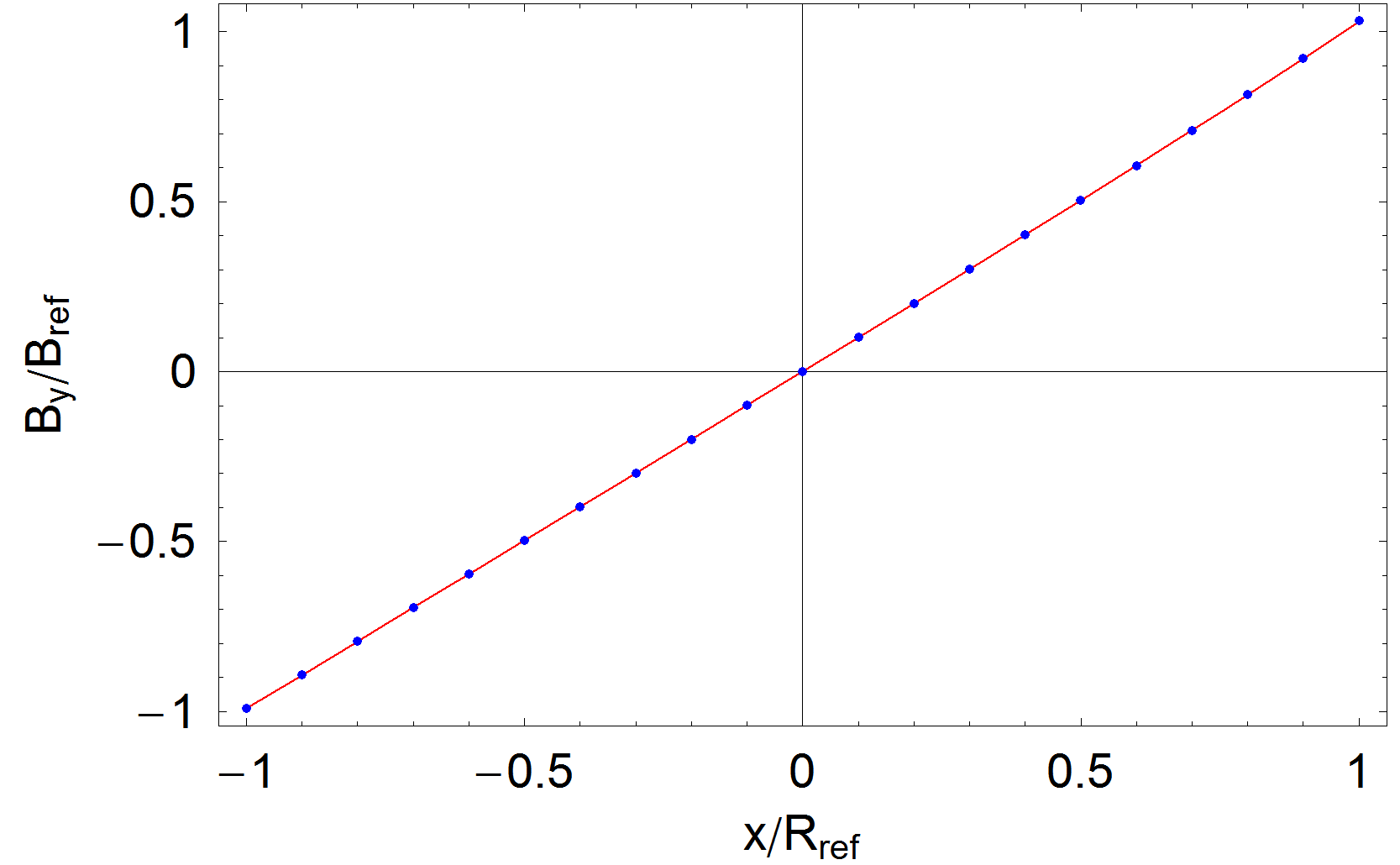}
\hspace{0.05\linewidth}
\includegraphics[width=0.45\linewidth]{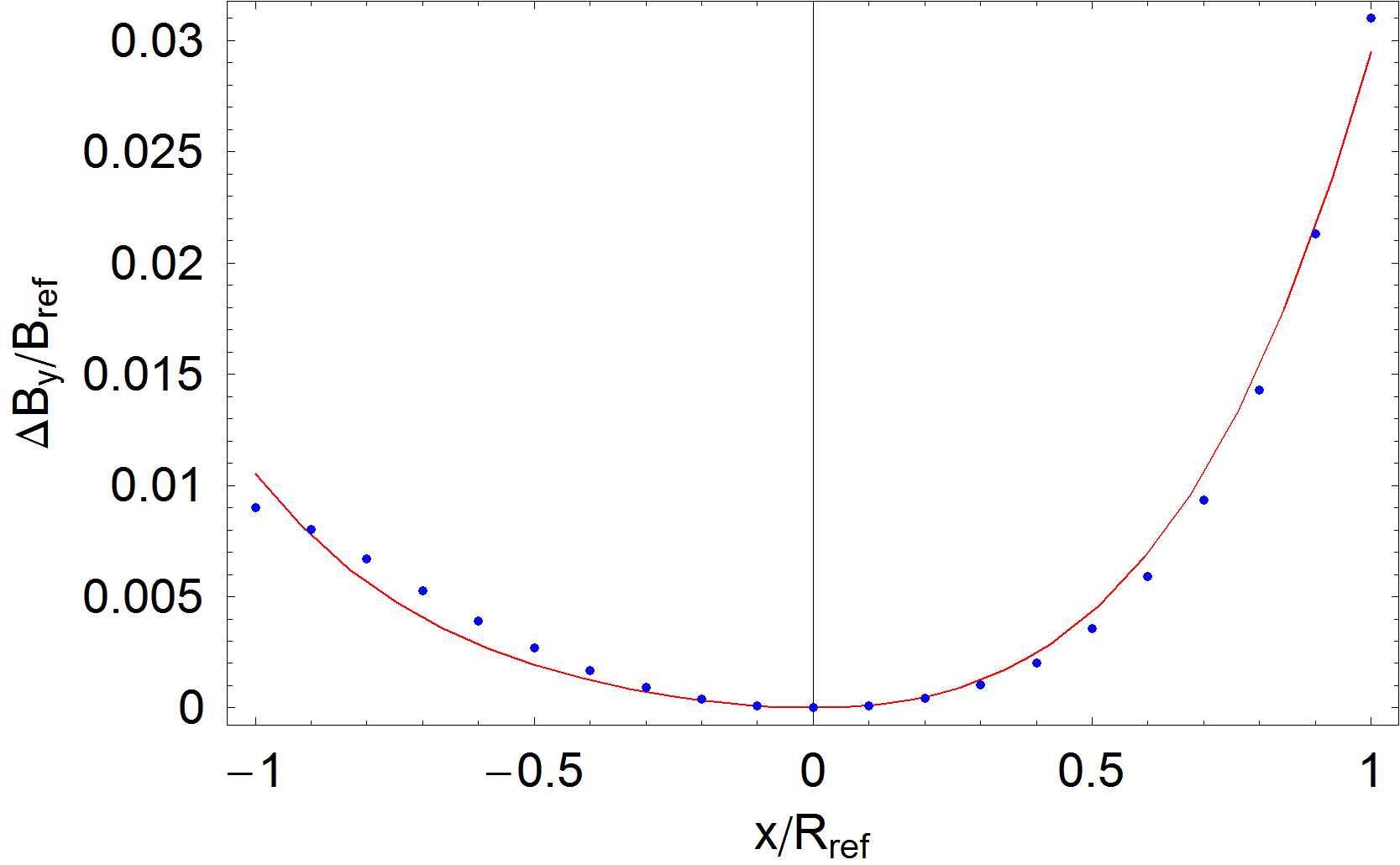}
\caption{Measured (points) and fitted (line) field in a quadrupole
with higher-order multipole errors of order 3, 4, 5 and 6.  Multipoles
up to order 5 are fitted.  Left: total field.  Right: deviation from
quadrupole field.
\label{fig:constructedfield5}}
\end{figure}

However, in general, multipoles of all orders are present, while our fit
is based on a finite number of multipoles.  If we try to fit the data
in our illustrative case using multipoles up to order 5 only (i.e. omitting
the order 6 multipole that is present), then we see that there is an
impact on the accuracy with which we determine the lower-order multipoles.
This can be seen in the final column of Table~\ref{tab:constructedfield}:
there is even an error in the value that we determine for the quadrupole
strength.  When we plot the fit against the field data, we see that there
is some small residual deviation between the data and the fit: this is to
be expected, since the function we are using to obtain the fit does not
match exactly the function used to generate the data.  Although not visible
in the total field, plotted in Fig.~\ref{fig:constructedfield5}, the
difference between the fit and the data is apparent in the plot of the
deviation from the quadrupole field. 

Our concern is that the presence of higher-order multipoles has affected
the accuracy with which we determine the lower-order multipoles, even
down to the quadrupole field strength.  This can have significant implications
for beam dynamics: the effect of a linear focusing error in a beam line (from
some variation in the quadrupole strength) can be very different from the
effects of higher-order multipole errors.  For example, if one is measuring
the betatron tunes or the beta functions in a storage ring, these can be
very sensitive to linear focusing errors, and relatively insensitive to
higher order multipoles.  Determining the multipole coefficients using
a polynomial fit can lead to inaccurate predictions of the linear behaviour
of the beam line, depending on the higher-order multipoles present in the
magnets.

The problem is that we have based our fit on monomials, i.e., powers of
$x$.  Our fit is a sum of these monomials, with coefficients determined
from the data.  However, it is possible to obtain a fit to data generated
using one monomial, with a different monomial.  For example, if one
constructs data which is purely linear in $x$, then one can obtain a fit
using a monomial $x^3$ (even though the fit will not be as good
as one obtained using a monomial $x$).  Mathematically, the basis functions
we are using (monomials in this case) are not orthogonal: the coefficients
we determine depend on which set of basis functions we choose to use.
A more robust technique would use basis functions that are orthogonal, i.e.,
the coefficients we determine will be the same, no matter which set of
functions we choose.  Fortunately, there exists an appropriate set of
functions that provides an orthogonal basis for multipole fields.  We discuss
this basis in the following section, \ref{sec:multipolefitpolarbasis}.
The advantage of orthogonal basis functions is that the coefficients we
determine for different terms in the fit are \emph{independent} of which
terms we include in the fit; for example, the quadrupole strength that we
find in a particular magnet will be the same, irrespective of which
higher-order terms we include in the fit, and which higher-order terms are
actually present.

\subsection{Fitting multipoles: polar basis\label{sec:multipolefitpolarbasis}}

From Eq.~(\ref{eq:multipolepolarcoords}) we know that the field in a
multipole magnet can be written in polar coordinates as
\begin{equation}
B_\theta + i B_r = \sum_{n=1}^\infty C_n r^{n-1} e^{in\theta}.
\nonumber
\end{equation}
We see that if we make a set of measurements of $B_r$ and $B_\theta$ at
different values of $\theta$ and fixed radial distance $r$, then we can
obtain the coefficients $C_n$ by a discrete Fourier transform.

Suppose we make $M$ measurements of the field, at $\theta = \theta_m$,
where
\begin{equation}
\theta_m = 2\pi\frac{m}{M}, \qquad m = 0, 1, 2 \dots M-1.
\end{equation}
We write the measurement at $\theta = \theta_m$ as $B_m$; note that
$B_m$ is a complex number, whose real and imaginary parts are given by the
azimuthal and radial components of the field at $\theta = \theta_m$.

Now we construct, for a chosen integer $n^\prime$
\begin{equation}
\sum_{m=0}^{M-1} B_m e^{-2\pi i n^\prime \frac{m}{M}}
  = \sum_{m=0}^{M-1} \sum_{n=1}^{\infty} C_n r_0^{n-1} e^{2\pi i (n - n^\prime ) \frac{m}{M}},
\nonumber
\end{equation}
where $r_0$ is the radial distance at which the field measurements are made.
The summation over $m$ on the right-hand side vanishes, unless $n = n^\prime$.
Thus we can write
\begin{equation}
\sum_{m=0}^{M-1} B_m e^{-2\pi i n^\prime \frac{m}{M}}
  = M C_{n^\prime} r_0^{n^\prime-1}.
\nonumber
\end{equation}
If we relabel $n^\prime$ as $n$, then we see that the multipole coefficients
$C_n$ are given by
\begin{equation}
C_n = \frac{1}{M r_0^{n-1}} \sum_{m=0}^{M-1} B_m e^{-2\pi i n \frac{m}{M}}.
\label{eq:fittedmultipolepolar}
\end{equation}

The advantage of this technique over that in
Section~\ref{sec:multipolefitcartnbasis} is that the basis functions used
to construct the fit are of the form $e^{in\theta}$, for integer $n$.  These
functions are orthogonal: mathematically, this means that
\begin{equation}
\int_0^{2\pi} e^{i n\theta} e^{-i n^\prime \theta} \, d\theta
  = 2\pi \delta_{nn^\prime},
\nonumber
\end{equation}
where the Kronecker delta function $\delta_{nn^\prime} = 1$ if $n = n^\prime$,
and $\delta_{nn^\prime} = 0$ if $n \neq n^\prime$.  The important consequence
for us is that the value we determine for any given multipole using
Eq.~(\ref{eq:fittedmultipolepolar}) is independent of the presence of any other
multipoles, of higher or lower order.

A further advantage of using the polar basis instead of the Cartesian basis
comes from the dependence of the field on the radial distance.  Suppose that
the field data are measured (or obtained from a model) with accuracy $\Delta B_m$.
Then the accuracy in the multipole coefficients will be
\begin{equation}
\Delta C_n \approx \frac{\Delta B_m}{r_0^{n-1}}.
\nonumber
\end{equation}
We obtain better accuracy in the multipole coefficients if we choose the radius
$r_0$, on which the measurements are made, to be as large as possible.
Furthermore, the accuracy in the fitted field will be
\begin{equation}
\Delta B \approx \Delta C_n \left( \frac{r}{r_0} \right)^{n-1}.
\nonumber
\end{equation}
We obtain \emph{improved} accuracy in the field for $r<r_0$; but the accuracy
reduces quickly (particularly for higher-order multipoles) for $r>r_0$.
It is important to choose the radial distance $r_0$ large enough to enclose
all particles likely to pass through the magnet, otherwise results from
tracking may not be accurate.

\subsection{Multipole decomposition: some comments}

In this section, we have considered two techniques for deriving the multipole
components of two-dimensional magnetostatic fields.  We have seen that while
the multipole components can be obtained, in principle, from a simple
least-squares fit of a polynomial to the field components along one or
other of the coordinate axes, there are advantages to basing the fit on
field data obtained on a circle enclosing the origin, with as large a
radius as possible.  In the next section, we shall see how the idea of a
multipole expansion can be generalized to three dimensions, and how a
multipole decomposition can be performed in that case.  However, it is
worth pausing to consider in a little more detail some of the reasons
for wishing to represent a field as a set of (multipole) modes.

It is of course possible to represent a magnetic field using a set of
numerical field data, giving the three field components on points forming
a `mesh' covering the region of interest.  In some ways, this is a very
convenient representation, since it is the one usually provided directly
by a magnetic modelling code: further processing is usually required to
arrive at other representations.  However, while a numerical field map
in two dimensions is often a practical representation, in three dimensions
the amount of data in even a relatively simple magnet can become extremely
large, especially if a high resolution is required for the mesh.  A multipole
representation, on the other hand, provides the description of a magnetic
field as a relatively small set of coefficients, from which the field
components at any point can be reconstructed, using the basis functions.
In other words, a multipole representation is more `portable' than a
numerical field map.

Secondly, a representation based on a multipole expansion lends itself to
further manipulation in ways that a numerical field map does not.  For
example, any noise in the data (from measurement or computational errors)
can be `smoothed' by suppression of higher-order modes.  Conversely,
random errors can be introduced into data based on a model with perfect
symmetry by introducing multipole coefficients corresponding to `forbidden'
harmonics.  There will of course be issues surrounding the suppression
or enhancement of errors by adjusting the multipole coefficients; however,
one benefit of this approach is that for \emph{any} set of multipole
coefficients, the field is at least a physical field, in the sense of
satisfying Maxwell's equations.  The same will not usually
be true if, for example, a general smoothing algorithm is applied to a numerical field
map.

Finally, one of the main motivations for performing a multipole decomposition
of a field is to provide data in a format appropriate for many beam
dynamics codes.  Accuracy is one criterion often important for beam
dynamics codes: efficiency is another.  Characterization of a storage
ring frequently requires tracking of thousands of particles over hundreds
or thousands of turns, through a beam line that can easily consist of
hundreds of magnetic elements.  Numerical integration of the equations
of motion for a particle in a numerical field map is generally too slow
to be a practical method.  There are many techniques that can be used to
improve the efficiency of particle tracking in accelerators: one of the
most common is the `thin lens' method.  The dynamical effects of
dipoles and quadrupoles usually need to be represented with high accuracy.
Fortunately, for these magnets, it is possible to write down accurate
solutions to the equations of motion in closed form, allowing tracking
through a magnet of given length to be performed in a single step.  The
same is not true for sextupoles, or higher-order multipoles; however, it
is usually sufficiently accurate to represent such magnets by a model in
which the length of the magnet approaches zero, but where the integrated
strength (the multipole coefficient multiplied by the length) remains constant.
For such a `thin lens' it is possible to write down exact solutions to
the equations of motion, allowing tracking again to be performed in a
single step.  A quadrupole with higher-order multipole errors can be
represented as a `long' perfect quadrupole field, with a set of `thin'
multipoles at one end, or at the centre.  However, construction of such
a model for a tracking code requires a multipole decomposition of the
field obtained from a magnet modelling code.

We should emphasize that in our discussion of multipole decomposition, 
here and in Sections~\ref{sec:multipolefitcartnbasis} and \ref{sec:multipolefitpolarbasis}, we have made no clear distinction
between field data obtained from a computational model, or from
measurement of a real magnet.  Of course, it is much easier to obtain
the data required from a computational model: it is then quite
straightforward to perform the required decomposition to determine the
values of the various multipole coefficients.  Unfortunately, the data
do not include manufacturing errors, which can be very important.
Measurements provide more realistic data: however, many other issues
need to be addressed, including accuracy of field measurements, alignment
of the measurement instruments with respect to the magnet, etc.  Such
issues are beyond the scope of our discussion.

\section{Three-dimensional fields\label{sec:threedfields}}

In the previous sections, we have restricted ourselves to the case
where the magnetic field is independent of the longitudinal coordinate.
The multipole modes that we can use for such fields actually provide a
good description for many accelerator multipole magnets, even though
such magnets of course have finite length.  The ends or `fringe fields'
of dipoles, quadrupoles and so on, where the field strengths often vary
rapidly with longitudinal position, cannot be accurately represented by
two-dimensional fields; however, in many accelerators, only the fringe
fields of dipoles have a significant impact on the dynamics.

But there are cases where a full three-dimensional description of
a magnetic field is desirable, or even necessary.  For example, the fields
of insertion devices (wigglers and undulators) are often represented as
a sequence of short dipoles of alternating polarity; however, where the
period becomes small compared with the aperture, the three-dimensional
nature of the field can start to have effects that cannot be ignored.
There can even be cases where `conventional' multipoles designed for
special situations (for example, where very wide aperture is required,
and where the length of the magnet needs to be short, because of space
constraints) can have fringe fields that affect the dynamics to a
significant extent.

It is therefore of somewhat more than purely academic interest to consider
how two-dimensional multipole representations may be generalized to
three dimensions.  As usual, there are many different ways to approach
the problem: the method that is used will often depend on the problem to
be solved.  In the following sections, we describe two rather general
methods that may be of use in many situations arising in accelerators.
First, we consider a field expansion based on Cartesian modes.  While
this provides some nice illustrations, the Cartesian expansion does have
some disadvantages.  To address these disadvantages, we describe how a
field expansion based on polar coordinates can be performed.

\subsection{Cartesian modes\label{sec:threedcartesian}}

Consider the field given by
\begin{eqnarray}
B_x & = & -B_0 \frac{k_x}{k_y} \sin (k_x x) \sinh (k_y y) \sin (k_z z),
\label{eq:threedcartnbx} \\
B_y & = & B_0 \cos (k_x x) \cosh (k_y y) \sin (k_z z),
\label{eq:threedcartnby} \\
B_z & = & B_0 \frac{k_z}{k_y} \cos (k_x x) \sinh (k_y y) \cos (k_z z).
\label{eq:threedcartnbz}
\end{eqnarray}
As may easily be verified, this field satisfies
\begin{equation}
\curlop \vec{B} = 0.
\nonumber
\end{equation}
Furthermore, the equation
\begin{equation}
\divop \vec{B}  = 0
\nonumber
\end{equation}
is satisfied if
\begin{equation}
k_y^2 = k_x^2 + k_z^2. \label{eq:modenumbersconstraint}
\end{equation}
We conclude that, as long as the constraint (\ref{eq:modenumbersconstraint})
is satisfied, the fields (\ref{eq:threedcartnbx})--(\ref{eq:threedcartnbz})
provide solutions to Maxwell's equations in regions with constant permeability,
and static (or zero) electric fields.  Of course, it is possible to find similar
sets of equations but with different `phase' along each of the coordinate axes,
and with the hyperbolic trigonometric function appearing for the dependence
on $x$ or $z$, rather than $y$.  By superposing fields, with appropriate
variations on the form given by
Eqs.~(\ref{eq:threedcartnbx})--(\ref{eq:threedcartnbz}), it is possible to
construct quite general three-dimensional magnetic fields.  For example, a
slightly more general field than that given by
Eqs.~(\ref{eq:threedcartnbx})--(\ref{eq:threedcartnbz}) can be obtained simply
by superposing fields with different mode numbers and amplitudes:
\begin{eqnarray}
B_x & = & -\int\!\!\!\int \tilde{B}(k_x, k_z)
\frac{k_x}{k_y} \sin (k_x x) \sinh (k_y y) \sin (k_z z) \, dk_x \, dk_z,
\label{eq:threedcartnbxdkxdkz} \\
B_y & = & \int\!\!\!\int \tilde{B}(k_x, k_z)
\cos (k_x x) \cosh (k_y y) \sin (k_z z) \, dk_x \, dk_z,
\label{eq:threedcartnbydkxdkz} \\
B_z & = & \int\!\!\!\int \tilde{B}(k_x, k_z)
\frac{k_z}{k_y} \cos (k_x x) \sinh (k_y y) \cos (k_z z) \, dk_x \, dk_z.
\label{eq:threedcartnbzdkxdkz}
\end{eqnarray}
In this form, we see already how to perform a mode decomposition, i.e.,
how we can determine the coefficients $\tilde{B}(k_x, k_z)$ as functions
of the `mode numbers' $k_x$ and $k_z$.  If we consider in particular the
vertical field component on the plane $y = y_0$, then we have from
(\ref{eq:threedcartnbydkxdkz})
\begin{equation}
\frac{B_y}{\cosh (k_y y_0)} = \int\!\!\!\int \tilde{B}(k_x, k_z)
\cos (k_x x) \sin (k_z z) \, dk_x \, dk_z.
\nonumber
\end{equation}
Hence $\tilde{B}(k_x, k_z)$ may be obtained from an inverse Fourier
transform of $B_y(x,z) / \!\cosh (k_y y_0)$.  Given field data on a grid
over $x$ and $z$, we can then perform numerically an inverse discrete
Fourier transform, to obtain a set of coefficients $\tilde{B}(k_x, k_z)$.
Note that once we have obtained these coefficients, we can then
reconstruct all field components at all points in space.  This is an
important consequence of the strong constraints on the fields provided
by Maxwell's equations: in general, for a static field, if we know how
one field component varies over a two-dimensional plane, then we can
deduce how all the field components vary over all space (on and off the
plane).

Let us consider an example.  To keep things simple, we shall again work
with the case where the field is independent of one coordinate: now, however,
we shall assume that the fields are independent of the horizontal transverse,
rather than the longitudinal coordinate.  This may be a suitable model for a
planar wiggler or undulator with very wide poles.  The model may of course be
extended to include dependence of the fields on the horizontal transverse
coordinate: although our immediate example strictly deals with a
two-dimensional field, the extension to three dimensions is quite
straightforward.

Suppose that the mode amplitude function $\tilde{B}(k_x, k_z)$ has the form
\begin{equation}
\tilde{B}(k_x, k_z) = \delta(k_x) \tilde{B}(k_z),
\label{eq:deltamode}
\end{equation}
where $\delta(k_x)$ is the Dirac delta function.  The delta function has the
property that, for any function $f(k_x)$,
\begin{equation}
\int_{-\infty}^\infty \delta(k_x) f(k_x) \, dk_x = f(0).
\nonumber
\end{equation}
Using (\ref{eq:deltamode}) in Eqs.~(\ref{eq:threedcartnbxdkxdkz})--(\ref{eq:threedcartnbzdkxdkz}) gives
\begin{eqnarray}
B_x & = & 0,
\nonumber \\
B_y & = & \int \tilde{B}(k_z) \cosh (k_z y) \sin (k_z z) \, dk_z,
\nonumber \\
B_z & = & \int \tilde{B}(k_z) \sinh (k_z y) \cos (k_z z) \, dk_z.
\nonumber
\end{eqnarray}
There is no horizontal transverse field component, and the vertical and
longitudinal field components have no dependence on $x$: we have a
two-dimensional field.  In a plane defined by a particular value for the
vertical coordinate, $y = y_0$, the vertical field component is given by
\begin{equation}
B_y(z) = \cosh (k_z y_0) \int \tilde{B}(k_z) \sin (k_z z) \, dk_z.
\nonumber
\end{equation}
The mode amplitude function can be obtained from a Fourier transform of
the vertical component of the magnetic field on the plane $y = y_0$.
Usually, we will have a finite set of field data, obtained from a magnet
modelling code, or from measurements on a real device.

Suppose that we have a data set of $2M+1$ vertical field measurements,
taken at locations
\begin{equation}
y=y_0, \qquad z = \frac{m}{M}\hat{z},
\end{equation}
where $m$ is an integer in the range $-M \leq m \leq M$.  The field at
any point is given by
\begin{equation}
B_y(y,z) =  \sum_{m=-M}^{M} \tilde{B}_m \cosh (m k_z y) \sin (m k_z z),
\nonumber
\end{equation}
where
\begin{equation}
k_z = \frac{2\pi}{2\hat{z}}.
\nonumber
\end{equation}
Note that in this case, the field is antisymmetric about $z=0$, i.e.,
\begin{equation}
B_y(y,-z) = - B_y(y,z).
\nonumber
\end{equation}
The mode amplitudes $\tilde{B}_m$ are obtained by
\begin{equation}
\tilde{B}_m  = \frac{1}{\cosh (m k_z y_0)} \frac{1}{2M}
               \sum_{m^\prime=-M}^{M} B_y(y_0,z) \sin (m^\prime k_z z).
\nonumber
\end{equation}
Note that, because of the antisymmetry of the field
\begin{equation}
\tilde{B}_{-m} = -\tilde{B}_m.
\nonumber
\end{equation}

\begin{figure}
\centering
\includegraphics[width=0.45\linewidth]{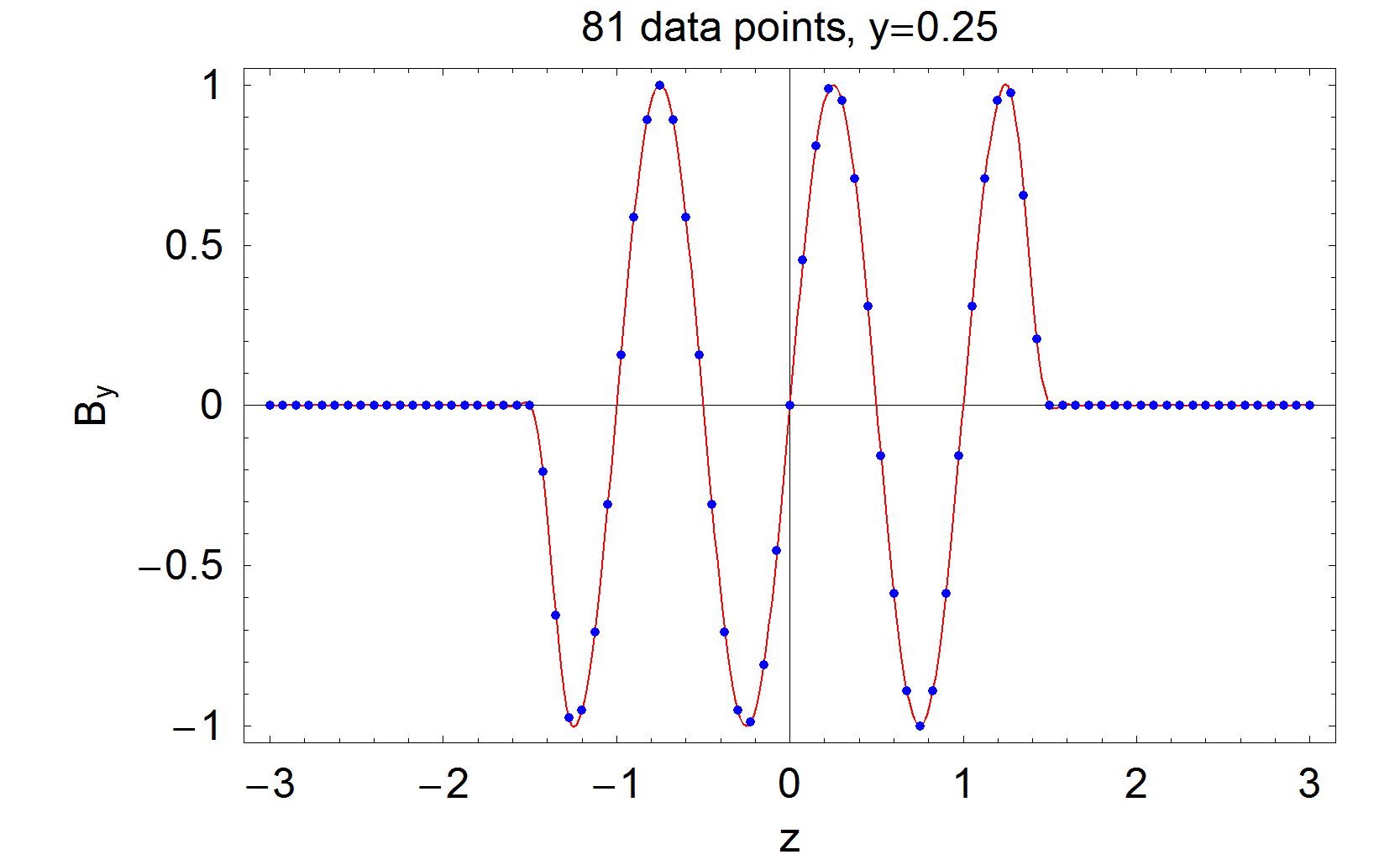}
\hspace{0.05\linewidth}
\includegraphics[width=0.45\linewidth]{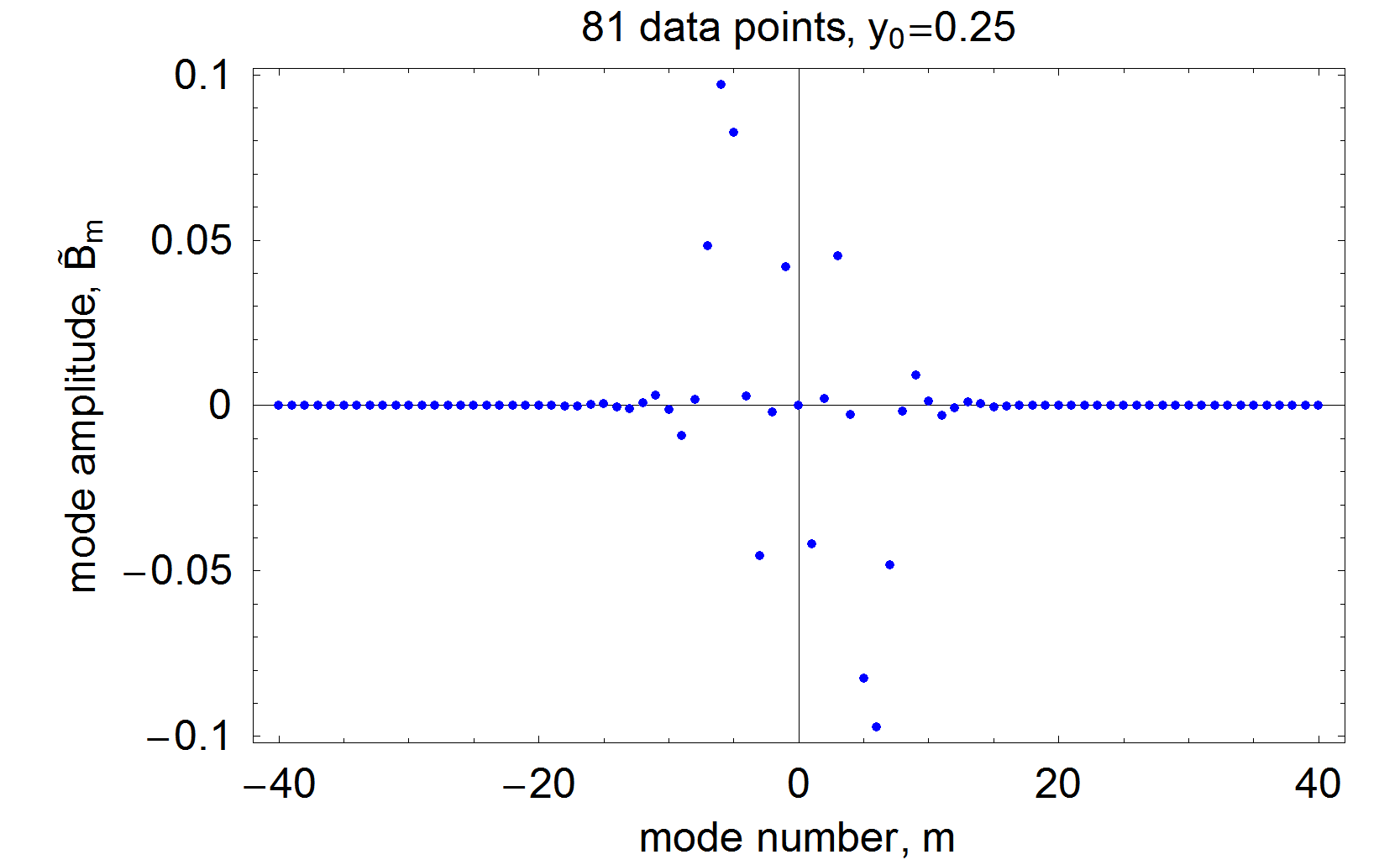}
\caption{Left: Field data (points) and fit (line) in a magnet with dependence
of the field on longitudinal coordinate $z$.  Right: Mode amplitudes.  The
field data consist of 81 measurements (or computations) at equally-spaced
points from $z=-3$ to $z=+3$, and $y=0.25$.
\label{fig:threedcartndata81}}
\end{figure}

As a specific numerical example, let us construct an `artificial' data
set along a line $y=y_0=0.25$, and with $\hat{z} = 3$.  The data are
constructed using a function that gives a sinusoidal variation in the field
along $z$ up to $|z|<1.25$; then a continuous and smooth (continuous first
derivative) fall-off to zero field for $|z|>1.5$.  For this
numerical example, we do not worry unduly about units: the reader may
assume lengths in cm, fields in kG, or any other preferred units.  Initially,
we take $M=40$, i.e., we assume we have 81 measurements of the field (or,
we have computed the field from a model at 81 equally-spaced points along $z$;
strictly speaking, because we are dealing with the case where the field is
antisymmetric in $z$, we need only half this number of field measurements or
computations).  The field `data', the fitted field (reconstructed using the
mode amplitudes) in $y=0.25$, and the mode amplitudes, are shown in
Fig.~\ref{fig:threedcartndata81}.

\begin{figure}
\centering
\includegraphics[width=0.45\linewidth]{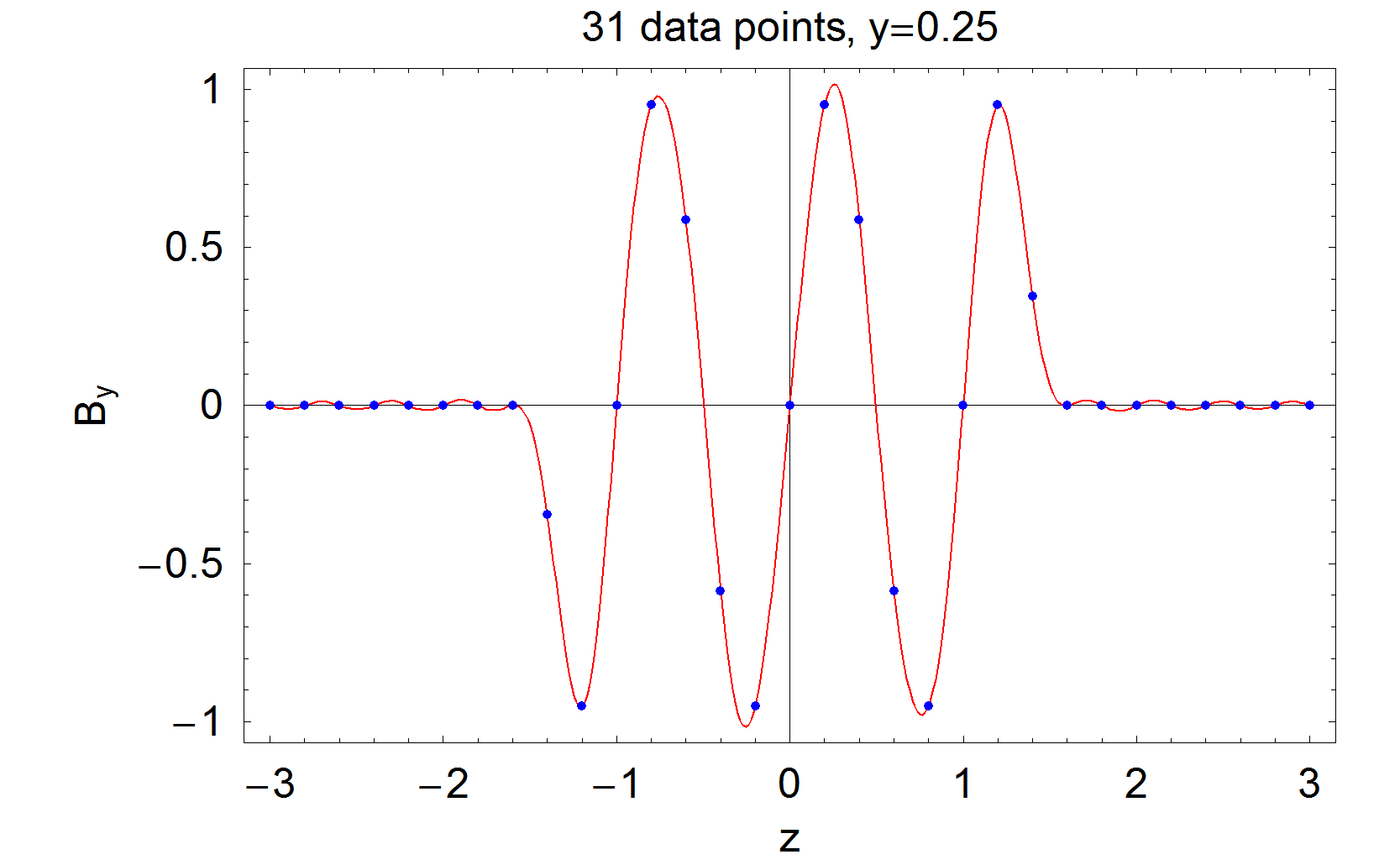}
\hspace{0.05\linewidth}
\includegraphics[width=0.45\linewidth]{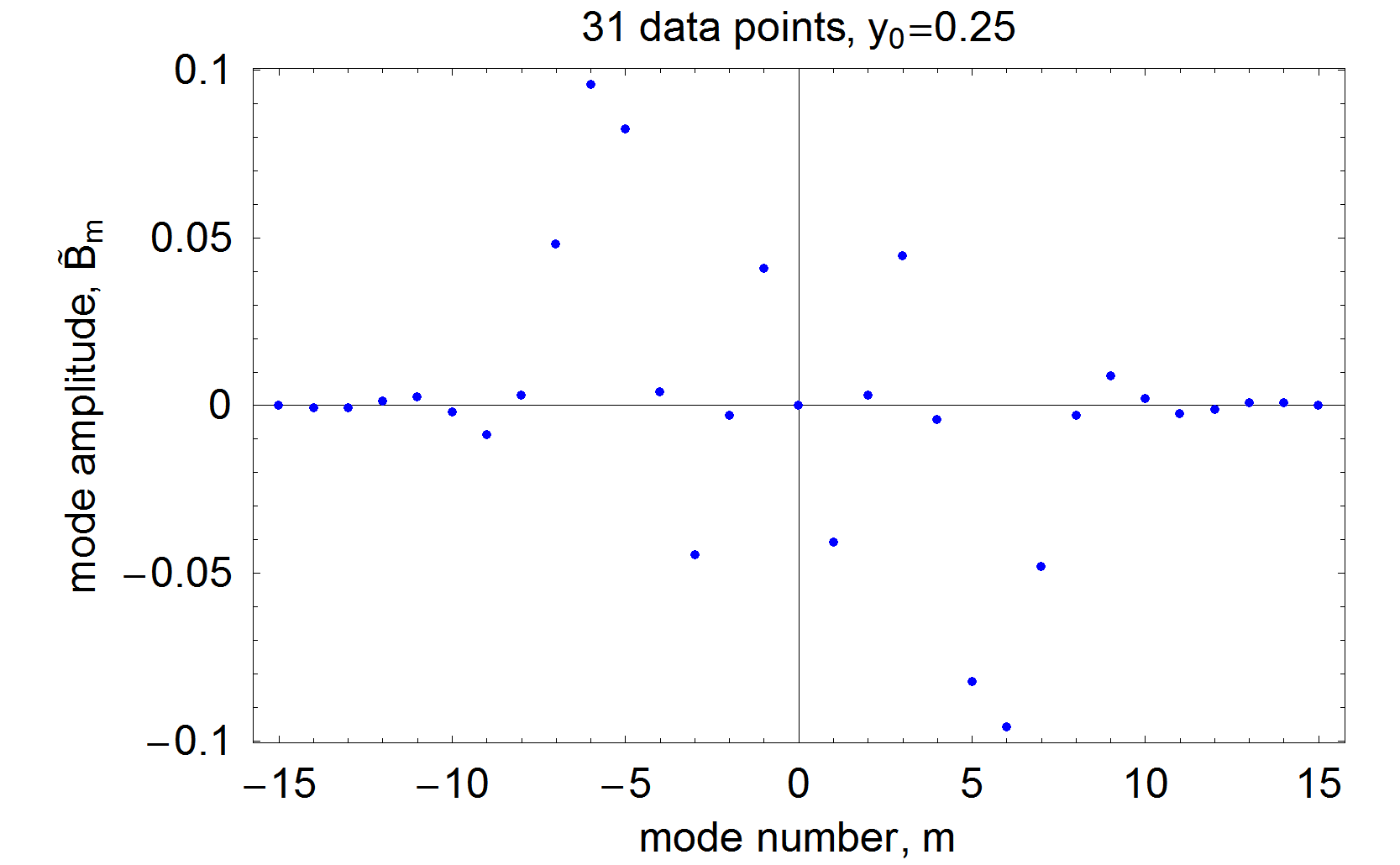}
\caption{Left: Field data (points) and fit (line) in a magnet with dependence
of the field on longitudinal coordinate $z$.  Right: Mode amplitudes.  The
field data consist of 31 measurements (or computations) at equally-spaced
points from $z=-3$ to $z=+3$, and $y=0.25$.
\label{fig:threedcartndata31}}
\end{figure}

It is interesting to compare with the situation where we have only 31 field
measurements or computations, i.e., $M=15$.  Using the same function that we
used to construct the data set with 81 data points, we produce the fit and
the mode amplitudes shown in Fig.~\ref{fig:threedcartndata31}.  Comparing
Figs.~\ref{fig:threedcartndata81} and \ref{fig:threedcartndata31}, we see
that in both cases the fitted field does pass exactly through all the
data points.  This is a necessary consequence of the fit, which is based on
a discrete Fourier transform of the data points.  However, using only 31
data points, there is a significant oscillation of the fitted field between
the data points in the region $1.5<|z|<3$, where the field is actually zero
(by construction).  This is a consequence of the fact that we have `truncated'
some modes with non-negligible amplitude.  The mode amplitudes in both cases
are the same for mode numbers $-15 \leq m \leq m$; but with 81 data points
we can determine amplitudes for a larger number of modes, which gives us a
more accurate interpolation between the data points. 

\begin{figure}
\centering
\includegraphics[width=0.45\linewidth]{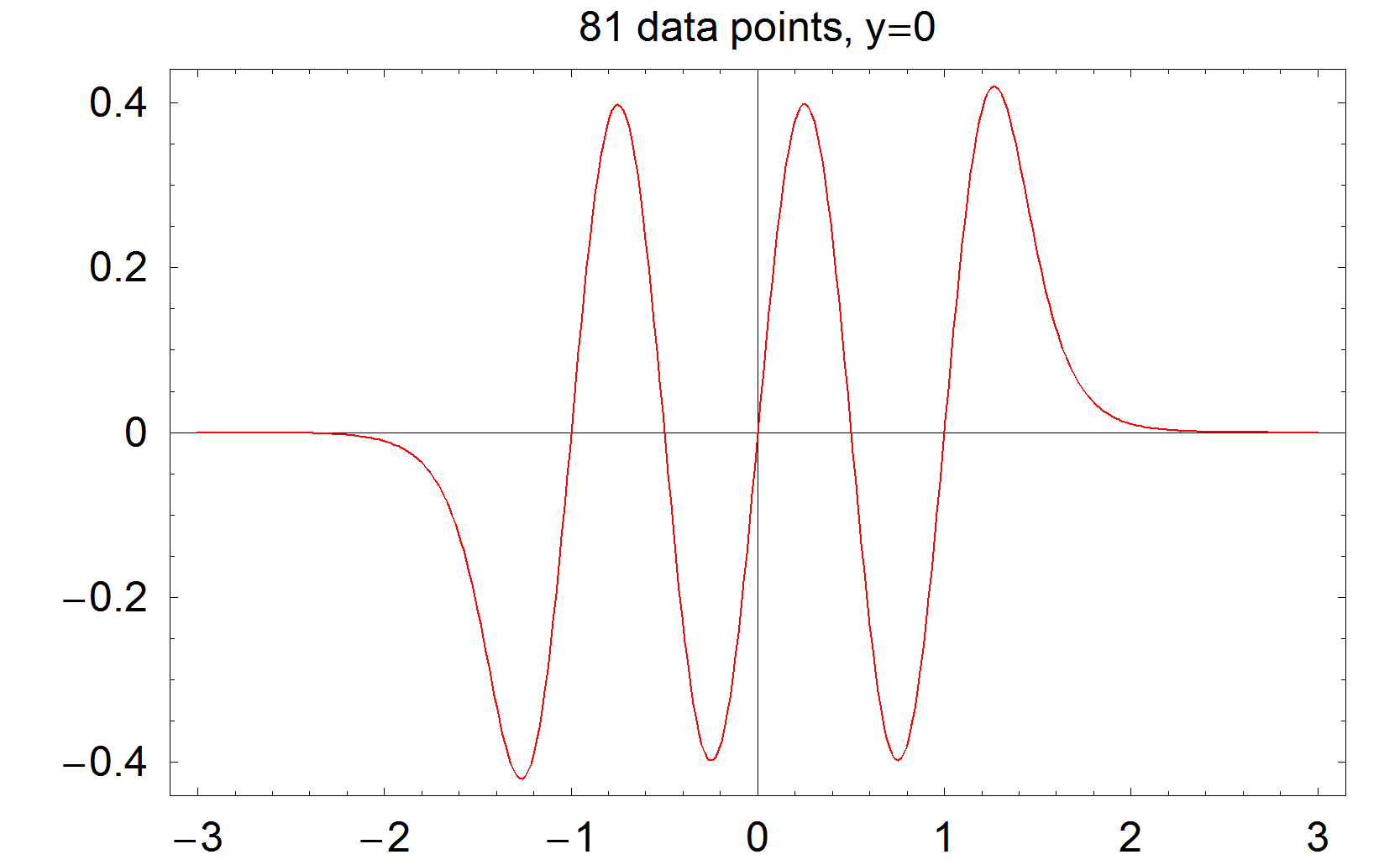}
\hspace{0.05\linewidth}
\includegraphics[width=0.45\linewidth]{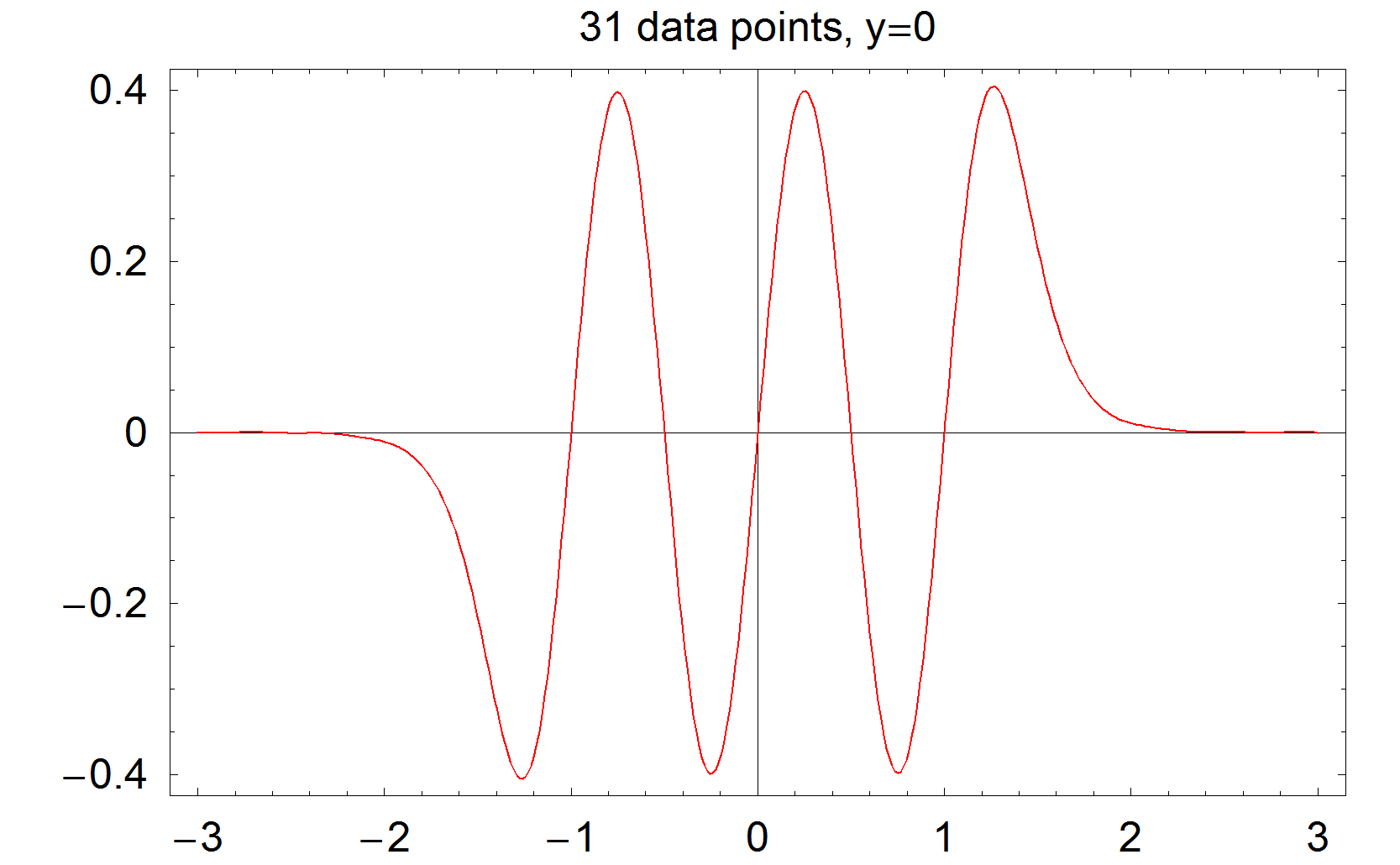}
\caption{Field on the plane $y = 0$ determined from fits to the data shown in
Figs.~\ref{fig:threedcartndata81} and \ref{fig:threedcartndata31}.
Left: fit determined from data set with 81 data points.  Right: fit
determined from data set with 31 data points.
\label{fig:threedcartnfity0}}
\end{figure}

Having obtained fits to the field in the plane $y = 0.25$, we can reconstruct
the field at any point, on or off the plane.  It is often of interest to look
at the mid-plane; usually, this is defined by $y = 0$.  In this plane, we do
not have any field data.  However, we can compare the field produced by the
fits with 81 data points and with 31 data points: these fields are shown in
Fig.~\ref{fig:threedcartnfity0}.  We see that the fit based on 31 data points
produces an essentially identical field on $y=0$ as the fit based on 81 data
points.  (The data points in each case are taken on the plane $y=0.25$.)  This
is a consequence of the `suppression' of higher-order modes, that arises
from the hyperbolic dependence of the field on the $y$ coordinate.

\begin{figure}
\centering
\includegraphics[width=0.45\linewidth]{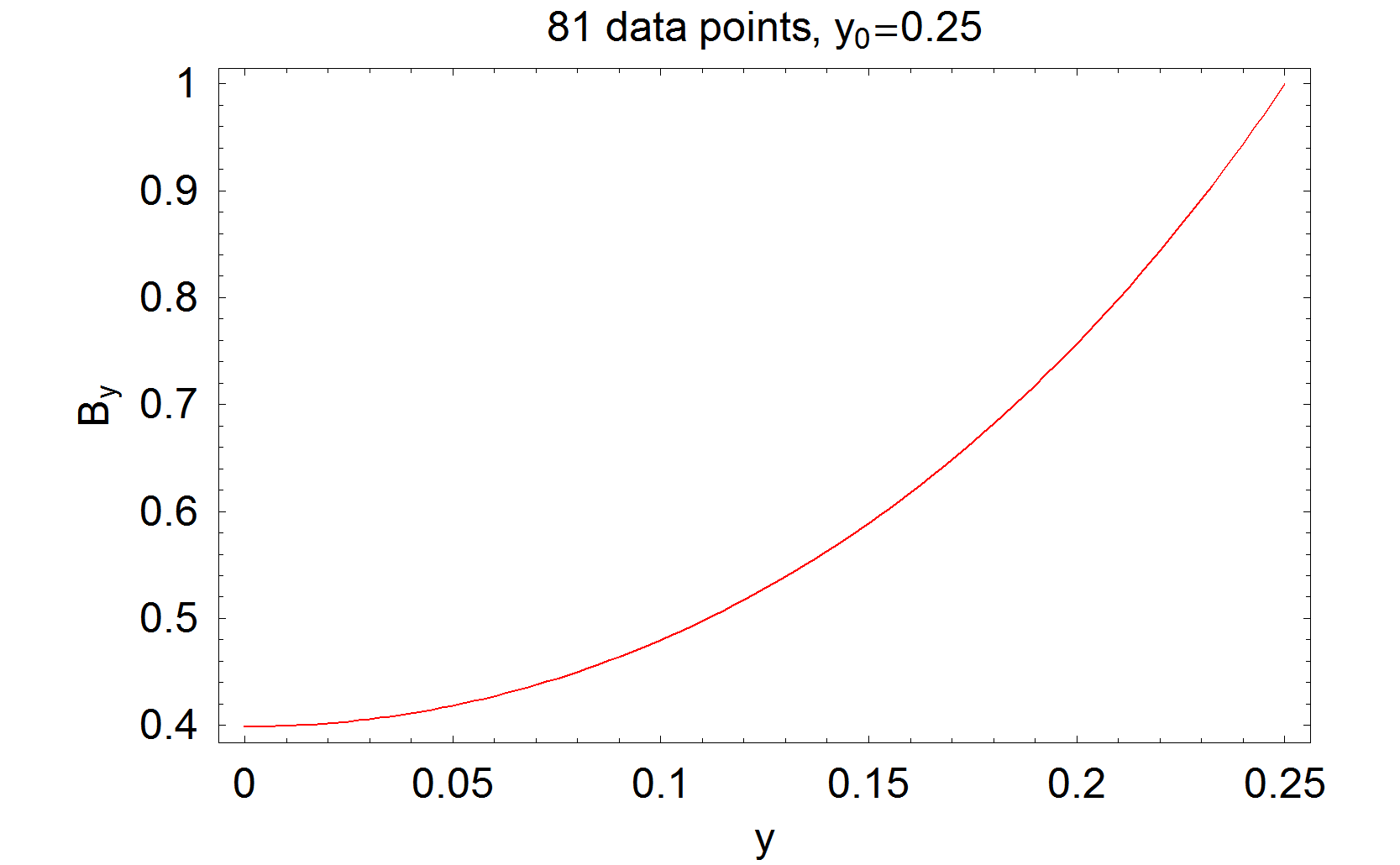}
\hspace{0.05\linewidth}
\includegraphics[width=0.45\linewidth]{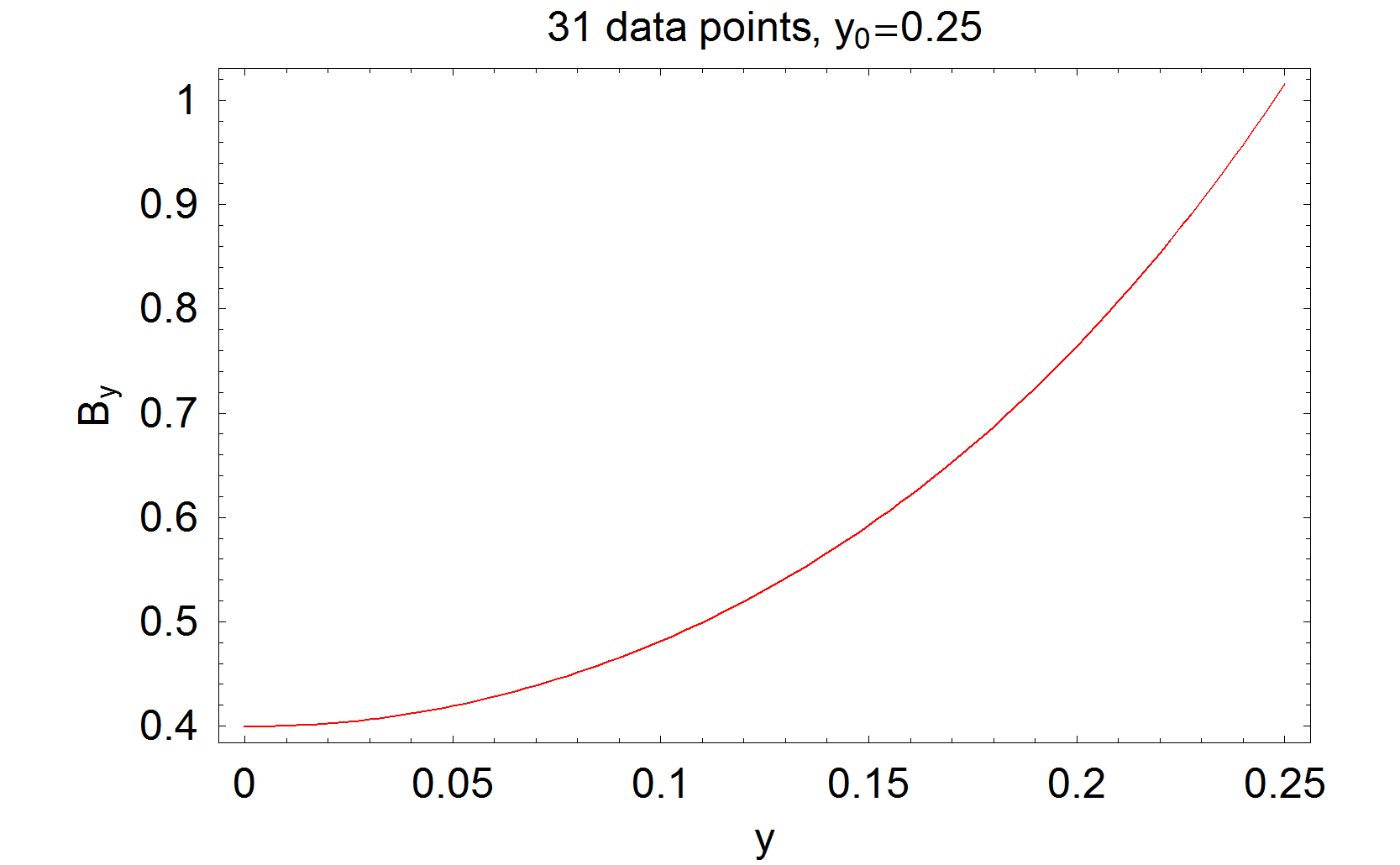}
\caption{Vertical field component as a function of $y$, for $z=0.25$.
The field is determined from fits to the data shown in
Figs.~\ref{fig:threedcartndata81} and \ref{fig:threedcartndata31}.
Left: fit determined from data set with 81 data points.  Right: fit
determined from data set with 31 data points.
\label{fig:threedcartnfitz025}}
\end{figure}

To emphasize the significance of the hyperbolic dependence of the field on
the vertical coordinate, we can look at the variation of $B_y$ with $y$, for
a given value of $z$.  We choose $z=0.25$, which corresponds to a peak in
the vertical field component as a function of $z$.  The variation of $B_y$
with $y$ for the two cases (fit based on 81 data points, and fit based on
31 data points) is shown in Fig.~\ref{fig:threedcartnfitz025}.

\begin{figure}
\centering
\includegraphics[width=0.45\linewidth]{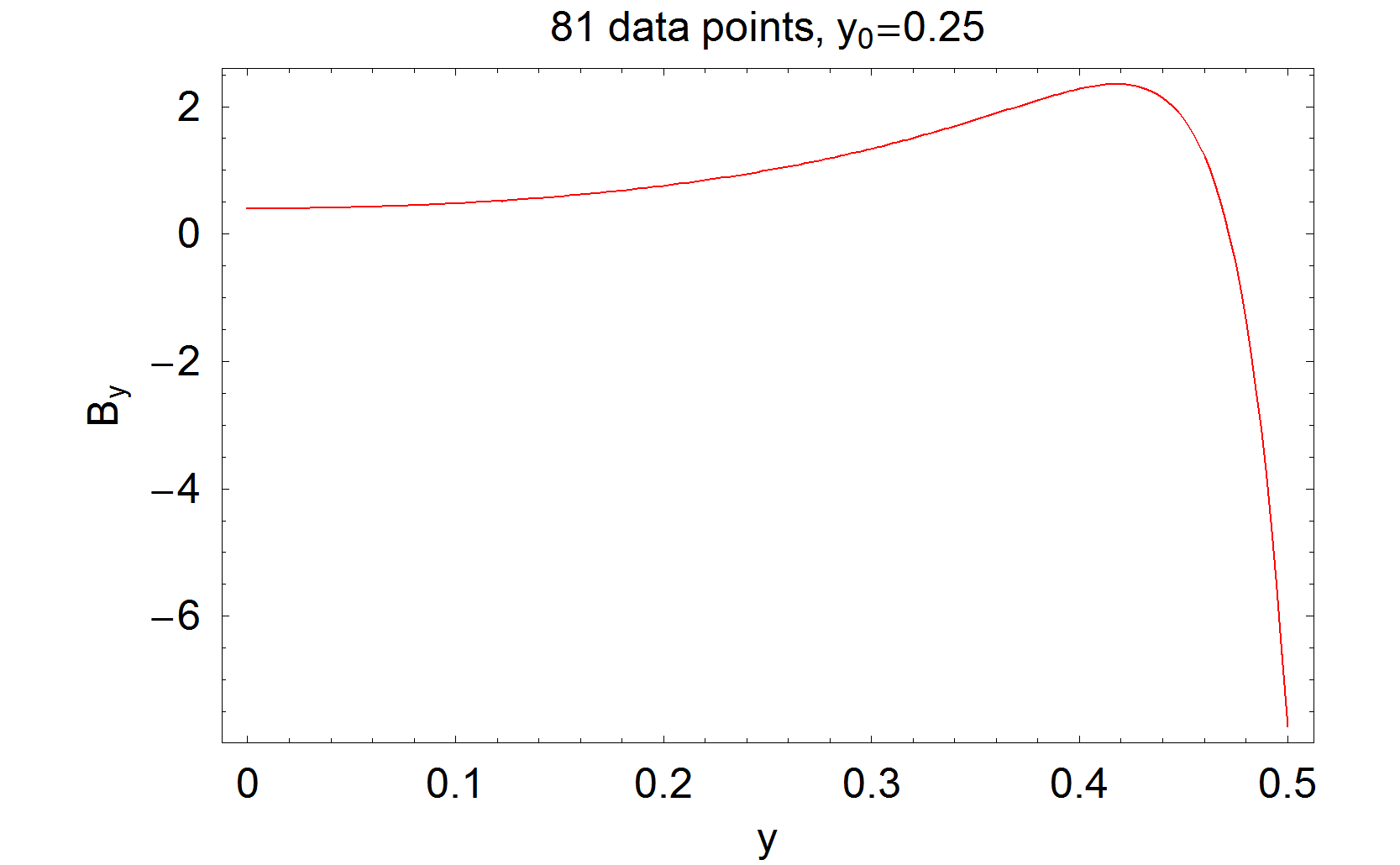}
\hspace{0.05\linewidth}
\includegraphics[width=0.45\linewidth]{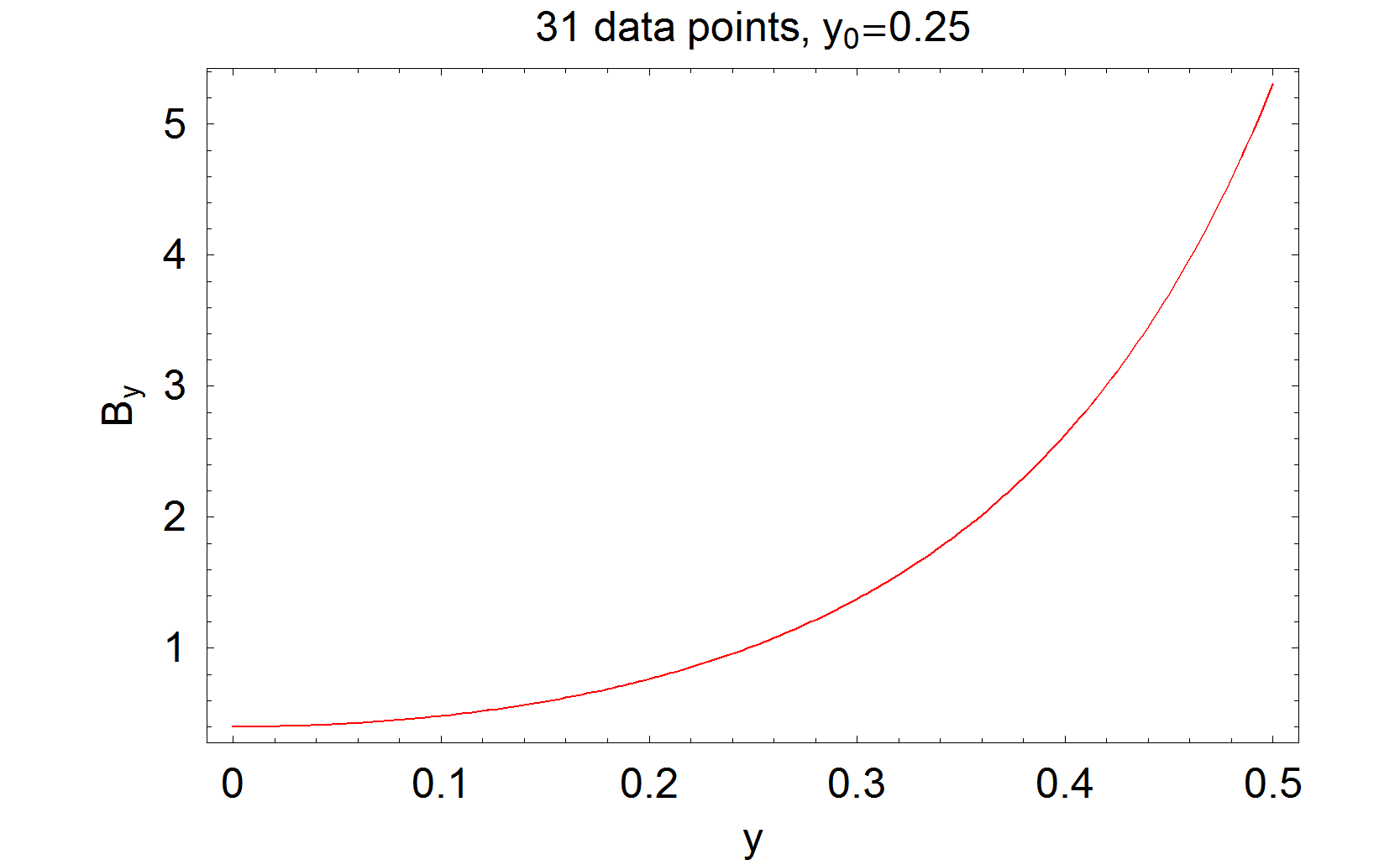}
\caption{Vertical field component as a function of $y$, for $z=0.25$.
The field is determined from fits to the data shown in
Figs.~\ref{fig:threedcartndata81} and \ref{fig:threedcartndata31}.
Left: fit determined from data set with 81 data points.  Right: fit
determined from data set with 31 data points.
\label{fig:threedcartnfitz025largey}}
\end{figure}

Up to $y=0.25$, the two fits give essentially the same field.  However,
if we try to extrapolate beyond this plane (the plane on which the fit
was performed), we see dramatically different behaviour.
Figure~\ref{fig:threedcartnfitz025largey} compares the vertical field
component obtained from the two fits (81 data points and 31 data points),
again at $z=0.25$, but now with a range of $y$ from 0 to 0.5.  In one
case (81 data points), the field increases to a maximum before dropping
rapidly.  In the other case (31 data points), the field increases
monotonically over the range.  The reason for the different behaviour
is the additional modes in the fit to the set of 81 data points.  These
higher order modes make only a small additional contribution to the field
for $|y|<0.25$; but for values of the vertical coordinate beyond this
value, because of the hyperbolic dependence of $y$, the contribution of
these modes becomes increasingly significant, and eventually, dominant.

The behaviour of the field fits for $|y|>0.25$ is a clear illustration
of why it is dangerous to extrapolate the fit beyond the region enclosed
by the plane of the fit.  In this case, because of the symmetry in the
vertical direction, the region enclosed is between the planes $y=-0.25$
and $y=+0.25$.  The `safe' region is also bounded in $z$, by
$z=-0.3$ and $z=+0.3$; because we use discrete mode numbers in $z$, the
fitted fields will in fact be periodic in $z$, and will repeat with
period $z=0.6$.  In general, there will be similar periodicity in $x$;
however, in this particular example, we analysed a field that was
independent of $x$, so the `safe' region of the fit is unbounded in
$x$.

\subsection{Cylindrical modes\label{sec:threedpolar}}

The Caretsian modes discussed in Section~\ref{sec:threedcartesian}
are often useful for describing fields in insertion devices, particularly
those that have weak variation of the field with $x$, and periodic
behaviour in $z$ (over some range): because the modes `reflect' the
geometry, it is often possible to achieve good fits to a given field
using a small number of modes.  To maximize the region over which the
fit is reliable, one needs to choose a plane with a value of $y$ as large
as possible, with $x$ and $z$ extending out as far as possible on this
plane.  For a planar undulator or wiggler, it is often possible to choose
a plane close to the pole tips in which $x$ in particular extends over
the entire vacuum chamber.

However, for other geometries, the Cartesian modes may not provide a
convenient basis.  For example, if the magnet has a circular aperture,
then the plane that provides the largest range in $x$ is the mid-plane,
$y=0$, and as $y$ increases, the available range in $x$ decreases.  To
base the fit on the Cartesian basis requires some compromise between
the range of reliability in the horizontal transverse and vertical
directions.

Fortunately, it is possible to choose an alternative basis for magnets
with circular aperture, in which the field fit can be based on the
surface of a cylinder inscribed through the magnet.  In that case, the
radius of the cylinder can be close to the aperture limit,
maximizing the range of reliability of the fit.  The appropriate
modes in this case are most easily expressed in cylindrical polar
coordinates.

\begin{figure}
\centering
\includegraphics[width=0.6\linewidth]{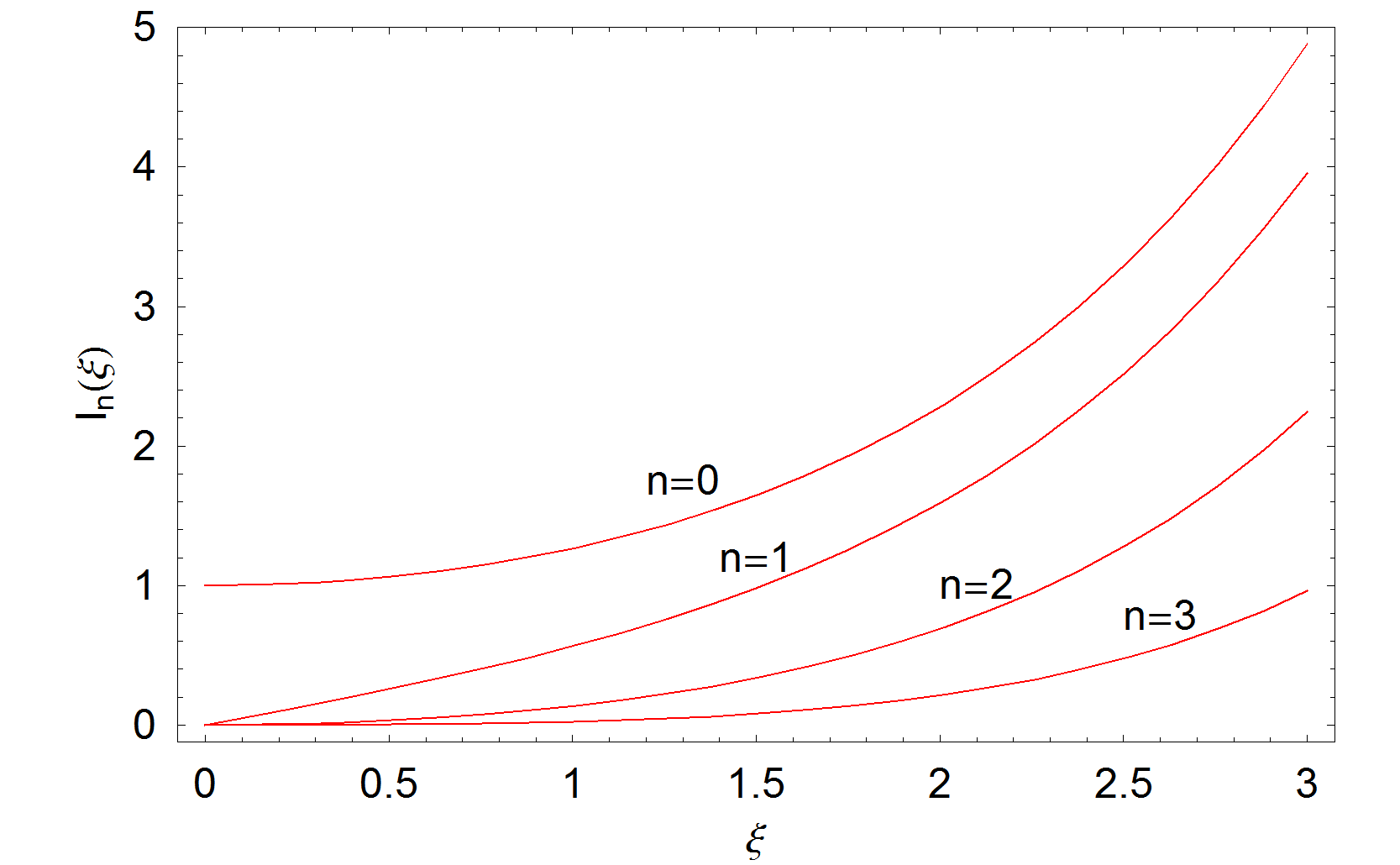}
\caption{Modified Bessel functions of the first kind, of order $n=0$
to $n=3$
\label{fig:modifiedbesselfunctions}}
\end{figure}

A field with zero divergence and curl (and hence satisfying Maxwell's
equations for static fields in regions with uniform permeability) is
given by
\begin{eqnarray}
B_r      & = & \int dk_z \sum_n \tilde{B}_n(k_z) \, I^\prime_n(k_z r)
               \sin (n\theta) \cos (k_z z), \label{eq:threedmodepolarbr} \\
B_\theta & = & \int dk_z \sum_n \tilde{B}_n(k_z) \, \frac{n}{k_z r} I_n(k_z r)
          \cos (n\theta) \cos (k_z z), \label{eq:threedmodepolarbtheta} \\
B_z      & = &-\int dk_z \sum_n \tilde{B}_n(k_z) \, I_n(k_z r)
          \sin (n\theta) \sin (k_z z). \label{eq:threedmodepolarbz}
\end{eqnarray}
Here, $I_n(k_z r)$ is the modified Bessel function of the first kind, of
order $n$.  Modified Bessel functions of the first kind for order $n=0$
to $n=3$ are plotted in Fig.~\ref{fig:modifiedbesselfunctions}.  For small
values of the argument $\xi$, the modified Bessel function of order $n$ has
the series expansion
\begin{equation}
I_n(\xi) = \frac{\xi^n}{2^n \Gamma\!(1+n)} + O(n+1).
\label{eq:besseliexpansion}
\end{equation}
For larger values of the argument, the modified Bessel functions $I_n(\xi)$
increase exponentially.  This is significant: it means that if we fit a field
to data on the surface of a cylinder of given radius, then residuals of the
fit will decrease exponentially within the cylinder towards $r=0$, and
increase exponentially outside the cylinder with increasing $r$.  The `safe'
region of the fit will be within the cylinder.

Note that Eqs.~(\ref{eq:threedmodepolarbr})--(\ref{eq:threedmodepolarbz})
may be generalized to include different `phases' in the azimuthal angle
$\theta$ and the longitudinal coordinate $z$.

An attractive feature of the polar basis is that it is possible to draw
a direct connection between the three-dimensional modes in this basis and
the multipole components in a two-dimensional field.  Consider a
mode amplitude $\tilde{B}_n(k_z)$ given (for some particular value of $n$) by
\begin{equation}
\tilde{B}_n(k_z) = 2^n \Gamma\!(1+n)C_n\frac{\delta(k_z)}{nk_z^{n-1}},
\label{eq:threedmultipolecoefficients}
\end{equation}
where $\delta (k)$ is the Dirac delta function, and $C_n$ is a constant.
Substituting these mode amplitudes into
Eqs.~(\ref{eq:threedmodepolarbr})--(\ref{eq:threedmodepolarbz}), using
the expansion (\ref{eq:besseliexpansion}), and performing the integral
over $k_z$ gives
\begin{eqnarray}
B_r      & = & \sum_n C_n r^{n-1} \sin (n\theta), \nonumber \\
B_\theta & = & \sum_n C_n r^{n-1} \cos (n\theta), \nonumber \\
B_z      & = & 0. \nonumber
\end{eqnarray}
Comparing with Eq.~(\ref{eq:multipolepolarcoords}), we see that this is a
multipole field of order $n$.  Thus a two-dimensional multipole field is
a special case of a three-dimensional field
(\ref{eq:threedmodepolarbr})--(\ref{eq:threedmodepolarbz}), with mode coefficient
given by Eq.~(\ref{eq:threedmultipolecoefficients}).

In general, the mode coefficients $\tilde{B}_n(k_z)$ may be obtained by
a Fourier transform of the field on the surface of a cylinder of given radius.
For example, it follows from Eq.~(\ref{eq:threedmodepolarbr}) that
\begin{equation}
\frac{B_r}{I^\prime_n(k_z r)} =
\int dk_z \sum_n \tilde{B}_n(k_z) \,
                 \sin (n\theta) \cos (k_z z). \nonumber
\end{equation}
An elegant feature of the polar basis, as compared to the Cartesian basis
discussed in Section~\ref{sec:threedcartesian}, is that the modes reflect
the real periodicity of the field in the angle coordinate $\theta$.  In the
Cartesian basis, the modes were periodic in $x$, although the field, in
general, would not have any periodicity in $x$.

Since the mode coefficients $\tilde{B}_n(k_z)$ are related to the multipole
coefficients in a two-dimensional field, we can use these coefficients to
extend the idea of a multipole to a three-dimensional field.  Strictly speaking,
the mode coefficients $\tilde{B}_n(k_z)$ are related to the field by a
two-dimensional Fourier transform; however, we can perform a one-dimensional
inverse Fourier transform (in the $z$ variable) to obtain a set of
functions which represent, in some sense, the `multipole components'
of a three-dimensional field as a function of $z$.  Here, we use the
term `multipole components' rather loosely, since a multipole field is
strictly defined only in the two-dimensional case (i.e., for a field that
is independent of the longitudinal coordinate).  A quantity that is perhaps
easier to interpret is the contribution to the field at any point made by
the mode coefficients $\tilde{B}_n(k_z)$ with a given $n$.  For $n=1$,
the field components $B_r$ and $B_\theta$ at any point in $z$ will behave as
for a dipole field; for $n=2$, $B_r$ and $B_\theta$ will behave as for a
quadrupole field, and so on.

As an illustrative example, we consider the field in a specific device:
the wiggler in a damping ring for TESLA (a proposed linear collider)
\cite{bib:teslatdr}.
This wiggler has a peak field of  1.6\,T and period 400\,mm; the total
length of wiggler in each of the TESLA damping rings would be over 400\,m.
The field in the wiggler has been studied extensively, because of concerns
that dynamical effects associated with the nonlinear components in the
field would limit the acceptance of the damping ring
\cite{bib:ilcdrconfigstudies}.  A model was
constructed for one quarter period of the magnet, which allowed the field
at any point within the body of the magnet to be computed.  Effects
associated with the ends of the wiggler were neglected, but could in
principle be included in the study.  By performing a mode decomposition 
using the techniques described above, it was possible to construct an
accurate dynamical model allowing fast tracking to characterize the
acceptance of the damping ring.  The methods used for the dynamical
analysis are beyond the scope of the present discussion; however, we
present the results of the analysis relating directly to the field, to
illustrate the methods described in this section.

\begin{figure}
\centering
\includegraphics[width=0.8\linewidth]{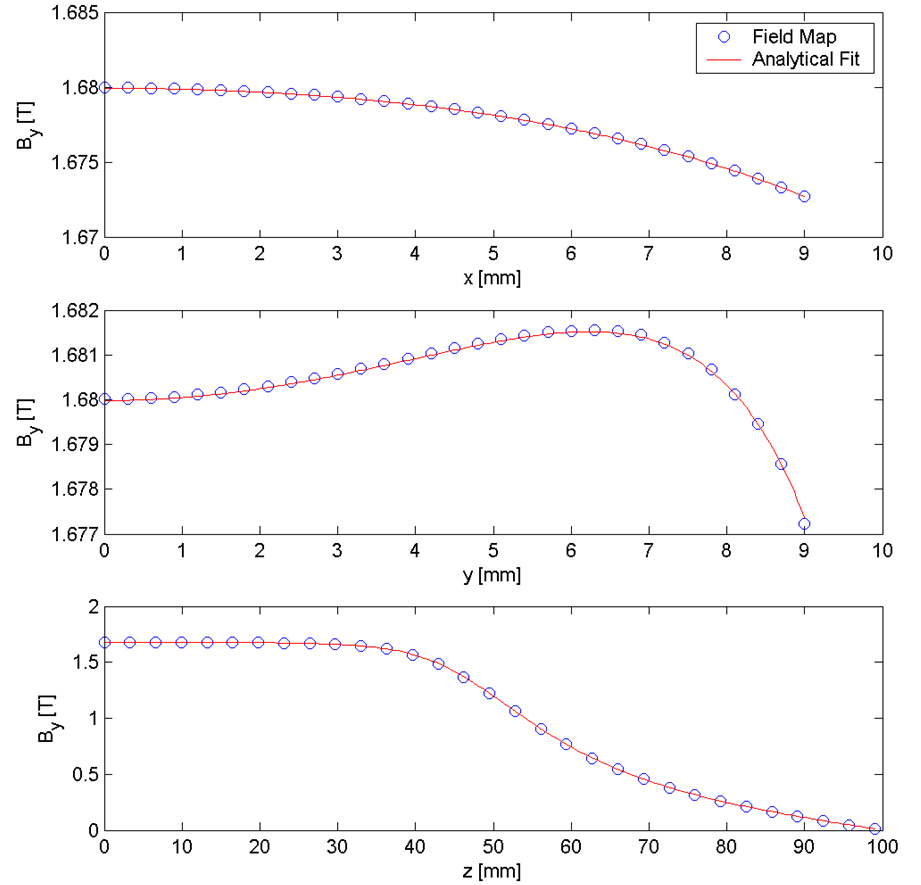}
\caption{Fit to the field of one quarter of one period of the TESLA
damping ring wiggler
\label{fig:teslafieldfit}}
\end{figure}

A model of the wiggler was used to compute the magnetic field on a
mesh of points bounded by a cylinder of radius 9\,mm, within one
quarter period of the wiggler.  Although all field components were
computed on the mesh, which covered the interior of the cylinder as
well as the surface, only the radial field component on the surface
of the cylinder was used to calculate the mode amplitudes.  The fit
can be validated by comparing the field `predicted' by the fit
with the field data (from the computational model) not used directly
in the fitting procedure.  A fit achieved using 7 azimuthal and 100
longitudinal modes is shown in Fig.~\ref{fig:teslafieldfit}.  Each
plot shows the variation of the vertical field as a function of one
Cartesian coordinate, with the other two coordinates fixed at zero.
In the vertical direction, the range shown is from the mid-plane of
the wiggler to close to the pole tip.  Note that the variation in
the field in the transverse ($x$ and $y$) directions is very small,
less than 0.1\% of the maximum field.  It appears from
Fig.~\ref{fig:teslafieldfit}, that there is very good agreement
between the fit (line) and the field data (circles) within the
cylinder on the surface of which the fit was performed.

\begin{figure}
\centering
\includegraphics[width=0.8\linewidth]{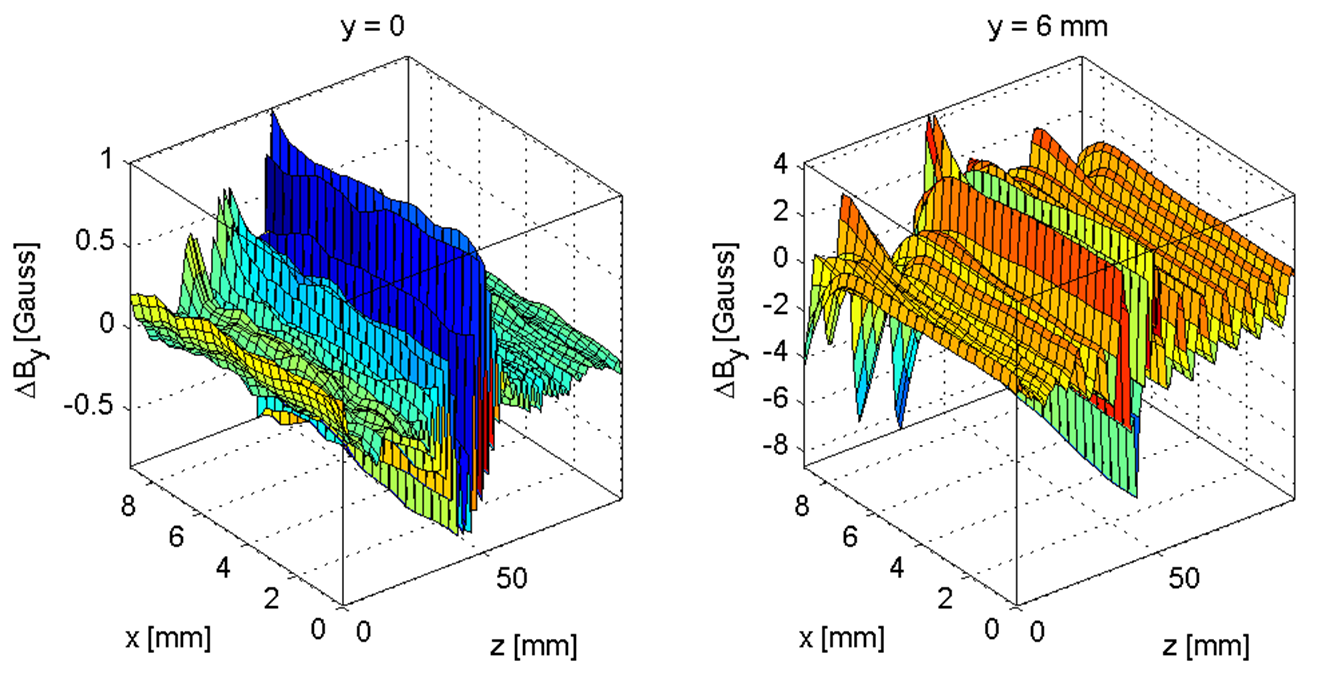}
\caption{Residuals of the fit to the field of one quarter of one
period of the TESLA damping ring wiggler
\label{fig:teslafieldfitresiduals}}
\end{figure}

The quality of the fit can be further illustrated by plotting the
residuals, i.e., the difference between the fitted field and the
field data.  The residuals for the vertical field component on two
horizontal planes, $y=$\,0\,mm and $y=$\,6\,mm are shown in
Fig.~\ref{fig:teslafieldfitresiduals}.  Note that to produced `smooth'
surface plots, we interpolate between the mesh points used in the
computational model.  On the mid-plane of the wiggler, the residuals
are less than 1 gauss (recall that the peak field is 1.6\,T); the
region shown in the left-hand plot in Fig.~\ref{fig:teslafieldfitresiduals}
lies entirely within the surface of the cylinder used in the fit.
On the plane $y=$\,6\,mm, the residuals are somewhat larger, and show
an exponential increase for large values of the horizontal transverse
coordinate: but note that for values of $x$ larger than about 6.7\,mm,
the points in the plot are \emph{outside} the surface of the cylinder
used for the fit.  In the longitudinal direction, the residuals appear
to be dominated by very high frequency modes: this suggests that it may
be possible to reduce the residuals still further by increasing the
number of longitudinal modes used in the fit.  However, this fit was
considered to be of sufficient quality to allow an accurate determination
of the effect of the wiggler on the beam dynamics to be made.

\begin{figure}
\centering
\begin{tabular}{cc}
\includegraphics[width=0.45\linewidth]{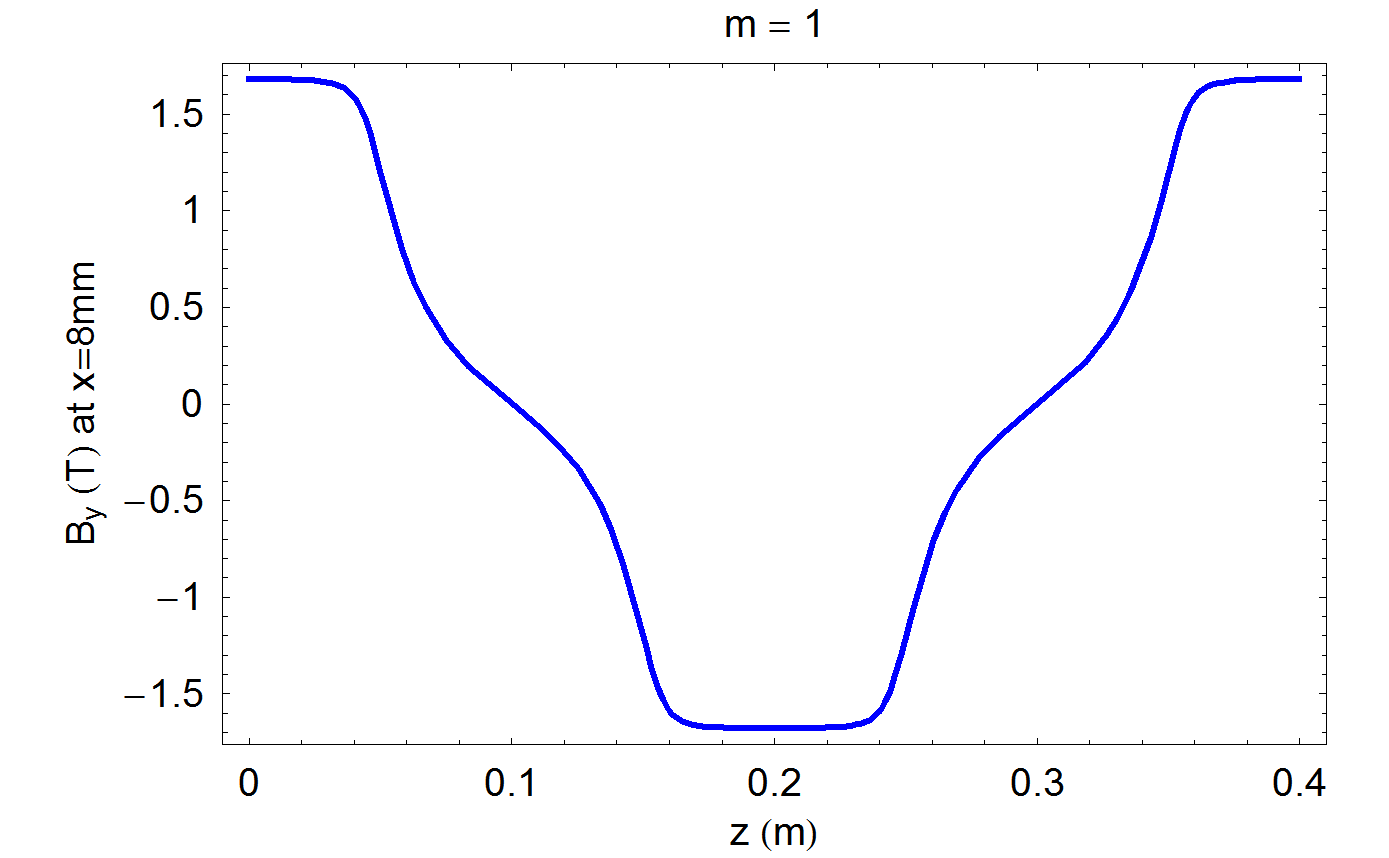} & 
\includegraphics[width=0.45\linewidth]{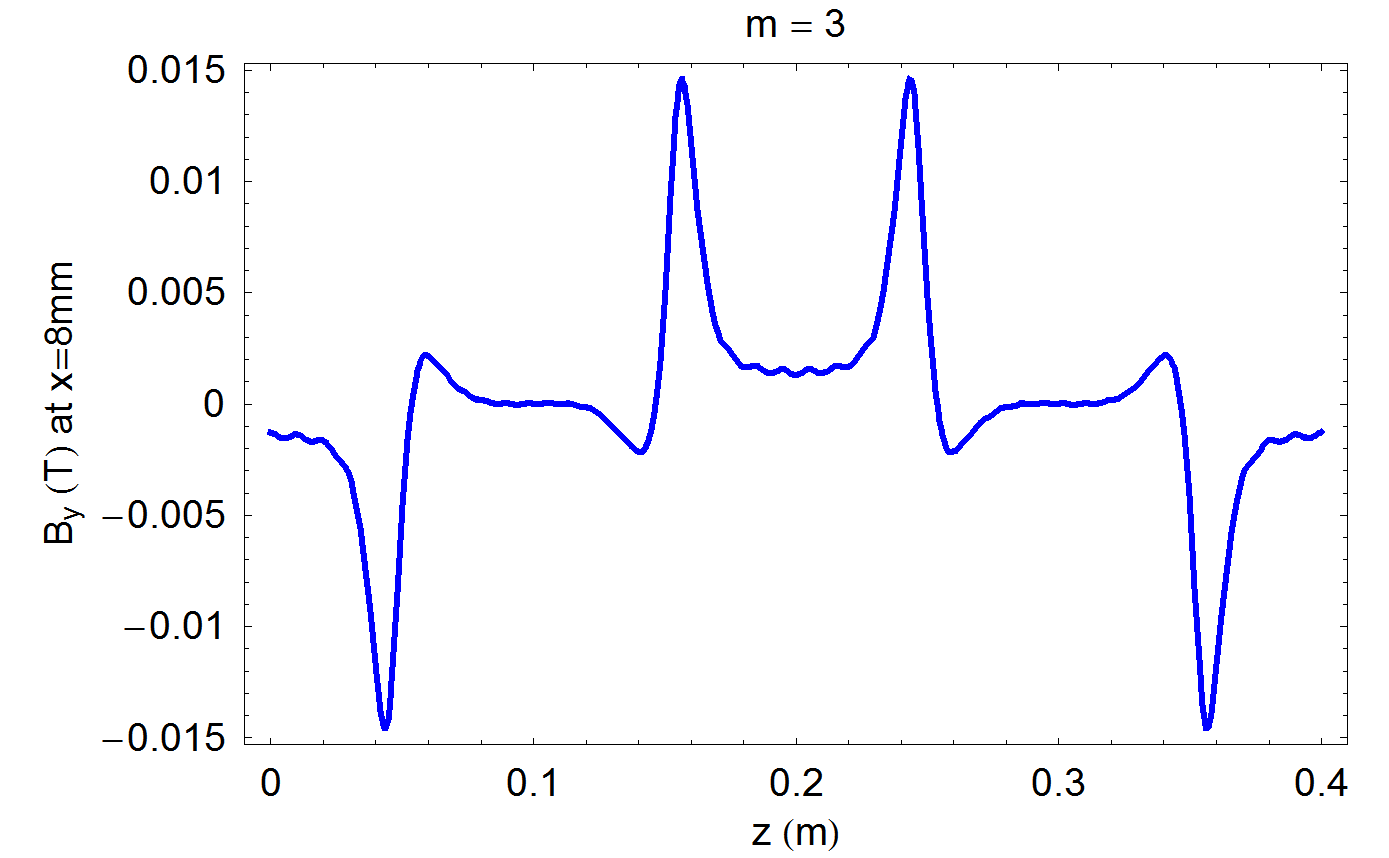} \\
\includegraphics[width=0.45\linewidth]{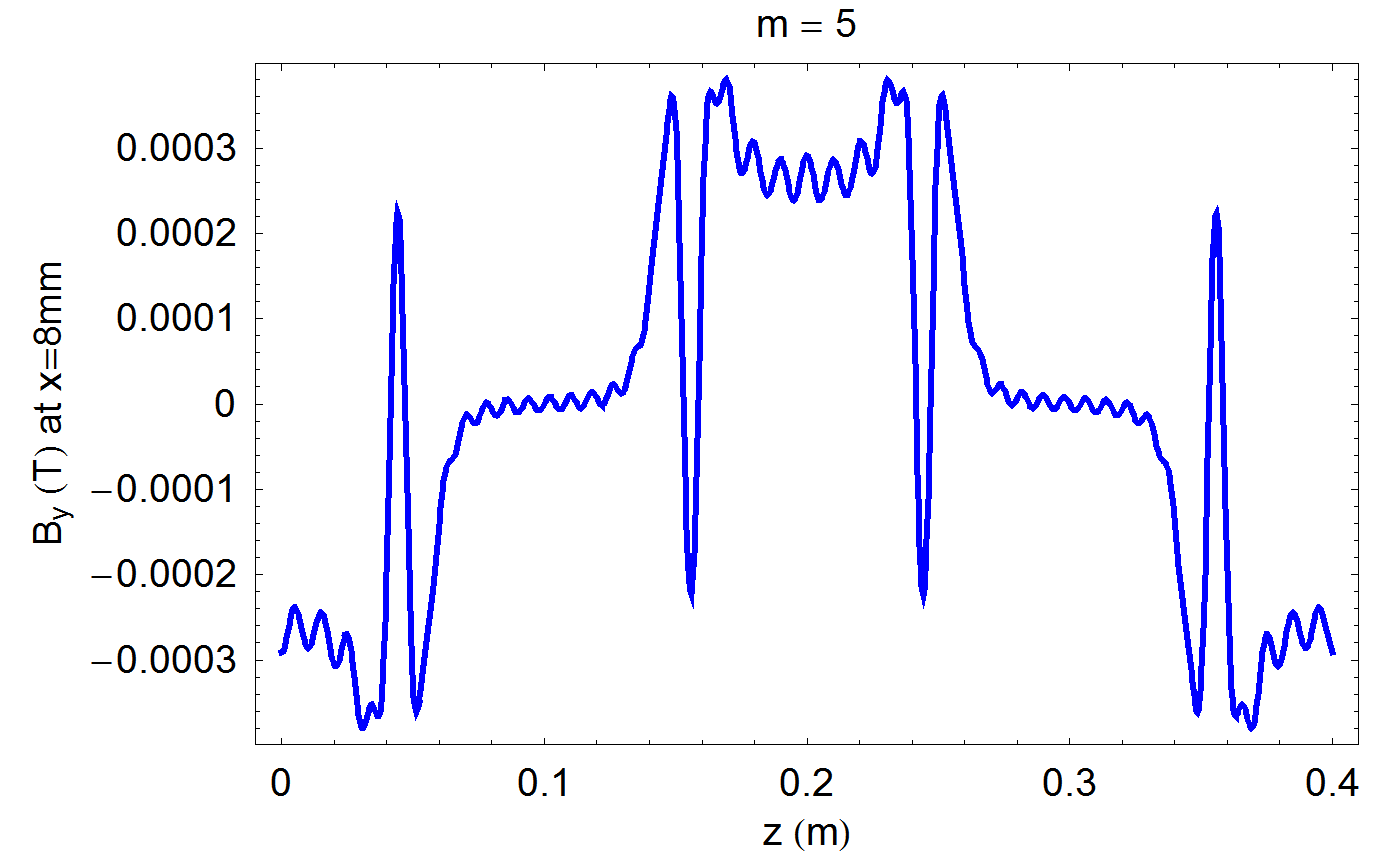} & 
\includegraphics[width=0.45\linewidth]{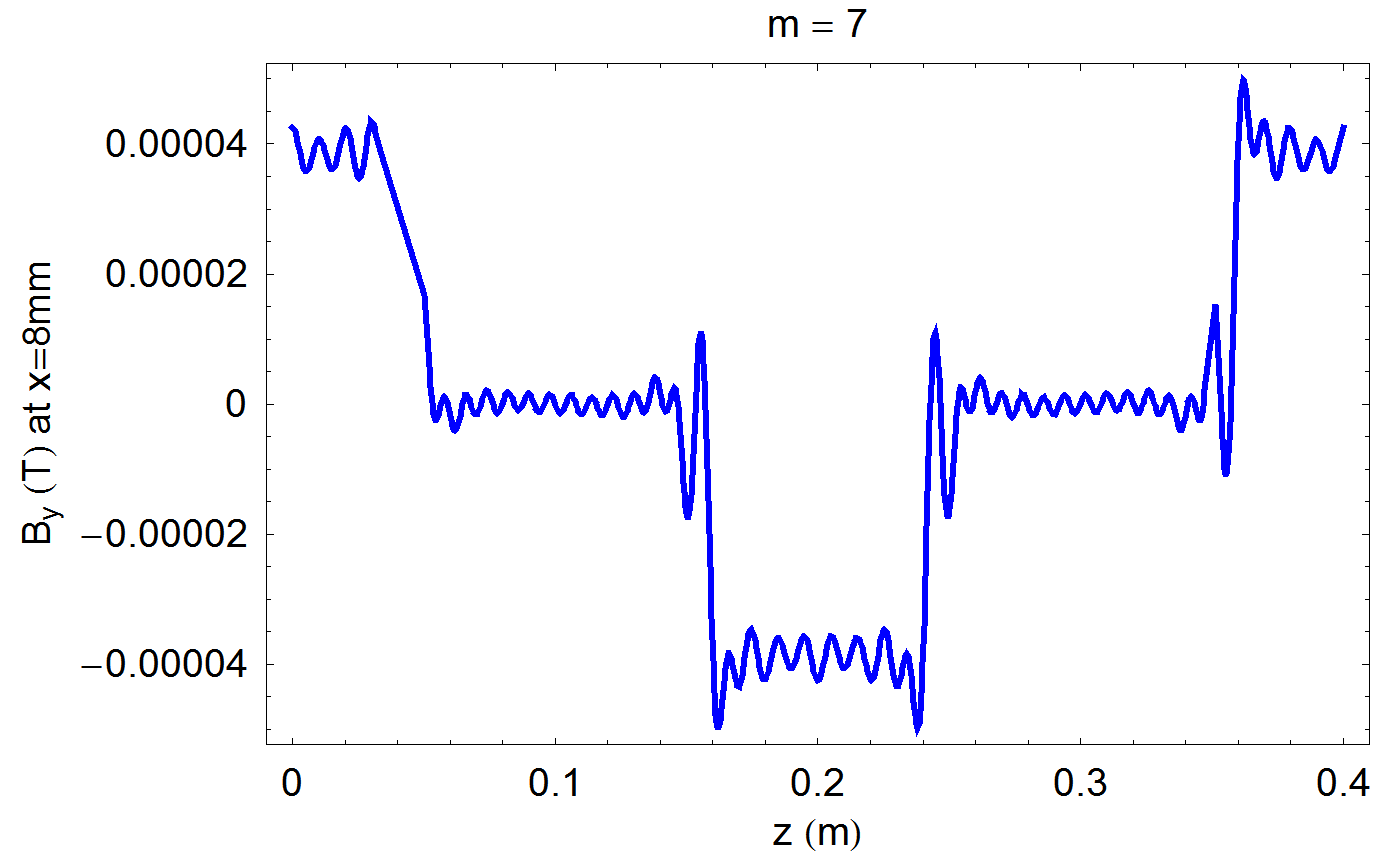}
\end{tabular}
\caption{Multipole contributions to the field in the TESLA damping
wiggler as a function of longitudinal position, at $x=$8\,mm and
$y=$0\,mm, for orders 1 (dipole), 3 (sextupole), 5 (decapole) and 7
\label{fig:teslawigglermultipolecontributions}}
\end{figure}

The tools used for study of the beam dynamics were based on the mode
coefficients determined by the fitting procedure.  Once a fit has been
obtained and shown to be of good quality, then, strictly speaking,
further analysis of the field is not required.  However, it is interesting
to compute, from the mode amplitudes, the contribution to the field in
the wiggler from different `multipole' components, as a function of
longitudinal position.  As described above, the contribution of a
multipole of order $n$ is obtained by a one-dimensional (in the
longitudinal dimension) inverse Fourier transform of the mode
amplitudes $\tilde{B}_n(k_z)$.  To obtain non-zero values for the
contributions from multipoles higher than order $n=$\,1 (dipole), we
need to choose non-zero values for either the $x$ or $y$ coordinates
at which we compute the field.  We choose (arbitrarily) $x=$\,8\,mm,
and $y=$\,0\,mm.  The contributions to the vertical field component
from multipoles of order 1 through 7 are shown in
Fig.~\ref{fig:teslawigglermultipolecontributions}.  Note that
multipoles of even order are forbidden by the symmetry of the wiggler
(see Section~\ref{sec:symmetryallowedforbidden}).  We see from
Fig.~\ref{fig:teslawigglermultipolecontributions} that the
dominant contribution by far is, as expected, the dipole component.
The sextupole component is not insignificant; the contributions of
higher order multipoles are extremely small, and the high-frequency
`oscillation' as a function of longitudinal position is probably
unphysical, and the result of noise in the fitting.

It is worth making a few final remarks about mode decompositions
for three-dimensional fields.  First, as already mentioned, in many
cases a full three-dimensional mode decomposition will not be
necessary.  While this does provide a detailed description of the
field in a form suitable for beam dynamics studies, three-dimensional
decompositions do rely on a large number of accurate and detailed field
measurements.  While such `measurements' may be conveniently
obtained from a model, it may be difficult or impractical to make
such measurements on a real magnet.  Fortunately, in many cases,
a two-dimensional field description in terms of multipoles is
sufficient.  Generally, a three-dimensional analysis only need be
undertaken where there are grounds to believe that the
three-dimensional nature of the field is likely to have a significant
impact on the beam dynamics.

Second, we have already emphasized that to obtain an accurate
description of the field within some region in terms of a mode
decomposition, the mode amplitudes should be determined by a fit
on a surface enclosing the region of interest.  Outside the region
bounded by the surface of the fit, the fitted field can be expected
to diverge exponentially from the real field.  However, in choosing
the surface for the fit, the geometry of the magnet will impose some
constraints.  A magnet with a wide rectangular aperture may lend
itself to a description using a Cartesian basis (fitting on the
surface of a rectangular box); a circular aperture, however, is more
likely to require use of a polar basis (fitting on the surface of
a cylinder with circular cross-section).  Both cases have been
described above.  It may be appropriate in other cases to perform
a fit on the surface of a cylinder with elliptical cross-section.
The basis functions in this case involve Mathieu functions.  For
further details, the reader is referred to work by Dragt~\cite{bib:dragt}
and by Dragt and Mitchell~\cite{bib:mitchell}.

\appendix
\section{The vector potential}

Our analysis of iron-dominated multipole magnets in
Section~\ref{sec:irondominatedmultipoles} was based on the magnetic
scalar potential $\varphi$.  The magnetic flux density can be derived from
a scalar potential
\begin{equation}
\vec{B} = -\gradop \varphi \nonumber
\end{equation}
in the case where the flux density has vanishing divergence and curl:
\begin{equation}
\divop \vec{B} = \curlop \vec{B} = 0. \nonumber
\end{equation}
More generally (in particular, where the flux density has
non-vanishing curl) one derives the magnetic flux density from a vector
potential $\vec{A}$, using
\begin{equation}
\vec{B} = \curlop \vec{A}. \label{eq:bequalcurla}
\end{equation}
Although we have not required the vector potential in our discussion
of Maxwell's equations for accelerator magnets, it is sometimes used
in analysis of beam dynamics.  In particular, descriptions of the
dynamics based on Hamiltonian mechanics generally use the vector
potential rather than the magnetic flux density or the magnetic scalar
potential.  We therefore include here a brief discussion of the
vector potential, paying attention to aspects relevant to the 
descriptions we have developed for two-dimensional and three-dimensional
magnet fields.

First, we note that the divergence of any curl is identically zero:
\begin{equation}
\divop \curlop \vec{V} \equiv 0, \nonumber
\end{equation}
for any differentiable vector field $\vec{V}$.
Thus, if we write $\vec{B} = \curlop \vec{A}$, then Maxwell's equation
(\ref{eq:maxwell2}):
\begin{equation}
\divop \vec{B} = 0, \nonumber
\end{equation}
is automatically satisfied.  Maxwell's equation (\ref{eq:maxwell4})
in uniform media (constant permeability), with zero current and static
electric fields gives
\begin{equation}
\curlop \vec{B} = \mu \vec{J}, \label{eq:curlbequalsmuj}
\end{equation}
where $\vec{J}$ is the current density.  This leads to the equation
for the vector potential:
\begin{equation}
\curlop \curlop \vec{A}
  \equiv \gradop(\divop \vec{A}) - \nabla^2 \vec{A}
  = \mu \vec{J}.
\label{eq:curlcurlaequalsmuj}
\end{equation}
Equation~(\ref{eq:curlcurlaequalsmuj}) is a second-order differential equation
for the vector potential in a medium with permeability $\mu$, where the
current density is $\vec{J}$.  This appears harder to solve than the
first-order differential equation for the magnetic flux density,
Eq.~(\ref{eq:curlbequalsmuj}).  However, Eq.~(\ref{eq:curlcurlaequalsmuj})
may be simplified significantly, if we apply an appropriate \emph{gauge
condition}.  To understand what this means, recall that the magnetic
flux density is given by the curl of the vector potential, and that the
curl of the gradient of any scalar field is identically zero.  Thus, we
can add the gradient of a scalar field to a vector potential, and obtain
a new vector potential that gives the same flux density as the old one.
That is, if
\begin{equation}
\vec{B} = \curlop \vec{A}, \nonumber
\end{equation}
and
\begin{equation}
\vec{A}^\prime = \vec{A} + \gradop \psi, \label{eq:gaugetransformation}
\end{equation}
for an arbitrary differentiable scalar field $\psi$, then
\begin{equation}
\curlop \vec{A}^\prime
  = \curlop \vec{A} + \curlop \gradop \psi
  = \curlop \vec{A} = \vec{B}. \nonumber
\end{equation}
In other words, the vector potential $\vec{A}^\prime$ leads to exactly
the same flux density as the vector potential $\vec{A}$.  Since the
dynamics of a given system are determined by the fields rather than the
potentials, either $\vec{A}^\prime$ or $\vec{A}$ is a valid choice for
the description of the system.  Equation~(\ref{eq:gaugetransformation}) is
known as a \emph{gauge transformation}.  The consequence of having the
freedom to make a gauge transformation means that the vector potential
for any given system is not uniquely defined: given some particular
vector potential, it is always possible to make a gauge tranformation
without any change in the physical observables of a system.  The analogue
in the case of electric fields, of course, is that the `zero' of the
electric scalar potential can be chosen arbitrarily: only \emph{changes}
in potential (i.e., energy) are observable, so given some particular
scalar potential field, it is possible to add a constant (that is,
a quantity independent of position) and obtain a new scalar potential
that gives the same physical observables as the original scalar potential.

For magnetostatic fields, we can use a gauge transformation to simplify
Eq.~(\ref{eq:curlcurlaequalsmuj}).  Suppose we have obtained a vector
potential $\vec{A}$ for some particular physical system.  Define a scalar
field $\psi$, which satisfies
\begin{equation}
\nabla^2 \psi = -\divop \vec{A}. \label{eq:psidef}
\end{equation}
Then define
\begin{equation}
\vec{A}^\prime = \vec{A} + \gradop \psi. \nonumber
\end{equation}
Since $\vec{A}^\prime$ and $\vec{A}$ are related by a gauge
transformation, they lead to the same magnetic flux density, and the
same physical observables for the system.  However, the divergence
of $\vec{A}^\prime$ vanishes:
\begin{equation}
\divop \vec{A}^\prime
  = \divop \vec{A} + \divop \gradop \psi
  = -\nabla^2 \psi + \nabla^2 \psi = 0, \nonumber
\end{equation}
where we have used Eq.~(\ref{eq:psidef}).  Thus, given any vector potential,
we can make a gauge transformation to find a new vector potential that
gives the same magnetic flux density, but has vanishing divergence.
The \emph{gauge condition}
\begin{equation}
\divop \vec{A} = 0 \label{eq:coulombgaugecondition}
\end{equation}
is known as the \emph{Coulomb gauge}.  It is possible to work with other
gauge conditions (for example, for time-dependent electromagnetic fields
the Lorenz gauge condition is often more appropriate); however, for our
present purposes, the Coulomb gauge leads to a simplification of
Eq.~(\ref{eq:curlcurlaequalsmuj}), which now becomes
\begin{equation}
\nabla^2 \vec{A} = - \mu \vec{J}. \label{eq:delsquaredaequalsminusmuj}
\end{equation}
Equation~(\ref{eq:delsquaredaequalsminusmuj}) is Poisson's equation for a
vector field.  Note that despite being a second-order differential equation,
it is in a sense simpler than Maxwell's equation (\ref{eq:maxwell4}), since
we have `decoupled' the components of the vectors; that is, we have a set
of three uncoupled second-order differential equations, where each equation
relates a component of the vector potential to the corresponding component
of the current density.  Equation~(\ref{eq:delsquaredaequalsminusmuj}) has the
solution
\begin{equation}
\vec{A}(\vec{r}) = -\frac{\mu}{4\pi}
  \int \frac{\vec{J}(\vec{r}^{\,\prime})}{|\vec{r} - \vec{r}^{\,\prime}|} \, d^3 r^\prime.
  \nonumber
\end{equation}
In this form, we see that the potential at a point in space is inversely
proportional to the distance from the source.

Now, consider the potential given by
\begin{equation}
A_x = 0, \quad A_y = 0, \quad A_z = -\textrm{Re}\, \frac{C_n(x+iy)^n}{n}.
\label{eq:multipolevectorpotential}
\end{equation}
Taking derivatives, we find that
\begin{eqnarray}
\frac{\partial A_z}{\partial x} & = & -\textrm{Re}\, C_n(x+iy)^{n-1}, \nonumber \\
\frac{\partial A_z}{\partial y} & = & \textrm{Im}\, C_n(x+iy)^{n-1}. \nonumber
\end{eqnarray}
Then, since $A_x$ and $A_y$ are zero, we have
\begin{equation}
\vec{B} = \curlop \vec{A}
        = \left( \frac{\partial A_z}{\partial y},
                    -\frac{\partial A_z}{\partial x}, 0 \right). \nonumber \\
\end{equation}
Hence
\begin{equation}
B_y + i B_x = C_n (x + iy)^{n-1},
\end{equation}
which is just the multipole field.  Thus
Eq.~(\ref{eq:multipolevectorpotential}) is a potential that gives a
multipole field.  Note also that, since $A_z$ is independent of $z$,
this potential satisfies the Coulomb gauge condition
(\ref{eq:coulombgaugecondition}):
\begin{equation}
\divop \vec{A} = 
\frac{\partial A_x}{\partial x} + 
\frac{\partial A_y}{\partial y} + 
\frac{\partial A_z}{\partial z} = 0. \nonumber
\end{equation}

An advantage of working with the vector potential in the Coulomb gauge
is that, for multipole fields, the transverse components of the vector
potential are both zero.  This simplifies, to some extent, the Hamiltonian
equations of motion for a particle moving through a multipole field.
However, note that the longitudinal component $B_z$ of the magnetic flux
density is zero in this case.  To generate a solenoidal field, with
$B_z$ equal to a non-zero constant, we need to introduce non-zero
components for $A_x$, or $A_y$, or both.  For example, a solenoid
field with flux density $B_\textrm{sol}$ may be derived from the
vector potential:
\begin{equation}
A_x = -\frac{1}{2} B_\textrm{sol} y, \quad
A_y =  \frac{1}{2} B_\textrm{sol} x. \nonumber
\end{equation}

\begin{figure}
\centering
\includegraphics[width=0.55\linewidth]{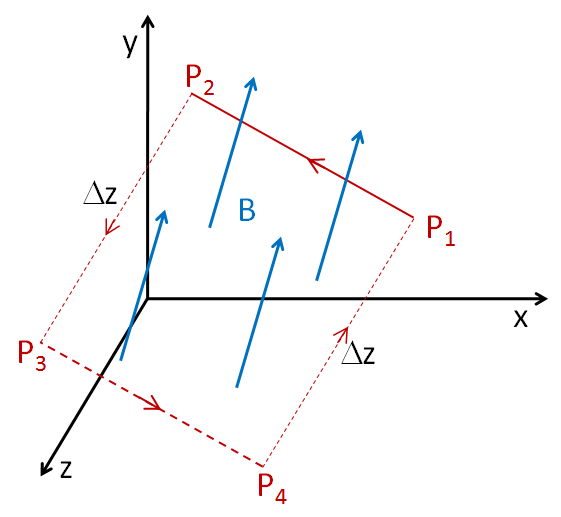} 
\caption{Interpretation of the vector potential in a two-dimensional
magnetic field (i.e., a field that is independent of $z$).  The change
in the vector potential between $P_1$ and $P_2$ is equal to the flux
of the magnetic field through the loop $P_1P_2P_3P_4$, divided by
$\Delta z$.
\label{fig:vectorpotentialinterp}}
\end{figure}

Let us return for a moment to the case of multipole fields.  If we work in
a gauge in which the transverse components of the vector potential are
both zero, then the field components are given by
\begin{equation}
B_y = -\frac{\partial A_z}{\partial x}, \quad
B_x =  \frac{\partial A_z}{\partial y}. \nonumber
\end{equation}
From these expressions, we see that if we take any two points with the
same $y$ coordinate, then the difference in the vector potential between
these two points is given by the `flux' passing through a line between
these points:
\begin{equation}
\Delta A_z = -\int B_y \, dx. \nonumber
\end{equation}
Similarly for any two points with the same $x$ coordinate
\begin{equation}
\Delta A_z = \int B_x \, dy. \nonumber
\end{equation}
In general, for a field that is independent of $z$, and working in a
gauge where $A_x = A_y = 0$, we can write
\begin{equation}
\Delta A_z = \frac{\Delta \Phi}{\Delta z}, \label{eq:twodflux}
\end{equation}
where $\Delta A_z$ is the change in the vector potential between two 
points $P_1$ and $P_2$ in a given plane $z = z_0$; and $\Delta \Phi$ is
the magnetic flux through a rectangular `loop' with vertices $P_1$,
$P_2$, $P_3$ and $P_4$: see Fig.~\ref{fig:vectorpotentialinterp}.
$P_3$ and $P_4$ are points obtained by transporting $P_1$ and $P_2$
a distance $\Delta z$ parallel to the $z$ axis.
Equation~(\ref{eq:twodflux}) can also be obtained by applying Stokes's
theorem to the loop $P_1P_2P_3P_4$, with the relationship
(\ref{eq:bequalcurla}) between $\vec{B}$ and $\vec{A}$:
\begin{equation}
\int \vec{A} \cdot d\vec{l} = \int \curlop \vec{A} \cdot d\vec{S}
= \int \vec{B} \cdot d\vec{S}, \nonumber
\end{equation}
hence
\begin{equation}
A_z(P_2)\Delta z - A_z(P_1)\Delta z = \Delta \Phi. \nonumber
\end{equation}

Finally, we give the vector potentials corresponding to three-dimensional
fields.  In the Cartesian basis, with the field given by 
Eqs.~(\ref{eq:threedcartnbx})--(\ref{eq:threedcartnbz}), a possible vector
potential (in the Coulomb gauge) is
\begin{eqnarray}
A_x & = & 0, \nonumber \\
A_y & = & B_0 \frac{k_z}{k_x k_y} \sin (k_x x) \sinh (k_y y) \cos (k_z z), \nonumber \\
A_z & = &-B_0 \frac{1}{k_x} \sin (k_x x) \cosh (k_y y) \sin (k_z z). \nonumber
\end{eqnarray}
In the polar basis, with the field given by Eqs.~(\ref{eq:threedmodepolarbr})--(\ref{eq:threedmodepolarbz}), a possible
vector potential is
\begin{eqnarray}
A_r      & = & -\frac{r}{m} I_m\!(k_z r)\cos (n\theta) \sin (k_z z), \nonumber \\
A_\theta & = & 0, \nonumber \\
A_z      & = & -\frac{r}{2m} I_m^\prime\!(k_z r)\cos (n\theta) \sin (k_z z). \nonumber
\end{eqnarray}
However, note that this potential does not satisfy the Coulomb gauge
condition.

\end{document}